\documentclass{aa}  

\usepackage{graphicx}
\usepackage{graphics,psfrag}
\usepackage{graphicx,psfrag}
\usepackage{morefloats}
\usepackage{txfonts}
\usepackage{tipa}
\usepackage{amsmath}
\usepackage{mathrsfs}
\usepackage{natbib,twoopt}
\usepackage[breaklinks=true]{hyperref} 
\bibpunct{(}{)}{;}{a}{}{,}             
\makeatletter
  \newcommandtwoopt{\citeads}[3][][]{\href{http://adsabs.harvard.edu/abs/#3}%
    {\def\hyper@linkstart##1##2{}%
     \let\hyper@linkend\@empty\citealp[#1][#2]{#3}}}
  \newcommandtwoopt{\citepads}[3][][]{\href{http://adsabs.harvard.edu/abs/#3}%
    {\def\hyper@linkstart##1##2{}%
     \let\hyper@linkend\@empty\citep[#1][#2]{#3}}}
  \newcommandtwoopt{\citetads}[3][][]{\href{http://adsabs.harvard.edu/abs/#3}%
    {\def\hyper@linkstart##1##2{}%
     \let\hyper@linkend\@empty\citet[#1][#2]{#3}}}
  \newcommandtwoopt{\citeyearads}[3][][]%
    {\href{http://adsabs.harvard.edu/abs/#3}
    {\def\hyper@linkstart##1##2{}%
     \let\hyper@linkend\@empty\citeyear[#1][#2]{#3}}}
\makeatother
\begin{document}


\title{Statistical properties of Fourier-based time-lag estimates}

   \author{A. Epitropakis
          \inst{1}
          \and
          I. E. Papadakis
          \inst{1,2}
          }

\institute{Department of Physics and Institute of Theoretical and Computational Physics, University of Crete, 71003 Heraklion, Greece
\and
IESL, Foundation for Research and Technology-Hellas, GR-71110 Heraklion, Crete, Greece}
\authorrunning{A. Epitropakis \& I. E. Papadakis}
\titlerunning{Statistical properties of Fourier-based time-lag estimates}
\date{Received .. ...... 2015; accepted .. ...... 2016}

 
  \abstract
   {The study of X-ray time-lag spectra in active galactic nuclei (AGN) is currently an active research area, since it has the potential to illuminate the physics and geometry of the innermost region (i.e. close to the putative super-massive black hole) in these objects. To obtain reliable information from these studies, the statistical properties of time-lags estimated from data must be known as accurately as possible.}
   {We investigated the statistical properties of Fourier-based time-lag estimates (i.e. based on the cross-periodogram), using evenly sampled time series with no missing points. Our aim is to provide practical `guidelines' on estimating time-lags that are minimally biased (i.e. whose mean is close to their intrinsic value) and have known errors.}
   {Our investigation is based on both analytical work and extensive numerical simulations. The latter consisted of generating artificial time series with various signal-to-noise ratios and sampling patterns/durations similar to those offered by AGN observations with present and past X-ray satellites. We also considered a range of different model time-lag spectra commonly assumed in X-ray analyses of compact accreting systems.}
   {Discrete sampling, binning and finite light curve duration cause the mean of the time-lag estimates to have a smaller magnitude than their intrinsic values. Smoothing (i.e. binning over consecutive frequencies) of the cross-periodogram can add extra bias at low frequencies. The use of light curves with low signal-to-noise ratio reduces the intrinsic coherence, and can introduce a bias to the sample coherence, time-lag estimates, and their predicted error.}
   {Our results have direct implications for X-ray time-lag studies in AGN, but can also be applied to similar studies in other research fields. We find that: a) time-lags should be estimated at frequencies lower than $\approx1/2$ the Nyquist frequency to minimise the effects of discrete binning of the observed time series; b) smoothing of the cross-periodogram should be avoided, as this may introduce significant bias to the time-lag estimates, which can be taken into account by assuming a model cross-spectrum (and not just a model time-lag spectrum); c) time-lags should be estimated by dividing observed time series into a number, say $m$, of shorter data segments and averaging the resulting cross-periodograms; d) if the data segments have a duration $\gtrsim20\,\mathrm{ks}$, the time-lag bias is $\lesssim15\%$ of its intrinsic value for the model cross-spectra and power-spectra considered in this work. This bias should be estimated in practise (by considering possible intrinsic cross-spectra that may be applicable to the time-lag spectra at hand) to assess the reliability of any time-lag analysis; e) the effects of experimental noise can be minimised by only estimating time-lags in the frequency range where the sample coherence is larger than $1.2/(1+0.2m)$. In this range, the amplitude of noise variations caused by measurement errors is smaller than the amplitude of the signal's intrinsic variations. As long as $m\gtrsim20$, time-lags estimated by averaging over individual data segments have analytical error estimates that are within 95\% of the true scatter around their mean, and their distribution is similar, albeit not identical, to a Gaussian.}


   \maketitle
%

\section{Introduction} \label{sec1}

The study of time-lags as a function of temporal frequency between X-ray light curves in different energy bands has frequently been used in the last two decades to probe the emission mechanism and geometry of the innermost region in active galactic nuclei \citep[AGN; e.g.][]{2001ApJ...554L.133P,2004MNRAS.348..783M,2006MNRAS.372..401A,2008MNRAS.388..211A,2009ApJ...700.1042S}  and X-ray binaries \citep[XRBs; e.g.][]{1989Natur.342..773M,1996MNRAS.280..227N,1999ApJ...510..874N}. In the last few years, time-lag studies have revealed that `soft' X-ray variations in AGN are delayed with respect to `hard' X-ray variations at frequencies higher than $\approx10^{-4}\,\mathrm{Hz}$ \citep[e.g.][]{2009Natur.459..540F,2010MNRAS.401.2419Z,2011MNRAS.412...59Z,2011MNRAS.416L..94E,2013MNRAS.431.2441D}. These time-lags are commonly referred to as soft lags, and have been observed in $\approx20$ sources so far. They have attracted considerable attention, since they are thought to be direct evidence of X-ray reverberation. Further credence to such a scenario came with recent discoveries of time-lags between the Fe K$\alpha$ emission line ($\approx5-7\,\mathrm{keV}$) and the X-ray continuum \citep[e.g.][]{2012MNRAS.422..129Z,2013MNRAS.428.2795K,2013MNRAS.430.1408K,2013MNRAS.434.1129K,2013ApJ...767..121Z,2014MNRAS.440.2347M}, as well as between the so-called Compton hump ($\approx10-30\,\mathrm{keV}$) and the X-ray continuum \citep[e.g.][]{2014ApJ...789...56Z,2015MNRAS.446..737K} in a few AGN.

The study of X-ray reverberation time-lags can provide valuable geometrical and physical information on the X-ray source and reflector, since they should depend on, for example, their typical size, relative distance, proximity to the central black hole (BH), the mass and spin of the BH, and the inclination of the system. To obtain this information, the statistical properties of time-lags estimated from observed light curves must be known as accurately as possible. For example, one must know their bias (i.e. how accurately their mean approximates the intrinsic time-lag spectrum), error, and probability distribution. The later is necessary if one wishes to fit the observed time-lag spectra with theoretical models. To the best of our knowledge, such a detailed investigation has not yet been performed.

Results from preliminary studies along these lines have been presented by \citet{2013MNRAS.435.1511A} and \citet{2014A&ARv..22...72U}. In this paper we report the results of a more detailed study regarding the statistical properties of Fourier-based time-lag estimates, based on both analytical work and extensive numerical simulations. Our primary goals are: a) to investigate whether the frequently used time-lag estimates are indeed reliable estimates of the intrinsic time-lag spectrum, b) to study the effects of light curve sampling patterns and duration, as well as measurement errors, on the statistical properties of these estimates, and c) to provide practical guidelines which ensure estimates that are minimally biased, have know errors, and follow a Gaussian distribution. The latter property would be desirable, for example, in the case of model fitting using traditional $\chi^2$ minimisation techniques. Our work should have direct impact on current time-lag studies in the area of AGN X-ray timing analyses. We believe that our results can apply equally well to all scientific fields where similar techniques are employed to search for delays between two observed time-varying signals.

\section{Definitions} \label{sec2}

Consider a continuous bivariate random time series $\{\mathscr{X}(t),\mathscr{Y}(t)\}$. We assume that the mean values ($\mu_\mathscr{X}$ and $\mu_\mathscr{Y}$), as well as the auto-covariance functions ($R_\mathscr{X}(\tau)$ and $R_\mathscr{Y}(\tau)$; $\tau$ is the so-called lag) of the individual time series are finite and time-independent (i.e. they are stationary random processes). A random function that is frequently used to quantify the correlation between two time series in the time domain is the cross-covariance function (CCF),
\noindent
\begin{equation} \label{eq1}
R_{\mathscr{XY}}(\tau)\equiv\mathrm{E}\{[\mathscr{Y}(t)-\mu_\mathscr{Y}][\mathscr{X}(t+\tau)-\mu_\mathscr{X}]\},
\end{equation}
\noindent
where $\mathrm{E}$ is the expectation operator. We assume that the CCF depends on $\tau$ only, i.e. the CCF does not vary with time. Defined as above, $R_{\mathscr{XY}}(\tau)$ is the CCF with $\mathscr{Y}(t)$ leading $\mathscr{X}(t)$. The Fourier transform of the CCF,
\noindent
\begin{equation} \label{eq2}
\noindent
h_{\mathscr{XY}}(\nu)\equiv\int_{-\infty}^{\infty}R_{\mathscr{XY}}(\tau)\mathrm{e}^{-\mathrm{i}2\pi\nu\tau}\mathrm{d}\tau,
\end{equation}
\noindent
defines the cross-spectrum (CS) of the bivariate processes. The CCF is not necessarily symmetric about $\tau=0$, and hence the CS is generally a complex number that can be written as
\noindent
\begin{equation} \label{eq3}
h_{\mathscr{XY}}(\nu)=c_{\mathscr{XY}}(\nu)-\mathrm{i}q_{\mathscr{XY}}(\nu)=|h_{\mathscr{XY}}(\nu)|\mathrm{e}^{\mathrm{i}\phi_{\mathscr{XY}}(\nu)}.
\end{equation}
\noindent
The real functions $c_{\mathscr{XY}}(\nu)$ and $\{-q_{\mathscr{XY}}(\nu)\}$ represent the real, $\Re[h_{\mathscr{XY}}(\nu)]$, and imaginary, $\Im[h_{\mathscr{XY}}(\nu)]$, parts of the CS, respectively. The function $c_{\mathscr{XY}}(\nu)$ is an even function of $\nu$, while $q_{\mathscr{XY}}(\nu)$ is an odd function of $\nu$ \citep[][P81 hereafter]{Priestley:81}. The quantity $|h_{\mathscr{XY}}(\nu)|$ is the CS amplitude, and $\phi_{\mathscr{XY}}(\nu)$ is the phase-lag spectrum, which is defined as
\noindent
\begin{equation} \label{eq4}
\phi_{\mathscr{XY}}(\nu)\equiv\mathrm{arg}[h_{\mathscr{XY}}(\nu)]=\mathrm{arctan}\left[-\frac{q_{\mathscr{XY}}(\nu)}{c_{\mathscr{XY}}(\nu)}\right].
\end{equation}
\noindent
The phase-lag spectrum represents the average phase shift between sinusoidal components of the two time series with frequency $\nu$. Since it is defined modulo $2\pi$, it is customary to define its principal value in the interval $(-\pi,\pi]$, which is the convention adopted in the present work as well. The time-lag spectrum is defined as
\noindent
\begin{equation} \label{eq5}
\tau_{\mathscr{XY}}(\nu)\equiv\frac{\phi_{\mathscr{XY}}(\nu)}{2\pi\nu},
\end{equation}
\noindent
and represents the average temporal delay between sinusoidal components of the two time series with frequency $\nu$. Given the definition of the CCF by Eq.\,\ref{eq1}, a positive time-lag value at a frequency $\nu$ indicates that, on average, the sinusoidal component of $\mathscr{X}(t)$ lags behind the respective component of $\mathscr{Y}(t)$.

Another statistic that is often used to study correlations between two random processes in Fourier space is the so-called coherence function,
\noindent
\begin{equation} \label{eq6}
\gamma_{\mathscr{XY}}^2(\nu)\equiv\frac{|h_{\mathscr{XY}}(\nu)|^2}{h_\mathscr{X}(\nu)h_\mathscr{Y}(\nu)},
\end{equation}
\noindent
where $h_\mathscr{X}(\nu)$ and $h_\mathscr{Y}(\nu)$ denote the power-spectral density functions (PSDs) of the time series $\mathscr{X}(t)$ and $\mathscr{Y}(t)$, respectively (the PSD is defined as the Fourier transform of the auto-covariance function of a random process). The coherence function is interpreted as the correlation coefficient between two processes at frequency $\nu$, as it measures the degree of linear correlation between them at each frequency. As with ordinary correlation coefficients, $0\le\gamma^2_{\mathscr{XY}}(\nu)\le1$, where $\gamma^2_{\mathscr{XY}}(\nu)=1$ indicates a perfect (linear) correlation, and $\gamma^2_{\mathscr{XY}}(\nu)=0$ implies the absence of any correlation at frequency $\nu$.

\subsection{The effects of binning and discrete sampling} \label{subsec21}

In practice, the data correspond to values of a single realisation of a random process that are recorded over a finite time interval, $T$ (the duration). The recording is typically performed at regular time intervals, $\Delta t_{\mathrm{sam}}$ (the sampling period).

Consider a pair of observed time series $\{x(t_r),y(t_r)\}$, where $t_r=r\Delta t_{\mathrm{sam}}$, $r=1,2,\ldots,N$, and $N$ is the total number of points. Let us denote by $\mathscr{X}(t)$ and $\mathscr{Y}(t)$ the intrinsic, continuous time series, which we assume are stationary random processes (henceforth, the term intrinsic will always refer to $\{\mathscr{X}(t),\mathscr{Y}(t)\}$). The observed time series $\{x(t_r),y(t_r)\}$ correspond to a particular, finite realisation of a discrete version of the intrinsic process, which we denote by $\{X(t_s),Y(t_s)\}$, where $t_s=s\Delta t_{\mathrm{sam}}$ and $s=0,\pm1,\pm2,\ldots$. In addition, light curves\footnote{We will henceforth use the terms light curves and time series interchangeably, since the former is used in astronomy to denote flux observations as a function of time.} in astronomy are the result of binning the intrinsic signal over time bins of size $\Delta t_{\mathrm{bin}}$.\footnote{In this work we considered the case of evenly sampled light curves for which the sampling period, $\Delta t_{\mathrm{sam}}$, is an integer multiple of the time bin size, $\Delta t_{\mathrm{bin}}$.} In this case, the relation between $X(t_s)$ and $\mathscr{X}(t)$ is given by
\noindent
\begin{equation} \label{eq7}
X(t_s)=\frac{1}{\Delta t_{\mathrm{bin}}}\int_{t_s-(\Delta t_{\mathrm{bin}}/2)}^{t_s+(\Delta t_{\mathrm{bin}}/2)}\mathscr{X}(t)\mathrm{d}t,
\end{equation}
\noindent
with an identical relation holding between $Y(t_s)$ and $\mathscr{Y}(t)$.

It is rather straight-forward to show that $\{X(t_s),Y(t_s)\}$ and $\{\mathscr{X}(t),\mathscr{Y}(t)\}$ have the same mean values, but different CCFs and CS. As we show in Appendix \ref{appa}, the CS of the discrete process, $h_{XY}(\nu)$, which is defined only at frequencies $|\nu|\le1/2\Delta t_{\mathrm{sam}}$, is given by 
\noindent
\begin{equation} \label{eq8}
h_{XY}(\nu)=\sum_{k=-\infty}^{\infty}h_{\mathscr{XY}}\left(\nu+\frac{k}{\Delta t_{\mathrm{sam}}}\right)\mathrm{sinc}^2\left[\pi\left(\nu+\frac{k}{\Delta t_{\mathrm{sam}}}\right)\Delta t_{\mathrm{bin}}\right].
\end{equation}
\noindent
In other words, at each frequency $\nu$ within the aforementioned interval, $h_{XY}(\nu)$ is the superposition of the intrinsic CS values, $h_{\mathscr{XY}}(\nu)$, at frequencies $\nu,\nu\pm(1/\Delta t_{\mathrm{sam}}),\nu\pm(2/\Delta t_{\mathrm{sam}}),\ldots$. This is entirely analogous to the so-called aliasing effect in the case of the PSD. However, although PSDs are always positive, hence aliasing in this case always implies transfer of `power' from higher to the sampled frequencies, the effects of aliasing on $h_{XY}(\nu)$ cannot be predicted without a priori knowledge of $h_{\mathscr{XY}}(\nu)$. Aliasing affects both $c_{XY}(\nu)$ and $q_{XY}(\nu)$, which are not necessarily positive at all frequencies. The situation is further complicated by the fact that while $c_{XY}(\nu)$ is an even function of $\nu$, $q_{XY}(\nu)$ is an odd function of $\nu$. Therefore, aliasing should affect these two functions in different ways.

Equation\,\ref{eq8} shows that aliasing is reduced if the light curves are binned, since in this case the values of $h_{\mathscr{XY}}(\nu)$ that are aliased in the sampled frequency range are suppressed by the $\mathrm{sinc}^2$ function, whose argument depends on the bin size, $\Delta t_{\mathrm{bin}}$. Since the $\mathrm{sinc}^2$ function asymptotically approaches unity in the limit whereby its argument goes to zero, if the observed time series are not binned (i.e. when $\Delta t_{\mathrm{bin}}\rightarrow0$) the $\mathrm{sinc}^2$ term in the right-hand side of Eq.\,\ref{eq8} vanishes, and hence the effects of aliasing on $h_{XY}(\nu)$ are maximised.

\section{Estimation of cross-spectra; the cross-periodogram} \label{sec3}

The discrete Fourier transform (DFT) of a time series $x(t_r)$ is defined as 
\noindent
\begin{equation} \label{eq9}
\zeta_x(\nu) \equiv\sqrt{\frac{\Delta t_{\mathrm{sam}}}{N}}\sum_{r=1}^{N}\left[x(t_r)-\overline{x}\right]\mathrm{e}^{-\mathrm{i}2\pi\nu r\Delta t_{\mathrm{sam}}}, 
\end{equation}
\noindent
where $\overline{x}$ is the sample mean of $x(t_r)$. It is customary to estimate DFTs at the following set of frequencies: $\nu_p=p/N\Delta t_{\mathrm{sam}}$, where $p=1,2,\ldots,N/2$, and $\nu_{\mathrm{Nyq}}=1/2\Delta t_{\mathrm{sam}}$ is the so-called Nyquist frequency, which corresponds to the highest frequency that can be probed for a given sampling period, $\Delta t_{\mathrm{sam}}$ (the symbol $\nu_p$ will henceforth always stand for this particular set of frequencies). The cross-periodogram of two time series is defined as
\noindent
\begin{equation} \label{eq10}
I_{xy}(\nu_p)\equiv\zeta_x(\nu_p)\zeta_y^{*}(\nu_p),
\end{equation}
\noindent
where the asterisk $^{*}$ denotes complex conjugation. The cross-periodogram is used in practice as an estimator of the CS. Given Eqs.\,\ref{eq4} and \ref{eq5}, it seems reasonable to use the real and imaginary parts of $I_{xy}(\nu_p)$ and accept
\noindent
\begin{align}
\label{eq11}
\hat{\phi}_{xy}(\nu_p) &\equiv\mathrm{arctan}\left\{\frac{\Im[I_{xy}(\nu_p)]}{\Re[I_{xy}(\nu_p)]}\right\},\hspace{0.1cm}\rm{and} \\
\label{eq12}
\hat{\tau}_{xy}(\nu_p) &\equiv\frac{\hat{\phi}_{xy}(\nu_p)}{2\pi\nu_p},
\end{align}
\noindent
as estimators of the phase- and time-lag spectrum, respectively. 

\subsection{The bias of cross-spectral estimators} \label{subsec31}

As we show in Appendix \ref{appb},
\noindent
\begin{equation} \label{eq13}
\mathrm{E}[I_{xy}(\nu_p)]=\int_{-\nu_{\mathrm{Nyq}}}^{\nu_{\mathrm{Nyq}}}h_{XY}(\nu')F_N(\nu'-\nu_p)\mathrm{d}\nu',
\end{equation}
\noindent
where $h_{XY}(\nu)$ is the CS of $\{X(t_s),Y(t_s)\}$ (as defined by Eq.\,\ref{eq8}). The function $F_N(\nu'-\nu_p)$, which is called the Fej\'{e}r kernel, has a large peak at $\nu'=\nu_p$ of magnitude $N\Delta t_{\mathrm{sam}}$, and decays to zero as $|\nu'|\rightarrow\infty$. Subsequent peaks appear at $\nu'\approx\nu_p\pm3/2\Delta t_{\mathrm{sam}},\nu_p\pm5/2\Delta t_{\mathrm{sam}},\ldots$, while the zeros of the function occur at $\nu'=\nu_p\pm1/N\Delta t_{\mathrm{sam}},\nu_p\pm2/N\Delta t_{\mathrm{sam}},\ldots$. As $N\rightarrow\infty$, $F_N(\nu'-\nu_p)\rightarrow\delta(\nu'-\nu_p)$ (the Dirac $\delta$-function). Therefore, in the limit $N\rightarrow\infty$ the right-hand side of Eq.\,\ref{eq13} converges to $h_{XY}(\nu_p)$. The cross-periodogram is therefore an asymptotically (when $N\rightarrow\infty$) unbiased estimate of the modified (owing to the effects of binning and discrete sampling) intrinsic CS.

Since $h_{XY}(\nu)$ is {not equal to $h_{\mathscr{XY}}(\nu)$, it follows that $\phi_{XY}(\nu)$ ($\tau_{XY}(\nu)$) will generally not be equal to $\phi_{\mathscr{XY}}(\nu)$ ($\tau_{\mathscr{XY}}(\nu)$) either. Even if aliasing and binning effects are minimal (so that we can assume that $\phi_{XY}(\nu)\approx\phi_{\mathscr{XY}}(\nu)$), in practice it is difficult to predict whether the duration $T=N\Delta t_{\mathrm{sam}}$ of a given pair of observed time series will be sufficiently long for the effects of the convolution of $h_{XY}(\nu)$ with the Fej\'{e}r kernel (Eq.\,\ref{eq13}) to be significant of not, unless $h_{XY}(\nu)$ is known a priori. We define the `bias' of the cross-periodogram as $b_I(\nu_p)\equiv \mathrm{E}[I_{xy}(\nu_p)]-h_{\mathscr{XY}}(\nu_p)$ (henceforth, the term bias for a statistical estimator will always refer to the difference between its mean and intrinsic value). Even if $b_I(\nu_p)$ was known, and hence could be used to `correct' $I_{xy}(\nu_p)$, it would still not be straightforward to determine the bias of $\hat{\phi}_{xy}(\nu_p)$ or $\hat{\tau}_{xy}(\nu_p)$, since $\mathrm{E}\{\mathrm{arg}[I_{xy}(\nu_p)]\}$ is not necessarily equal to $\mathrm{arg}\{\mathrm{E}[I_{xy}(\nu_p)]\}$. 

To quantify the bias of time-lag estimates based on the cross-periodogram, we performed an extensive number of simulations, which we describe below. The characteristics of the simulated time series (i.e. time bin size, sampling rate, duration, intrinsic CS, and PSDs) are representative of those observed in AGN X-ray light curves. However, most of our results should apply to any time series observed in practice (see discussion in Sect.\,\ref{sec10}).

\section{Simulating correlated random processes} \label{sec4}

The simulations we performed are based on the procedure outlined by \citet{1995A&A...300..707T}\footnote{We note that \citet{2013MNRAS.433..907E} have more recently presented an improved method for producing artificial light curves. Since we are not interested in the probability distribution of the synthetic time series in our study, we used the Timmer \& Koenig method to minimise the time needed to perform the simulations.} to generate artificial realisations of a discrete stationary random process with a specified model PSD, $h_{\mathscr{X}}(\nu)$, number of points, $N$, sampling period, $\Delta t_{\mathrm{sam}}$, and mean count-rate, $\mu_{\mathscr{X}}$. We assumed a model PSD of the form
\noindent
\begin{equation} \label{eq14}
h_{\mathscr{X}}(\nu)=\mu_{\mathscr{X}}^2\frac{A\nu^{-1}}{1+(\nu/\nu_{\mathrm{b}})}.
\end{equation}
\noindent
This function describes a power law with a low-frequency slope of $-1$ which smoothly `bends' to a slope of $-2$ at frequencies above the bend-frequency, $\nu_{\mathrm{b}}$, as in the case of inferred X-ray PSDs of most AGN. We chose an amplitude value of $A=0.01$, and assumed that $\nu_{\mathrm{b}}=2\times10^{-4}\,\mathrm{Hz}$. Furthermore, we set $N=10.24\times10^6$ and $\Delta t_{\mathrm{sam}}=1\,\mathrm{s}$. By construction, the intrinsic PSD of the generated light curves is discrete, and has non-zero values only at frequencies $\nu_j=j/N\Delta t_{\mathrm{sam}}$, where $j=\pm1,\pm2,\ldots,\pm N/2$.

For each simulated light curve we also created a corresponding `partner' with a mean count-rate $\mu_{\mathscr{Y}}$, assuming a specified model phase-lag spectrum, $\phi_{\mathscr{XY}}(\nu)$. This was achieved by multiplying the DFT of the first light curve by the factor $(\mu_{\mathscr{Y}}/\mu_{\mathscr{X}})\mathrm{e}^{-\mathrm{i}\phi_{\mathscr{XY}}(\nu)}$. The light curves thus have the same PSD shape, and can be considered as realisations of a discrete process whose intrinsic CS is equal to $h_{\mathscr{XY}}(\nu)=(\mu_{\mathscr{Y}}/\mu_{\mathscr{X}})h_{\mathscr{X}}(\nu)\mathrm{e}^{\mathrm{i}\phi_{\mathscr{XY}}(\nu)}$. Since the CS, in effect, represents the average product of the Fourier component amplitudes of each light curve, and this amplitude is proportional to the PSD, the CS is, by construction, non-zero only at the same frequencies where the PSD is non-zero as well. The light curve pairs generated following this procedure will henceforth be referred to as original realisations.

\subsection{The model time-lag spectra} \label{subsec41}

We considered three different model time-lag spectra:
\noindent
\begin{enumerate}
\item Constant delays: In this case we assumed that $\phi_{\mathscr{XY}}(\nu)=2\pi\nu d$, where $d=10\,\mathrm{s}$, $150\,\mathrm{s}$ and $550\,\mathrm{s}$ is a constant delay between the light curve pairs (henceforth, experiments CD1, CD2, and CD3, respectively).
\item Power law delays: In this case we assumed that $\phi_{\mathscr{XY}}(\nu)=(2\pi\nu)B\nu^{-\beta}$, where $B$ is the normalisation and $\beta$ the power-law index of the corresponding model time-lag spectrum. We considered the values $\{B,\beta\}=\{0.001,1\}$, $\{0.01,1\}$, $\{0.1,1\}$, $\{0.01,0.5\}$, $\{0.01,1.5\}$ (henceforth, experiments PLD1, PLD2, PLD3, PLD4, and PLD5, respectively).
\item Top-hat response functions: In this case we assumed that $\phi_{\mathscr{XY}}(\nu)=\mathrm{arg}[1+f\mathrm{e}^{\mathrm{i}2\pi\nu t_0}\mathrm{sinc}(\pi\nu\Delta)]$. This phase-lag spectrum is expected if the intrinsic time series are related by the following equation: $\mathscr{X}(t)=\mathscr{Y}(t)+\int_{-\infty}^{\infty}\mathscr{Y}(t-t')\Psi(t')\mathrm{d}t'$, where $\Psi(t)$ (the so-called response function) is a simple so-called top-hat function, i.e. it has a constant value of $f/\Delta$ in the interval $|t-t_0|\le\Delta/2$ and zero otherwise ($\Delta$ is the width of the top-hat and $t_0$ its centroid). We considered the model parameter values $\{f,t_0,\Delta\}=\{0.2,200\,\mathrm{s},200\,\mathrm{s}\}$ and $\{0.2,2000\,\mathrm{s},2000\,\mathrm{s}\}$ (henceforth, experiments THRF1 and THRF2, respectively).
\end{enumerate}

\subsection{The sampling patterns of the simulated light curves} \label{subsec42}

For each model time-lag spectrum listed above, we created 30 light curve pairs as per the specifications given in Sect.\,\ref{sec4}. To simulate light curves encountered in practice, the original realisations need to be properly `chopped', binned, and sampled.

Most of the data in X-ray astronomy are provided by satellites in low-Earth orbit with a typical orbital period of $\approx96\,\mathrm{min}$ and bin size of $16\,\mathrm{s}$, such as \textit{ASCA}, \textit{RXTE}, \textit{Suzaku} and \textit{NuSTAR}. Light curves obtained from observations with these satellites are `affected' by periodic Earth occultations of a target during every orbit. As a result, they contain gaps that are typically $\approx1-3\,\mathrm{ks}$ long. They are hence not appropriate for Fourier analysis using the `standard' techniques we considered in this work. One can bin the data at one orbital period to acquire evenly sampled light curves with no missing points. Such light curves can be used to probe variability on long time-scales. On the other hand, the data usually contain an appreciable number of continuously sampled segments, with a duration typically $\lesssim3\,\mathrm{ks}$. These segments can, in principle, be used to probe variability on short timescales.

Contrary to low-Earth orbit satellites, \textit{XMM-Newton} observations result in continuously sampled light curves up to $\approx120\,\mathrm{ks}$ long owing to its highly elliptical orbit. We can use such a pair of light curves to compute the cross-periodogram and time-lag estimates directly, or `chop' them into segments of shorter duration, calculate the cross-periodogram for each segment and then bin the resulting estimates at certain frequencies to estimate the time-lags (see Sect.\,\ref{sec6} for a more detailed discussion on this issue).

The purpose of the simulations we performed is to study the sampling properties of time-lags when estimated using light curves with durations and sampling patterns similar to those described above. To this end, we adopted the following strategy:
\noindent
\begin{enumerate}
\item Chop each of the original realisations into 100 parts and bin them at $100\,\mathrm{s}$. This process generates 3000 light curve pairs with $T=102.4\,\mathrm{ks}$ and $\Delta t_{\mathrm{sam}}=\Delta t_{\mathrm{bin}}=100\,\mathrm{s}$ (LS102.4 light curves, hereafter).
\item Chop each of the LS102.4 light curves into 2/5/10/20 segments. This process generates 6000/15000/30000/60000 light curve pairs with $T=40.8/20.4/10.2/5.1\,\mathrm{ks}$ and $\Delta t_{\mathrm{sam}}=\Delta t_{\mathrm{bin}}=100\,\mathrm{s}$ (LS40.8/20.4/10.2/5.1 light curves, hereafter).
\item Chop each of the original realisations into $33\times96=3168$ parts and bin them at $16\,\mathrm{s}$. This process generates 95040 light curves with $T=3.2\,\mathrm{ks}$ and $\Delta t_{\mathrm{sam}}=\Delta t_{\mathrm{bin}}=16\,\mathrm{s}$ (LS3.2 light curves, hereafter).
\item Chop each of the original realisations into 33 parts and bin them at $5760\,\mathrm{s}$ ($\approx96\,\mathrm{min}$). This process generates 990 light curves with $T=305.3\,\mathrm{ks}$ and $\Delta t_{\mathrm{sam}}=\Delta t_{\mathrm{bin}}=5760\,\mathrm{s}$ (OB light curves, hereafter).
\end{enumerate}
\noindent

The reason for the original realisations being longer and with a finer sampling rate than the `final' light curves is to simulate the effects of binning and finite light curve duration on the bias of the time-lag estimates, $b_{\hat{\tau}}(\nu_p)$ (henceforth, the time-lag bias).

\section{The bias of the time-lag estimates in practice} \label{sec5}

To quantify $b_{\hat{\tau}}(\nu_p)$, we calculated $\hat{\tau}_{xy}(\nu_p)$ using Eq.\,\ref{eq12} for the 10 numerical experiments and each light curve type described in Sects.\,\ref{subsec41} and \ref{subsec42}. We then computed the sample mean at each frequency, $\langle\hat{\tau}_{xy}(\nu_p)\rangle$ (quantities in angle brackets will hereafter denote their sample mean). We chose to study the properties of $\hat{\tau}_{xy}(\nu_p)$ mainly, since the majority of works concerned with AGN X-ray timing studies use this estimator. In addition, we computed the quantity $\delta_{\hat{\tau}}(\nu_p)\equiv[\tau_{\mathscr{XY}}(\nu_p)-\langle\hat{\tau}_{xy}(\nu_p)\rangle]/\tau_{\mathscr{XY}}(\nu_p)$ to quantify the time-lag bias in terms of its intrinsic value (henceforth, the relative bias). This allows us to directly compare the time-lag bias obtained from different light curve types in the various numerical experiments we considered.

Figures\,\ref{figd1}--\ref{figd4} in Appendix \ref{appd} show $\delta_{\hat{\tau}}(\nu_p)$ for the LS102.4/LS3.2/OB (top rows; continuous black, red, and brown curves, respectively) and LS40.8/20.4/10.2/5.1 (bottom rows; continuous black, red, brown, and green curves, respectively) light curves. Each column in these figures corresponds to a different numerical experiment. The main results of the simulations we performed are summarised below.

\subsection{Light curve binning and aliasing} \label{subsec51}

A common feature in all mean sample time-lag spectra is that they decrease to zero at high frequencies, irrespective of the light curve sampling pattern, duration, and model time-lag spectrum. This decrease is due to the effects of aliasing and light-curve binning.

To demonstrate this issue, we considered experiment PLD2 (the results are similar for the other numerical experiments as well). We generated two additional ensembles of light curve pairs that have the same length as the LS102.4 light curves. The first ensemble was constructed using the original realisations of experiment PLD2, which were chopped into 3000 parts of length $102.4\,\mathrm{ks}$ each, and sampled (not binned) every $100\,\mathrm{s}$ (henceforth, the LS102.4-2 light curves). These light curves are still affected by aliasing, since the original realisations have a finer sampling rate of $1\,\mathrm{s}$. For the second ensemble, we constructed 30 new original realisations with the same length as the rest of the numerical experiments (i.e. $10.24\,\mathrm{Ms}$) and a sampling period of $100\,\mathrm{s}$. We subsequently chopped them into 3000 parts of length $102.4\,\mathrm{ks}$ each (henceforth, the LS102.4-3 light curves). Since the original realisations in this case have the same sampling rate as the LS102.4-3 light curves, they should not be affected by either binning or aliasing.

   \begin{figure}
   \centering
   \includegraphics[width=\hsize]{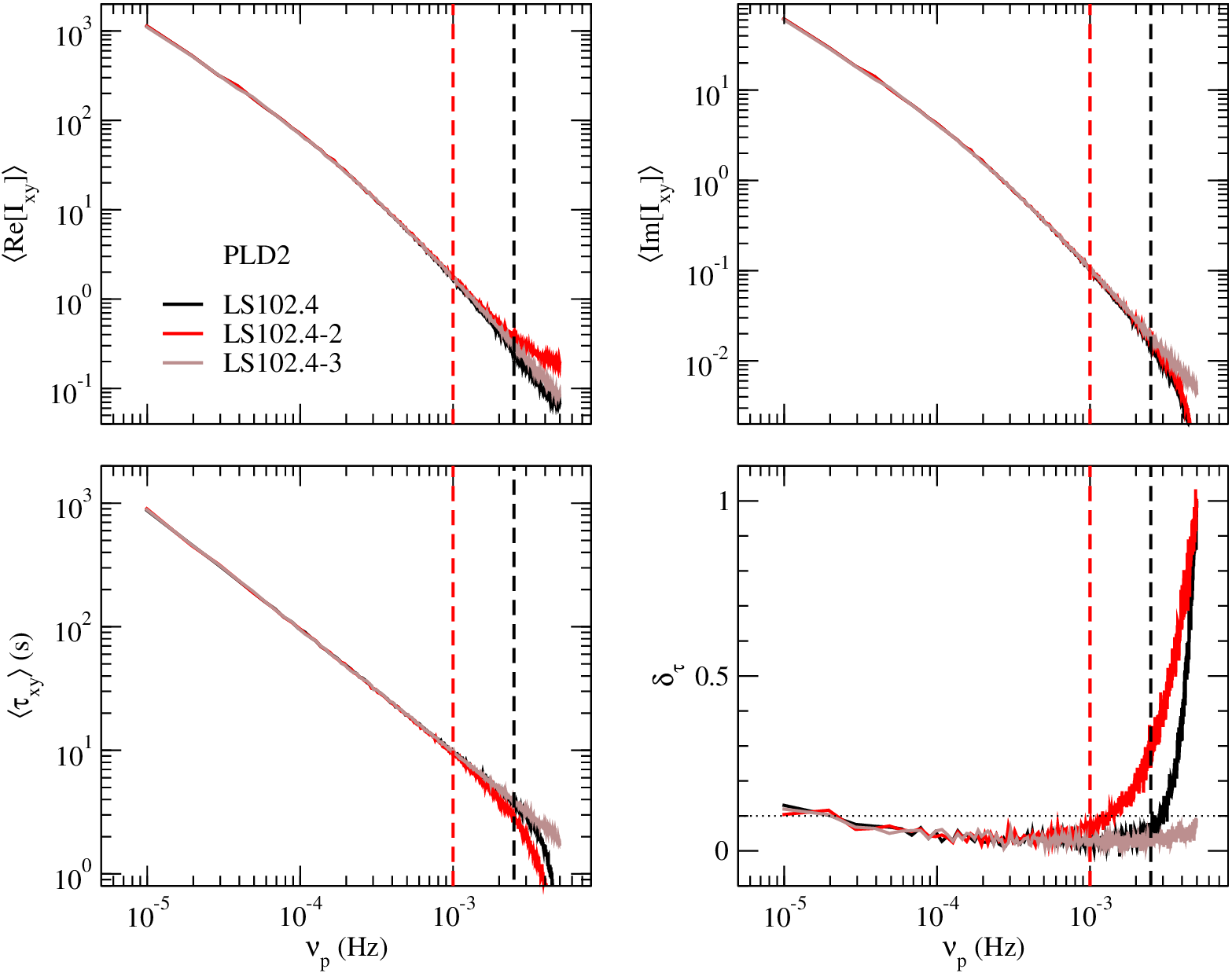}
      \caption{Sample mean of the real and imaginary parts of the cross-periodogram (top left and right panels, respectively), the sample mean time-lag spectrum and the relative time-lag bias (bottom left and right panel, respectively) in experiment PLD2. The vertical black and red dashed lines indicate $\nu_{\mathrm{Nyq}}/2$ and $\nu_{\mathrm{Nyq}}/5$, respectively. Above these frequencies, the LS102.4 (black curve) and LS102.4-2 (red curve) relative time-lag bias begins to noticeably increase (see text for more details). The horizontal dotted line in the bottom right panel, and in all subsequent $\delta_{\hat{\tau}}$ plots, indicates the 0.1 (i.e. 10\%) relative time-lag bias.}
      \label{fig1}
   \end{figure}

We then calculated $\langle\hat{\tau}_{xy}(\nu_p)\rangle$ and $\delta_{\hat{\tau}}(\nu_p)$ for these two additional light curve types. Figure\,\ref{fig1} shows $\langle\Re[I_{xy}(\nu_p)]\rangle$, $\langle\Im[I_{xy}(\nu_p)]\rangle$, $\langle\hat{\tau}_{xy}(\nu_p)\rangle$, and $\delta_{\hat{\tau}}(\nu_p)$ (top left and right, bottom left and right panels, respectively) for the LS102.4, LS102.4-2, and LS102.4-3 light curves (black, red, and brown curves, respectively).\footnote{The sample mean of the real and imaginary parts of the cross-periodograms shown in Fig.\,\ref{fig1}, and all similar subsequent figures, have been normalised by the factor $\mu_\mathscr{X}\mu_\mathscr{Y}$ (which is the common normalisation factor of the intrinsic CS used in our numerical experiments; see Sect.\,\ref{sec4}) to make them dimensionless and independent of the specific count-rates assumed to construct the simulated light curve pairs.} The sample mean time-lag spectra for the LS102.4-2 and LS102.4 light curves decrease to zero at frequencies higher than $\approx10^{-3}\,\mathrm{Hz}$ ($=\nu_{\mathrm{Nyq}}/5$) and $\approx2.5\times10^{-3}\,\mathrm{Hz}$ ($=\nu_{\mathrm{Nyq}}/2$), respectively. These frequencies are indicated by the vertical dashed lines in the same figure. This is not the case for the LS102.4-3 light curves, as shown by the brown curves in the bottom panels of Fig.\,\ref{fig1}. The time-lag bias of the sampled and binned light curves can be understood by the plots in the top panels of Fig.\,\ref{fig1}. On average, the gradients of the imaginary and real parts of the cross-periodogram enter steeper rates of decrease and increase, respectively, compared to their intrinsic values, around the frequencies indicated by the vertical dashed lines in Fig.\,\ref{fig1}. As a result, their ratio (which determines the phase-lag estimate) is decreased on average, hence the time-lag bias increases. This increase is most severe at high frequencies, where the effects of aliasing and light curve binning on the cross-periodogram are maximised. The time-lag bias owing to these effects is more pronounced for the LS102.4-2 light curves, since light curve binning suppresses the aliasing effect (see Sect.\,\ref{sec3}).

   \begin{figure}
   \centering
   \includegraphics[width=\hsize]{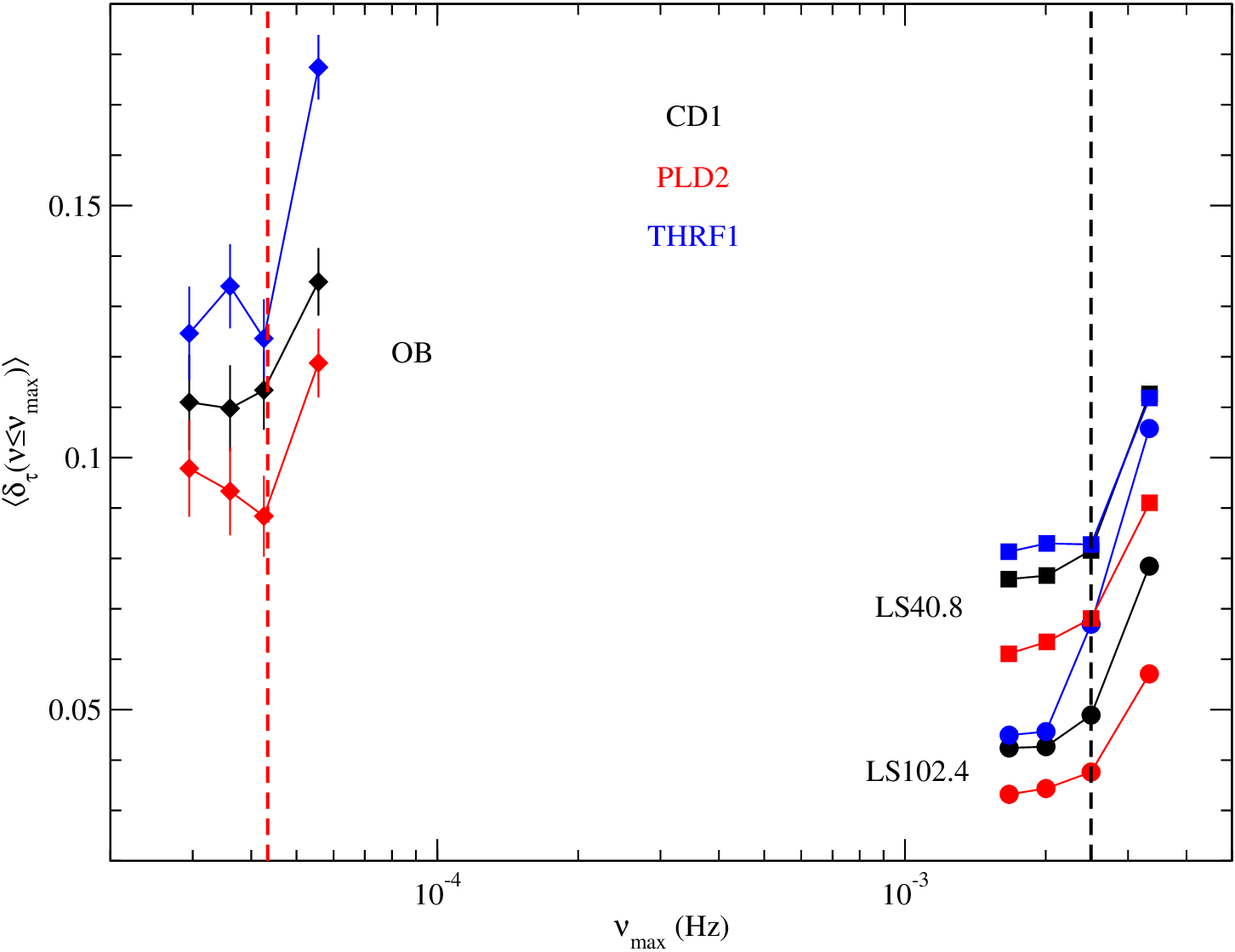}
      \caption{Mean relative time-lag bias over all frequencies below $\nu_{\mathrm{max}}$, plotted as a function of $\nu_{\mathrm{max}}$, for different light curve types in various numerical experiments. The left and right dashed vertical lines indicate $\nu_{\mathrm{Nyq}}/2$ for the OB- and LS-type light curves, respectively.}
      \label{fig2}
   \end{figure}

Figures\,\ref{figd1}--\ref{figd4} show that the onset of the high-frequency increase in the relative time-lag bias occurs at $\approx\nu_{\mathrm{Nyq}}/2$ for all light curve types, model time-lag spectra, and light curve durations. To further illustrate this effect, we define the function $\langle\delta_{\hat{\tau}}(\nu_p\le\nu_{\mathrm{max}})\rangle$ as the mean relative time-lag bias at all frequencies below $\nu_{\mathrm{max}}$. Figure\,\ref{fig2} shows $\langle\delta_{\hat{\tau}}(\nu_p\le\nu_{\mathrm{max}})\rangle$, evaluated at $\nu_{\mathrm{max}}=\nu_{\mathrm{Nyq}}/3$, $\nu_{\mathrm{Nyq}}/2.5$, $\nu_{\mathrm{Nyq}}/2$, and $\nu_{\mathrm{Nyq}}/1.5$, for the OB (diamonds), LS40.8 (squares), and LS102.4 (circles) light curves in experiments CD1 (black), PLD2 (red), and THRF1 (blue). The vertical red and black dashed lines indicate $\nu_{\mathrm{Nyq}}/2$ for the OB and LS102.4/40.8 light curves, respectively. At frequencies below $\nu_{\mathrm{Nyq}}/2$, the relative time-lag bias remains approximately constant, but increases markedly at higher frequencies (the same trend is observed in all numerical experiments we considered).

We conclude that time-lags should be estimated at frequencies $\lesssim\nu_{\mathrm{Nyq}}/2$ to minimise the effects of light-curve binning and aliasing on the time-lag bias.

\subsection{Finite light curve duration} \label{subsec52}

The time-lag estimates are biased even at frequencies $\lesssim\nu_{\mathrm{Nyq}}/2$. Figures\,\ref{figd1}--\ref{figd4} show that the relative time-lag bias is generally $\lesssim15\%$ at frequencies $\lesssim\nu_{\mathrm{Nyq}}/2$ for the LS102.4/40.8/20.4 and OB light curves, and larger for the rest. It is usually positive, in the sense that the estimates are, on average, smaller in absolute magnitude than their corresponding intrinsic values. The bias at intermediate/low frequencies is not due to light curve binning and/or aliasing (we note, for example, that $\delta_{\hat{\tau}}(\nu_p)$ at frequencies $\lesssim\nu_{\mathrm{Nyq}}/5$ in Fig.\,\ref{fig1} is identical for all three light curve types, which have the same length but different time bin sizes). This bias is due to the finite duration, $T$, of the light curves, or, technically speaking, due to the convolution of $h_{XY}(\nu_p)$ with the Fej\'{e}r kernel (see Eq.\,\ref{eq13}). 
   
To quantify the dependence of this bias on $T$, Fig.\,\ref{fig3} shows  $\langle\delta_{\hat{\tau}}(\nu_p\le\nu_{\mathrm{Nyq}}/2)\rangle$ as a function of $T$ for the LS5.1/10.2/20.4/40.8/102.4 and OB light curves (filled and open points, respectively) in all experiments that do not exhibit so-called phase-flipping (see Sect.\,\ref{subsec53} for details). The dashed red line indicates the best-fit relation to the LS-type data: $\langle\delta_{\hat{\tau}}(\nu_p\le\nu_{\mathrm{Nyq}}/2)\rangle=0.08(T/40.8\,\mathrm{ks})^{-1/2}$. This relation fits the LS-type data quite well. We therefore conclude that the mean relative time-lag bias decreases with increasing light curve duration as $1/\sqrt{T}$. However, this relation is not consistent with the points corresponding to the OB light curves. This is unexpected at first, since their duration ($\approx300\,\mathrm{ks}$) is larger than the duration of all LS-type light curves. This discrepancy arises because the frequency range probed by the OB light curves is significantly lower than by the LS-type light curves (owing to their longer duration and larger time bin size), as well as the fact that the relative time-lag bias increases at lower frequencies.

To illustrate this effect, we generated additional ensembles of light curve pairs that have the same time bin size ($100\,\mathrm{s}$) and a duration two, three, four, and five times longer than LS102.4 light curves in the case of experiment CD1 (the results are identical for the other numerical experiments as well). The solid lines in the top panel of Fig.\,\ref{fig4} show the relative time-lag bias at all frequencies below $\nu_{\mathrm{Nyq}}/2$ for all these light curves (their duration is listed next to each curve). As expected, the relative bias at each frequency decreases with increasing $T$. For a fixed $T$, the relative bias increases with decreasing frequency. The filled boxes in the same plot indicate the mean relative bias over the whole frequency range, plotted at the mean logarithmic frequency. The mean relative bias, over the full frequency range, decreases with increasing $T$. In the bottom panel of the same figure we plot the same results in the case of OB-type light curves with a duration of $\approx50-500\,\mathrm{ks}$. The relative bias at each frequency is almost identical to the previous case, despite the larger time bin size of the OB-type light curves. Although the mean relative bias decreases with increasing $T$ as before, owing to the larger time bin size in this case, we sample a lower frequency range, and hence the mean relative bias is larger than before (for the same $T$). We therefore conclude that the relative time-lag bias depends on $1/\sqrt{T}$, although the normalisation of this relation depends on the frequency range on which the time-lags are estimated; the lower the frequencies, the more biased the time-lag estimates will be. For the model CS we considered, our results suggest that, to obtain time-lag estimates with a relative bias $\lesssim15\%$, the light curves must have $T\gtrsim20\,\mathrm{ks}$.

   \begin{figure}
   \centering
   \includegraphics[width=\hsize]{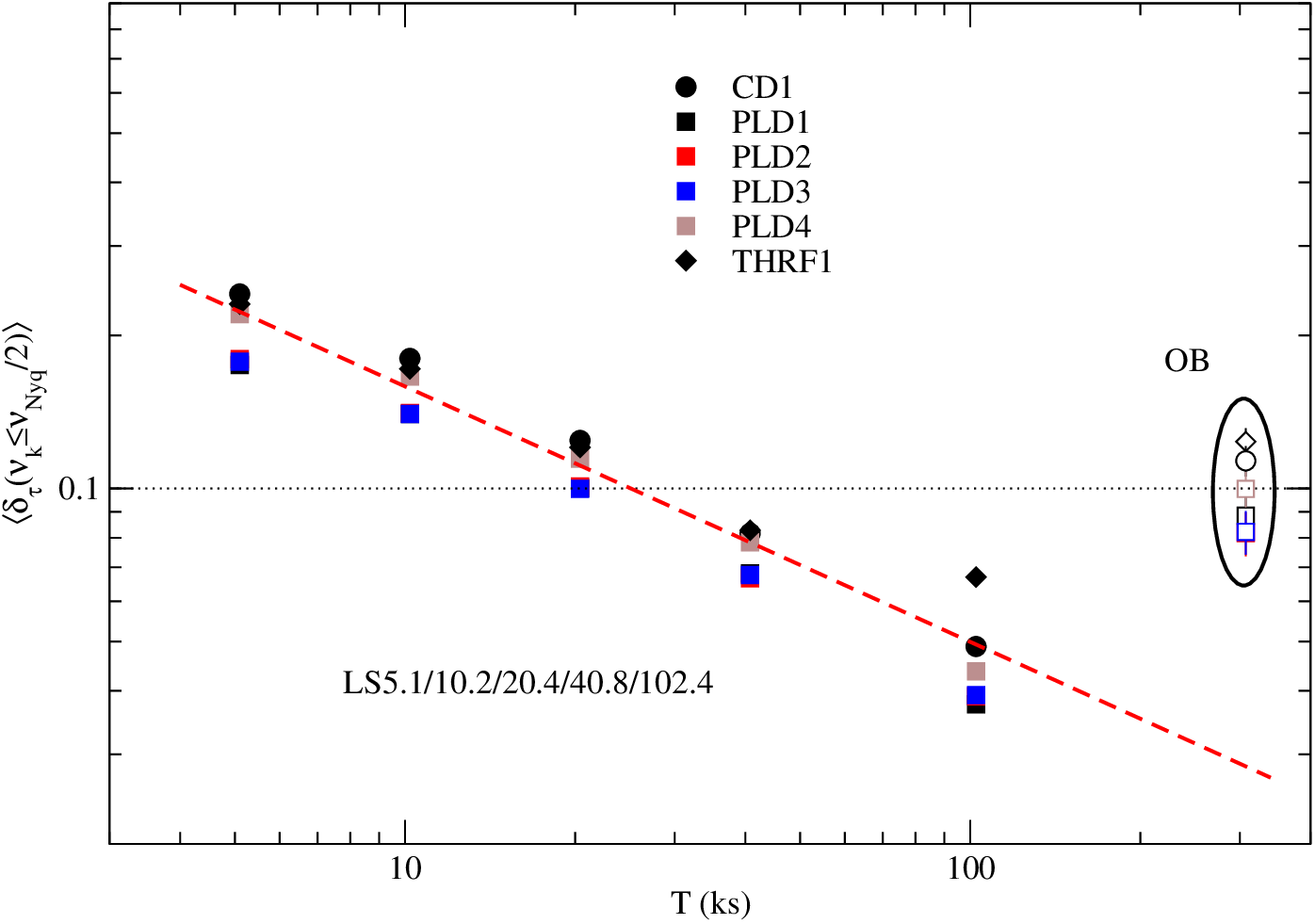}
      \caption{Mean relative time-lag bias, plotted as function of light curve duration for LS-type and OB light curves.}
      \label{fig3}
   \end{figure}

   \begin{figure}
   \centering
   \includegraphics[width=\hsize]{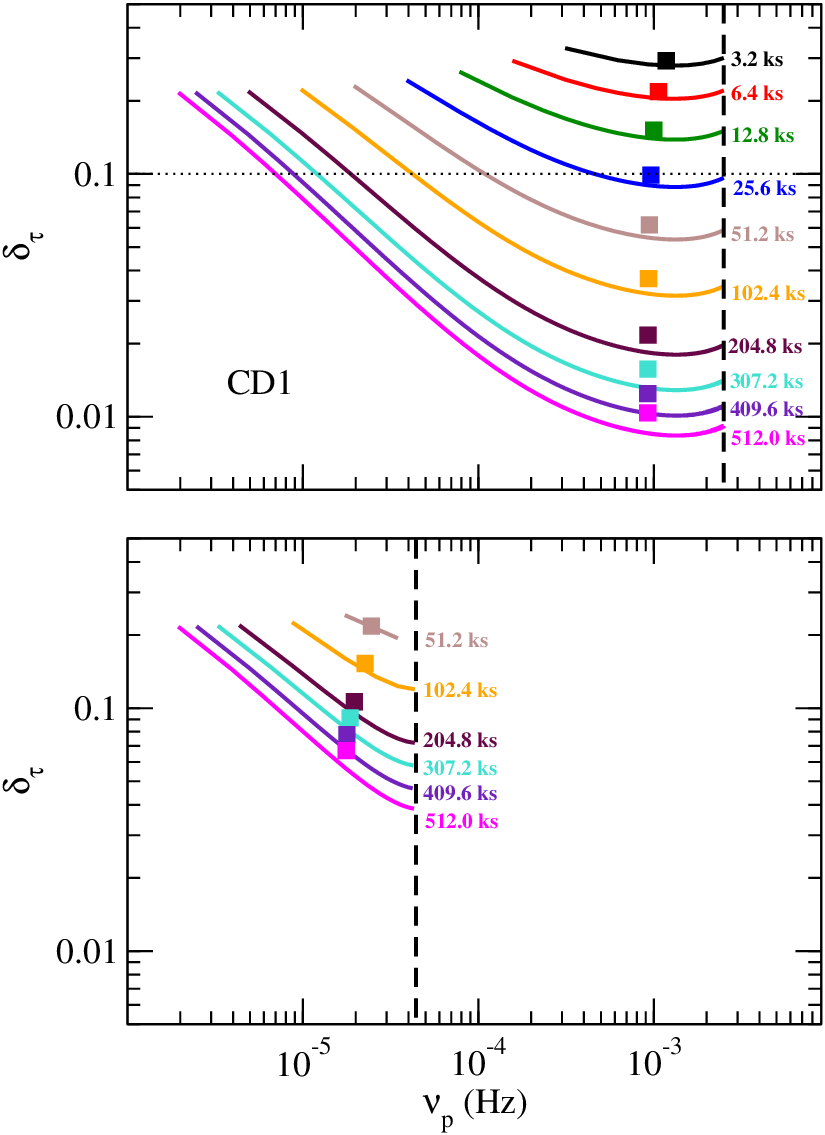}
      \caption{Relative time-lag bias for the LS- and OB-type light curves (top and bottom panel, respectively), for various durations in experiment CD1. Filled squares indicate the mean relative time-lag bias over the full sampled frequency range, evaluated at the mean logarithmic frequency. The dashed vertical line in the top and bottom panel indicate $\nu_{\mathrm{Nyq}}/2$ for the LS- and OB-type light curves, respectively.}
      \label{fig4}
   \end{figure}

\subsection{Phase-flipping} \label{subsec53}

Since the phase-lag spectrum is defined on the interval $(-\pi,\pi]$, there can be frequencies whereby this function will exceed the value of $\pi$ and `flip back' to the value of $-\pi$ (and vice-versa). This effect is known as phase-flipping (or phase-wrapping).

These kinds of events take place in experiments CD2, CD3, and PLD5 (see Figs.\,\ref{figd1} and \ref{figd3}). The top left panel of Fig.\,\ref{fig5} shows $\langle\hat{\tau}_{xy}(\nu_p)\rangle$ for the LS102.4 light curves in experiment CD3 (continuous black line), along with the corresponding model time-lag spectrum (black dashed line). The simulated light curves in this experiment are separated by a constant delay. As a result, the model phase-lag spectrum is a linearly increasing function of frequency, i.e. $\phi_{\mathscr{XY}}(\nu)=2\pi\nu d$. At $\nu=1/2d\approx9.1\times10^{-4}\,\mathrm{Hz}$ we get $\phi_{\mathscr{XY}}(\nu)=\pi$ and then, by definition, $\phi_{\mathscr{XY}}(\nu)$ `jumps' to the value $-\pi$. Since $\tau_{\mathscr{XY}}(\nu)=\phi_{\mathscr{XY}}(\nu)/2\pi\nu$, the model time-lag spectrum will likewise jump from $550\,\mathrm{s}$ to $-550\,\mathrm{s}$. Subsequent phase-flips occur at frequencies $\nu=j/2d$ ($j=2,3,\ldots$), where the model time-lag spectrum undergoes discontinuous jumps from positive to negative values with decreasing amplitude.

The mean sample time-lag spectrum also fluctuates from positive to negative values at these frequencies, although the transition is much smoother. To illustrate the reason for this behaviour, in the top right and bottom panels of Fig.\,\ref{fig5}, we plot the probability distribution of the phase-lag estimate at the following frequencies: $8.1\times10^{-4}\,\mathrm{Hz}$ (top right panel), $9.1\times10^{-4}\,\mathrm{Hz}$ and $1.0\times10^{-3}\,\mathrm{Hz}$ (bottom left and right panel, respectively). These frequencies are indicated by the vertical dashed lines in the top left panel of the same figure. The middle frequency is very close to the frequency where the first phase-flip occurs, while the other two are lower and higher than that frequency. The solid and dashed vertical lines in the top-right and bottom panels of Fig.\,\ref{fig5} show the model and sample mean phase-lag value at these frequencies, respectively. Since the phase-lag estimate is defined on the interval $(-\pi,\pi]$, the so-called wings of its distribution that exceed these boundaries (indicated by the `empty' histograms in the plots) are `folded back' into the allowed range. This causes the mean of the distribution to shift towards a value lower than that of the model phase-lag spectrum. This effect is most severe in the vicinity of frequencies where phase-flipping occurs, since, in this case, the mean of the phase-lag estimate is close to zero.

   \begin{figure}
   \centering
   \includegraphics[width=\hsize]{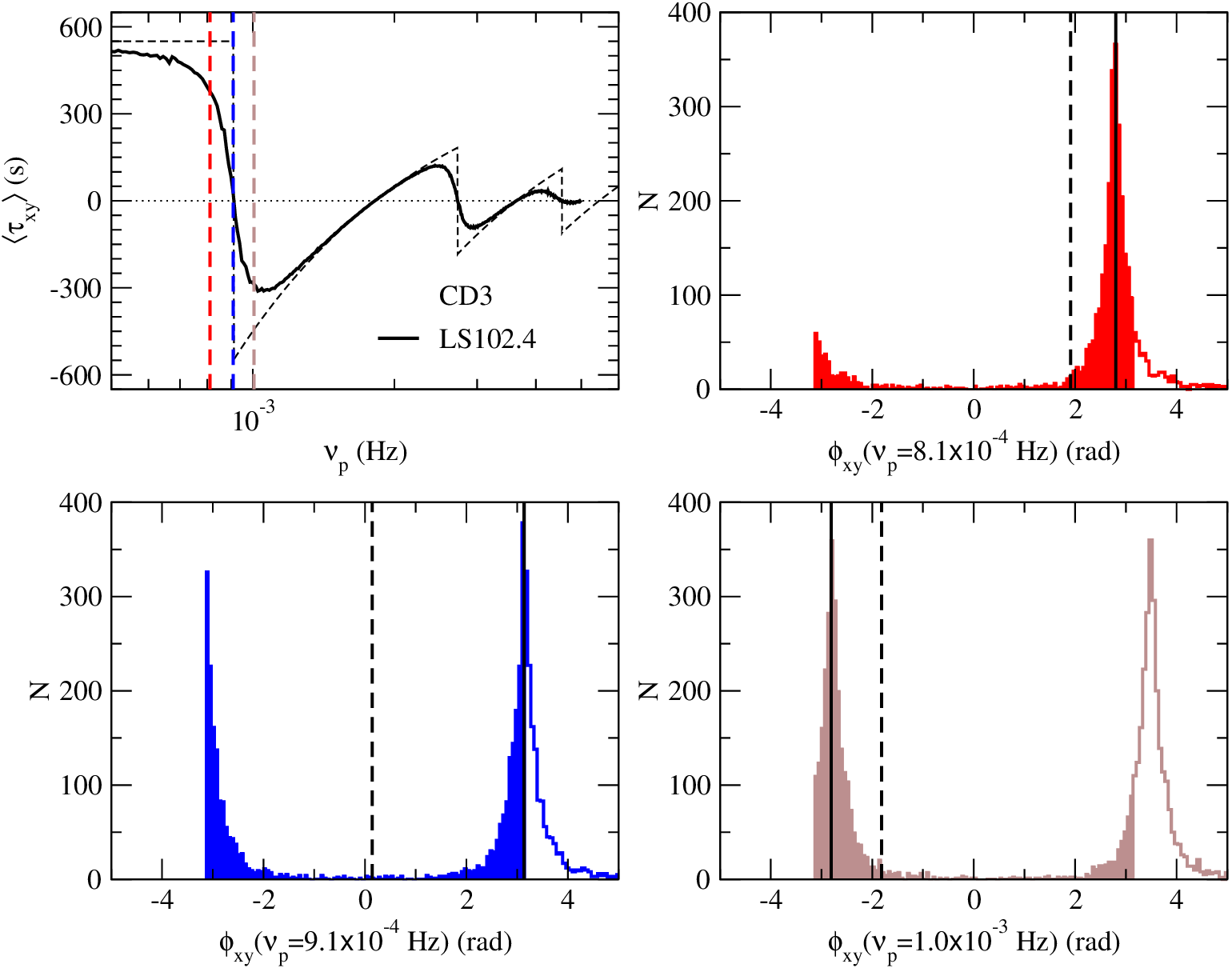}
      \caption{Top left panel: Mean sample time-lag spectrum for the LS102.4 light curves in experiment CD3 (continuous black line). The black dashed line indicates the model time-lag spectrum. Top right and bottom panels: The probability distribution of the phase-lag estimates at the frequencies indicated by the vertical dashed lines in the top-left panel. The solid and dashed vertical line in these panels indicate the model and sample mean phase-lag value at these frequencies, respectively.}
      \label{fig5}
   \end{figure}

Phase-flipping does not necessarily occur at increasingly higher frequencies. If, as in the case of experiments PLD4 and PLD5 for example, the intrinsic phase-lag spectrum increases (in absolute magnitude) with decreasing frequency, phase-flipping takes place at increasingly lower frequencies. This is seen in the bottom left panel of Fig.\,\ref{fig6}, where we plot $\langle\hat{\tau}_{xy}(\nu_p)\rangle$ for the LS102.4 light curves in experiment PLD5 (open black circles). The model time-lag spectrum for this experiment is $\tau_{XY}(\nu)=0.01\nu^{-1.5}$ (indicated by the continuous brown line). The corresponding phase-lag spectrum is $\phi_{\mathscr{XY}}(\nu)=(2\pi\nu)0.01\nu^{-1.5}$, and phase-flipping occurs at $(4\times10^{-4}/j^2)\,\mathrm{Hz}$ ($j=1,2,\ldots$). At frequencies close to $4\times10^{-4}\,\mathrm{Hz}$, where the first phase-flip occurs, the mean sample time-lag spectrum exhibits a jump which is smoother than the abrupt jump of the model time-lag spectrum, for exactly the same reasons we discussed above. This is not very clear in the plot shown in the bottom left panel of Fig.\,\ref{fig6}, but is evident in the plot of $\delta_{\hat{\tau}}(\nu_p)$ (bottom-right panel in the same figure).

The dashed blue line in the top panels of Fig.\,\ref{fig6} indicate the model real and imaginary parts of the CS. Interestingly, the sample mean of the real and imaginary parts of the cross-periodogram (indicated by the open circles in the top panels) are not biased at frequencies around $\approx4\times10^{-4}\,\mathrm{Hz}$. In fact, the argument of their ratio divided by the angular frequency, i.e. $\mathrm{arg}\{\mathrm{E}[\hat{h}_{XY}(\nu_p)]\}/2\pi\nu_p$, is very similar to the model time-lag spectrum, and exhibits a sharp jump at this frequency. This is a case where, for the reasons explained above, $\mathrm{arg}\{\mathrm{E}[\hat{h}_{XY}(\nu)]\}$ is not equal to $\mathrm{E}\{\mathrm{arg}[\hat{h}_{XY}(\nu)]\}$.

   \begin{figure}
   \centering
   \includegraphics[width=\hsize]{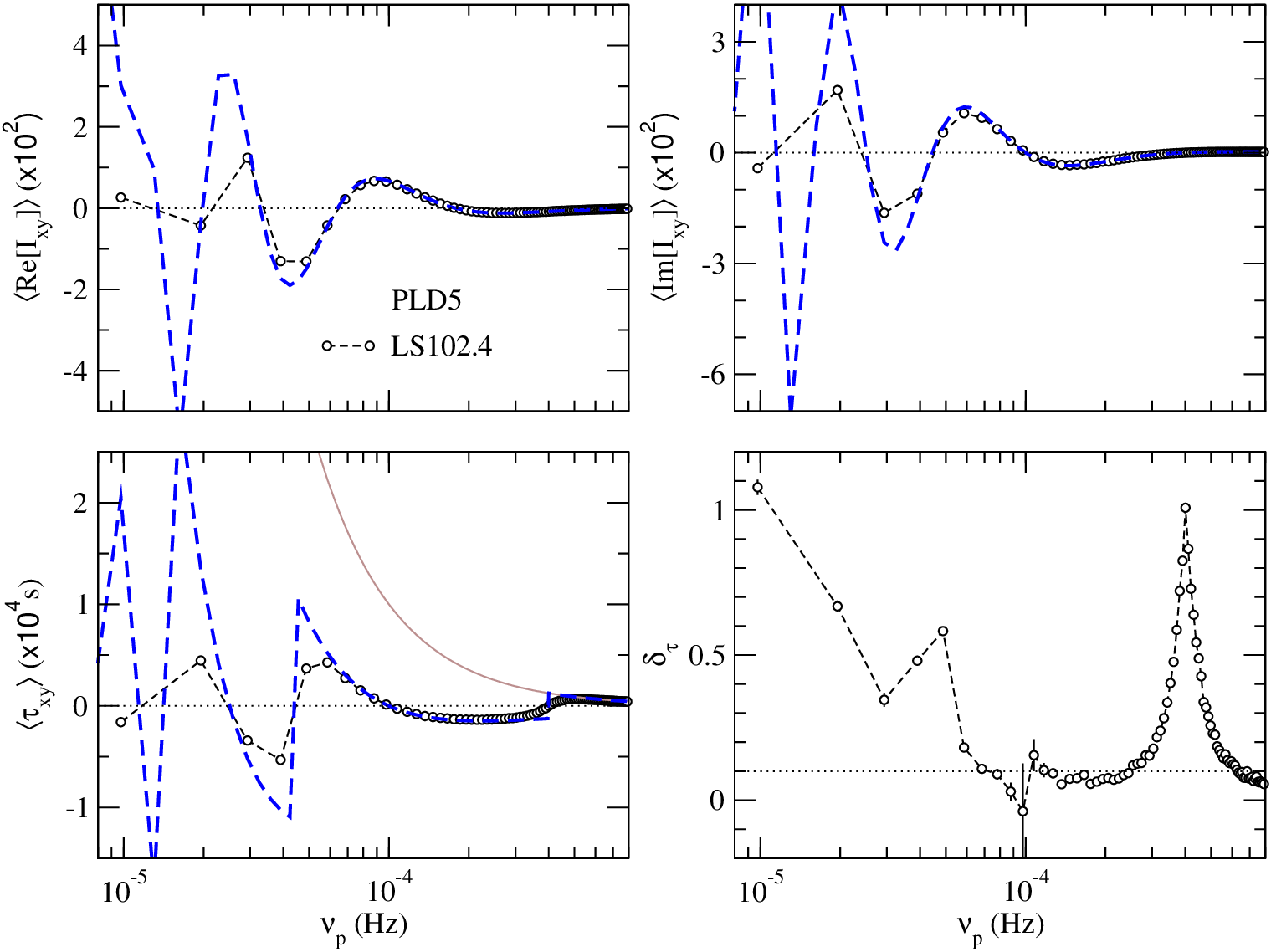}
      \caption{As in Fig.\,\ref{fig1}, for experiment PLD5. The blue dashed lines indicate the model CS (upper panels) and time-lag spectrum (lower left panel). The solid brown line in the bottom left panel indicates the model time-lag spectrum without taking the effects of phase-flipping into account.}
      \label{fig6}
   \end{figure}

The time-lag estimates at the two lowest frequencies are heavily biased owing to the bias of the real and imaginary parts of the cross-periodogram. This is reflected in the mean sample time-lag spectrum, which has a large relative bias at those frequencies (bottom right panel in Fig.\,\ref{fig6}). This bias originates from the convolution of the intrinsic CS with the Fej\'{e}r kernel, which causes the mean real and imaginary parts of the cross-periodogram to diverge from the model CS because of its rapid oscillatory behaviour at low frequencies.

\section{Smoothed/averaged cross-spectral estimators} \label{sec6}

The variance of the real and imaginary parts of the cross-periodogram is $\sim c^2_{XY}(\nu_p)$ and $\sim q^2_{XY}(\nu_p)$, respectively (P81). They are unknown, and do not depend on the number of points in the observed time series (i.e they will not decrease if we increase their duration). In practice, we usually average a certain number, say $m$, of consecutive cross-periodogram estimates, i.e.
\noindent
\begin{equation} \label{eq15}
\hat{h}_{xy}(\nu_k) \equiv\frac{1}{m}\sum_{p=(k-1)m+1}^{km}I_{xy}(\nu_p), \\
\end{equation}
\noindent
and accept $\hat{h}_{xy}(\nu_k)$ as the CS estimator at the frequencies $\nu_k=(1/m)\sum_p\nu_p$ ($k=1,2,\ldots, N/m$). The process of binning over consecutive frequencies is called smoothing.\footnote{There are many smoothing schemes for the cross-periodogram. The one defined by Eq.\,\ref{eq15} corresponds to a particular type of smoothing `window' called the Daniell, or rectangular, spectral window (see P81 for more details).} The real and imaginary parts of the smoothed CS estimator are correspondingly given by
\noindent
\begin{align}
\label{eq16}
\hat{c}_{xy}(\nu_k) &\equiv\Re[\hat{h}_{xy}(\nu_k)]=\frac{1}{m}\sum_{p}\Re[I_{xy}(\nu_p)], \\
\label{eq17}
-\hat{q}_{xy}(\nu_k) &\equiv\Im[\hat{h}_{xy}(\nu_k)]=\frac{1}{m}\sum_{p}\Im[I_{xy}(\nu_p)].
\end{align}
\noindent
An alternative procedure is to partition the available time series into $m$ shorter segments of duration $T/m$, compute the cross-periodogram for each segment, $I_{xy}^{(l)}(\nu_k)$ ($l=1,2,...,m$, and $\nu_k=k/(T/m)$, where $k=1,2,...,N/m$), and then average the different cross-periodogram values at each $\nu_k$:
\noindent
\begin{equation} \label{eq18}
\hat{h}_{xy}(\nu_k) \equiv\frac{1}{m}\sum_{l=1}^{m}I_{xy}^{(l)}(\nu_k),
\end{equation}
\noindent
with the real and imaginary parts estimated as in the case of the smoothed estimates (to differentiate between estimators that are smoothed or averaged over individual segments, we henceforth refer to the former as smoothed and the latter as averaged estimates).

One can show that the variance of the real and imaginary parts of the smoothed/averaged cross-periodogram (as well as their covariance) is inversely proportional to $m$. As the duration of the observed light curves increases (i.e. $N$ increases for a fixed $\Delta t_{\mathrm{sam}}$), $m$ can be increased proportionally without degrading the frequency resolution. In this sense, the variance of the smoothed/averaged estimates decreases with increasing $N$ (in the limit $N\rightarrow\infty$, it approaches zero).

We can use the real and imaginary parts of the smoothed/averaged CS estimates to construct the following estimators of the phase- and time-lag spectrum:
\noindent
\begin{align}
\label{eq19}
\hat{\phi}_{xy}(\nu_k) &\equiv\mathrm{arctan}\left[-\frac{\hat{q}_{xy}(\nu_k)}{\hat{c}_{xy}(\nu_k)}\right], \\
\label{eq20}
\hat{\tau}_{xy}(\nu_k) &\equiv\frac{\hat{\phi}_{xy}(\nu_k)}{2\pi\nu_k}.
\end{align}
\noindent
Their variance is given by the following asymptotic formulae \citep[P81;][]{1999ApJ...510..874N,Ben-Pier:11}
\noindent
\begin{align}
\label{eq21}
\mathrm{Var}[\hat{\phi}_{xy}(\nu_k)] &\sim\frac{1}{2m}\frac{1-\gamma^2_{XY}(\nu_k)}{\gamma^2_{XY}(\nu_k)}, \\
\label{eq22}
\mathrm{Var}[\hat{\tau}_{xy}(\nu)] &\sim\frac{\mathrm{Var}[\hat{\phi}_{xy}(\nu_k)]}{(2\pi\nu_k)^2}.
\end{align}
\noindent
Equations\,\ref{eq21} and \ref{eq22} highlight the importance of $\gamma^2_{XY}(\nu)$ in constructing a reliable time-lag estimator. The lower the coherence, the higher the variance of the phase- and time-lag estimates will be (this is expected, as it must be more `difficult' to detect delays when the two light curves are highly incoherent). The `natural' choice for the coherence estimator, as suggested by Eq.\,\ref{eq6}, is the following:
\noindent
\begin{equation} \label{eq23}
\hat{\gamma}^2_{xy}(\nu_k)\equiv\frac{\left\lvert\hat{h}_{xy}(\nu_k)\right\rvert^2}{\hat{h}_x(\nu_k)\hat{h}_y(\nu_k)}=\frac{\hat{c}_{xy}^2(\nu_k)+\hat{q}_{xy}^2(\nu_k)}{\hat{h}_x(\nu_k)\hat{h}_y(\nu_k)},
\end{equation}
\noindent
where $\hat{h}_x(\nu_k)$ and $\hat{h}_y(\nu_k)$ are the smoothed/averaged PSD estimators of the two observed time series. The variance of the coherence estimator is given by the asymptotic formula \citep[P81;][]{1997ApJ...474L..43V,1999ApJ...510..874N,Ben-Pier:11}
\noindent
\begin{equation} \label{eq24}
\mathrm{Var}[\hat{\gamma}^2_{xy}(\nu_k)]\sim\frac{2}{m}\gamma^2_{XY}(\nu_k)\left[1-\gamma^2_{XY}(\nu_k)\right]^2.
\end{equation}
\noindent
Since the intrinsic coherence, $\gamma^2_{XY}(\nu_k)$, is unknown, it is customary to replace it in Eqs.\,\ref{eq21} and \ref{eq24} by its estimate, $\hat{\gamma}^2_{xy}(\nu_k)$, to obtain a numerical value.

\subsection{Bias due to smoothing} \label{subsec61}

\begin{figure}
\centering
\includegraphics[width=\hsize]{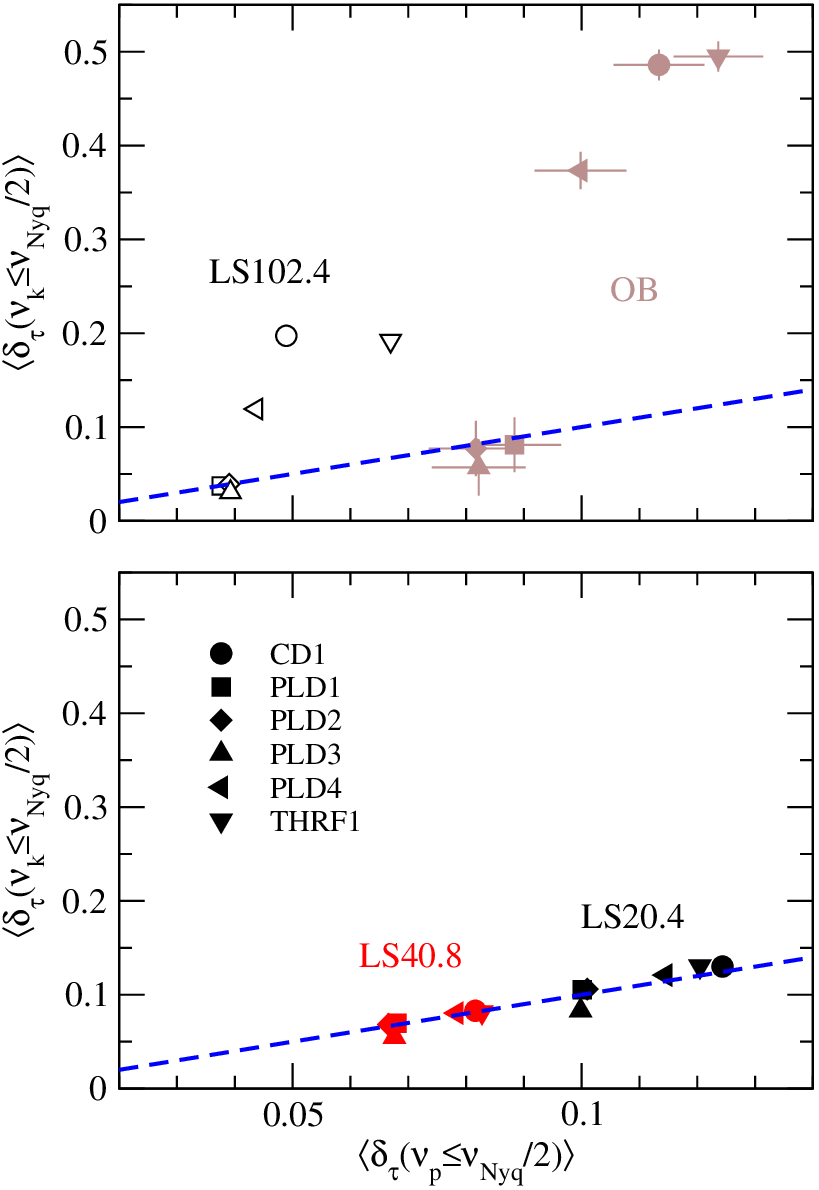}
      \caption{Mean relative bias of the $m=20$ smoothed (top panel) and averaged (bottom panel) time-lag estimates, plotted as a function of the relative bias of the non-smoothed/averaged time-lag estimates (the dashed lines show the one-to-one relation line) in various numerical experiments.}
      \label{fig7}
   \end{figure}

To investigate whether smoothing or averaging increases the bias of the cross-spectral estimates, we smoothed the real and imaginary parts of the cross-periodograms estimated from the LS102.4 and OB light curves (for all numerical experiments) using Eqs.\,\ref{eq16} and \ref{eq17} for $m=20$. We also used the cross-periodograms of the LS20.4/40.8 light curves, and averaged $m=20$ of them to produce the corresponding averaged estimates. We then computed, in both cases and for all experiments, $\hat{\tau}_{xy}(\nu_k)$ and $\hat{\gamma}^2_{xy}(\nu_k)$ using Eqs.\,\ref{eq20} and \ref{eq23}, as well as their corresponding sample mean values, and $\delta_{{\hat{\tau}}}(\nu_k)$. Figures\,\ref{figd5}--\ref{figd8} show $\delta_{\hat{\tau}}(\nu_k)$ for the smoothed (top panels; open white circles and filled brown squares for the LS102.4/OB light curves) and averaged estimates (bottom panels; continuous black and red curves for the LS20.4/40.8 light curves).

In general, smoothing increases the time-lag bias at low frequencies. The top panel in Fig.\,\ref{fig7} shows $\langle\delta_{\hat{\tau}}(\nu_k\le\nu_{\mathrm{Nyq}}/2)\rangle$ vs $\langle\delta_{\hat{\tau}}(\nu_p\le\nu_{\mathrm{Nyq}}/2)\rangle$ for the smoothed estimates (in the case of the numerical experiments that do not exhibit phase-flipping in the sampled frequency range). The bottom panel shows the same plot for the averaged estimates. The blue dashed line shows the one-to-one relation in each case. Although the bias of the averaged estimates falls on the expected one-to-one relation line, this is not always the case for the smoothed estimates. Significant, additional bias appears in the case of experiments CD1, PLD4, and THRF1. The intrinsic CS in these cases exhibits a prominent non-linear variation with frequency. As a result, linear smoothing introduces a bias to the real and imaginary parts of the smoothed cross-periodogram. This is demonstrated in Fig.\,\ref{fig8}, which shows $\langle\Re[\hat{h}_{xy}(\nu_k)]\rangle$ (top left panel), $\langle\Im[\hat{h}_{xy}(\nu_k)]\rangle$ (top right panel), $\langle\hat{\tau}_{xy}(\nu_k)\rangle$ (bottom left panel), and $\delta_{\hat{\tau}}(\nu_k)$ (bottom right panel) for the $m=20$ smoothed estimates of the LS102.4 light curves in experiment CD1. The intrinsic CS and time-lag spectrum are shown as blue dashed lines. At low frequencies ($\lesssim2\times10^{-4}\,\mathrm{Hz}$) where the intrinsic CS has the highest curvature (i.e. where its second derivative has the largest value), the bias of the smoothed cross-periodogram and time-lag estimate is maximised.
 
   \begin{figure}
   \centering
   \includegraphics[width=\hsize]{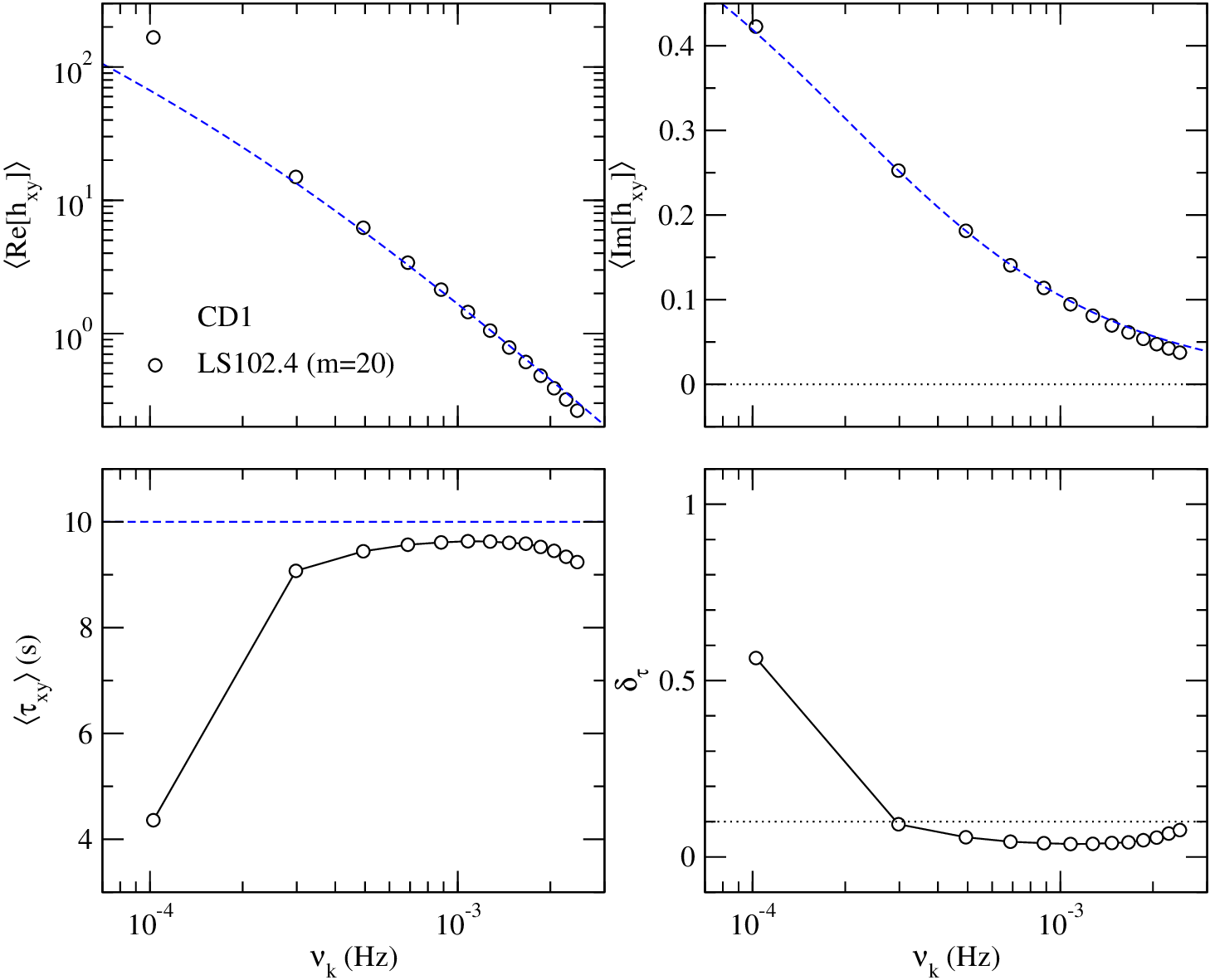}
      \caption{As in Figs.\,\ref{fig1} and \ref{fig6}, for the $m=20$ smoothed estimates obtained from the LS102.4 light curves in experiment CD1. The blue dashed lines indicate the model CS (upper panels) and time-lag spectrum (lower left panel).}
      \label{fig8}
   \end{figure}
   
The mean sample coherence at all frequencies is $\approx1$ in most, but not all, cases. Figure\,\ref{figd9} shows the mean sample coherence for experiments CD2, CD3, and PLD5 (filled circles and solid lines indicate the $m=20$ smoothed and averaged estimates, respectively). Despite the fact that, by construction, $\gamma^2_{\mathscr{XY}}(\nu)=1$ at all frequencies, the mean sample coherence is significantly smaller than unity in these cases. This is difficult to explain and predict. The coherence bias is a complicated function of the $\langle\hat{c}^2_{xy}(\nu_k)\rangle$ and $\langle\hat{q}^2_{xy}(\nu_k)\rangle$ bias, as well as the bias of the smoothed/averaged PSD estimates. For example, although the mean real part of the smoothed cross-periodogram is biased in experiment CD1 when $m=20$ (see the top left panel in Fig.\,\ref{fig8}), the mean sample coherence is not, presumably because it is counterbalanced by the bias of the smoothed PSD estimates.
 
\section{The effect of measurement errors} \label{sec7}

Every measured signal will inevitably be affected by experimental `noise'. To study the effect of measurement errors on the time-lag and coherence estimates, we considered the LS20.4 and LS40.8 light curve pairs in experiments CD1, PLD2, and THRF1. For these experiments, we increased the number of original realisations from 30 to 100. This increase in number of simulated light curves for each numerical experiment results in a significant increase in the computing time. For that reason, in this and the following section we consider only experiments CD1, PLD2, and THRF1. They do not exhibit phase-flipping effects, and are representative of the three categories of model phase-lag spectra we consider. In total, we thus ended up with $5\times10^4$ LS20.4 and $2\times10^4$ LS40.8 light curve pairs for each experiment.

Furthermore, we created five copies of each of these pairs to simulate the effects of different signal-to-noise ratio (S/N) combinations, $\{\mathrm{(S/N)}_x,\mathrm{(S/N)}_y\}$, for each experiment. The effects of measurement errors were simulated by adding a Gaussian random number with zero mean to each point of every light curve, which were uncorrelated both with each other, as well as with the intrinsic light curve values. The variance of the random numbers was chosen such that there were five different S/N combinations for each light curve pair in the aforementioned experiments: $\{\mathrm{(S/N)}_x,\mathrm{(S/N)}_y\}=\{3,3\}$, $\{9,3\}$, $\{18,3\}$, $\{9,9\}$, and $\{18,9\}$.

We then calculated the $m=10$, 20, 30, and 40 averaged cross-periodogram to estimate the time-lags and coherence according to Eqs.\,\ref{eq20} and \ref{eq23}, along with their error estimates according to Eqs.\,\ref{eq22} and \ref{eq24}. The number of averaged time-lag and coherence estimates were thus 5000 (2000), 2500 (1000), 1666 (666), and 1250 (500) for the LS20.4 (LS40.8) light curves in each experiment and every S/N combination. Figures\,\ref{figd10}--\ref{figd12} show $\delta_{\hat{\tau}}(\nu_k)$ (top row) and $\langle\hat{\gamma}^2_{xy}(\nu_k)\rangle$ (bottom row) for the LS20.4/40.8 light curves (solid and dashed lines, respectively) for all S/N combinations, in each of the three experiments and for $m=20$.

\subsection{The effect of measurement errors on the time-lag bias} \label{subsec71}

   \begin{figure}
   \centering
   \includegraphics[width=\hsize]{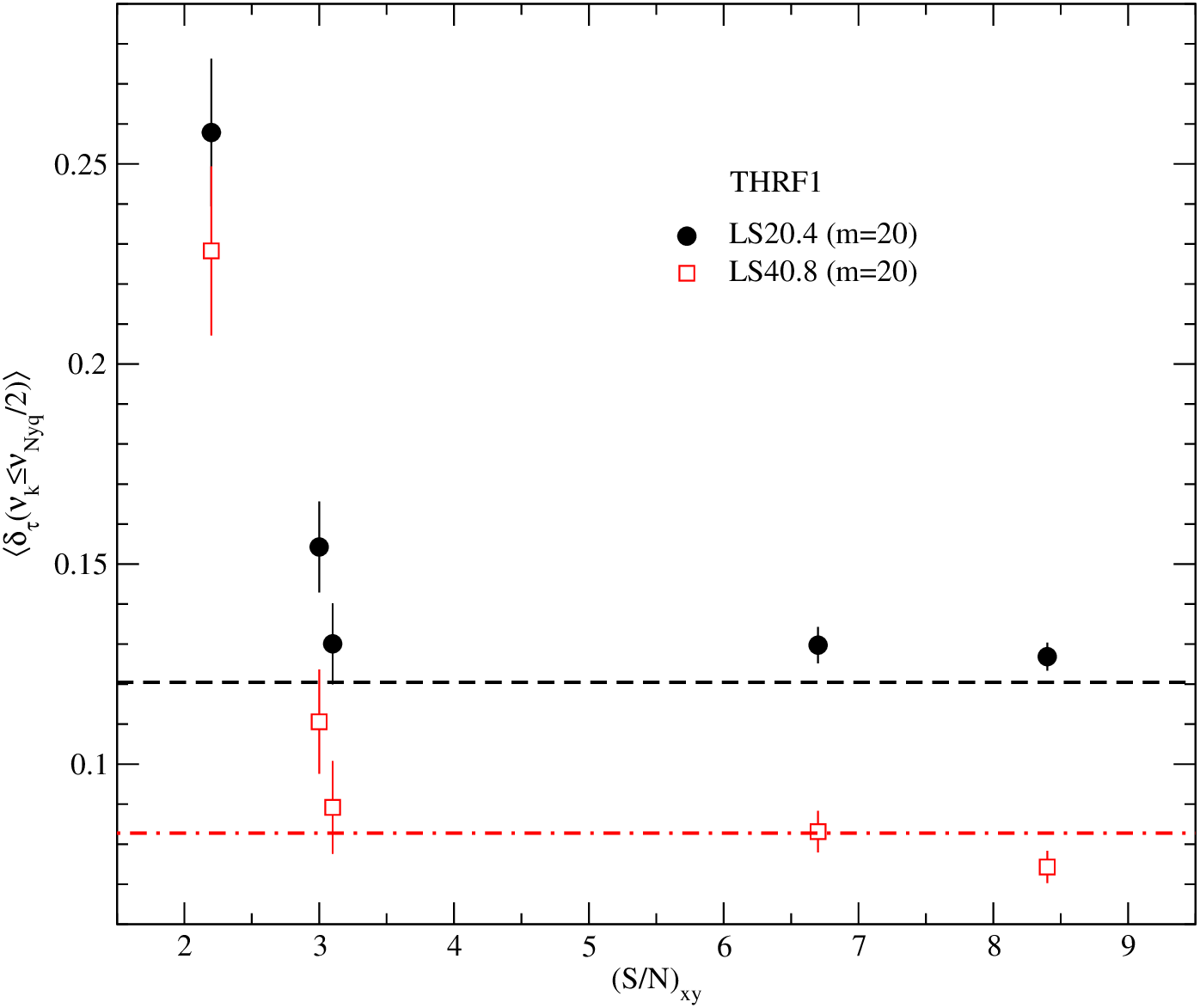}
      \caption{Relative bias of the averaged time-lag estimates for the LS20.4 and LS40.8 light curves, plotted as a function of $\mathrm{(S/N)}_{xy}$ in experiment THRF1. The horizontal dashed lines correspond to the value of the relative bias when $\mathrm{(S/N)}_{xy}\rightarrow\infty$.}
      \label{fig9}
   \end{figure}

A comparison between, for example, the top panels in Fig.\,\ref{figd10} and the bottom left panel in Fig.\,\ref{figd5} reveals that the time-lag estimates become increasingly biased at high frequencies (in the sense that they converge to zero) with decreasing S/N. As the S/N of the light curves increases, $\delta_{\hat{\tau}}(\nu_k)$ becomes consistent with its corresponding value when there is no noise present. To illustrate the effect of noise on the time-lag bias, we computed $\langle\delta_{\hat{\tau}}(\nu_k\le\nu_{\mathrm{Nyq}}/2)\rangle$ as a function of $\mathrm{(S/N)}_{xy}\equiv[\mathrm{(S/N)}_x^{-2}+\mathrm{(S/N)}_y^{-2}]^{-1/2}$ for experiment THRF1 (this quantity is a measure of the combined S/N of both light curves). The results are shown in Fig.\,\ref{fig9} for the LS20.4 and LS40.8 light curves. The horizontal dashed black and dotted-dashed red line correspond to the $\langle\delta_{\hat{\tau}}(\nu_k\le\nu_{\mathrm{Nyq}}/2)\rangle$ value when there is no noise (i.e. when $\mathrm{(S/N)}_{xy}\rightarrow\infty$) in the LS20.4 (filled black circles) and LS40.8 (open red squares) light curves, respectively. Clearly, the time-lag estimates are more biased when $\mathrm{(S/N)}_{xy}\lesssim3$. Identical trends are observed in the case of experiments CD1 and PLD2 as well.

This is a rather unexpected result, since the noise introduced to one light curve is independent of the noise introduced to the other. We would therefore not expect the time-lag bias to be affected by noise (although we would expect the resulting estimates to have a larger scatter around the mean, which is indeed the case; compare, for example, the scatter of the time-lags in the top panels in Fig.\,\ref{figd10} and in the bottom left panel of Fig.\,\ref{figd5}). As in the case of phase-flipping (see Sect.\,\ref{subsec53}), this bias arises because the phase-lag estimates are defined on the interval $(-\pi,\pi]$. As an example, in Fig.\,\ref{fig10} we show the probability distribution of phase-lag estimates for the LS40.8 light curves with $\mathrm{(S/N)}_x=\mathrm{(S/N)}_y=9$ in experiment THRF1 at three different frequencies: $6.1\times10^{-4}\,\mathrm{Hz}$, $1.2\times10^{-3}\,\mathrm{Hz}$, and $2.5\times10^{-3}\,\mathrm{Hz}$ (top, middle, and bottom panel, respectively). Filled black bars and open red bars indicate the distribution of the $m=10$ and $m=40$ averaged phase-lag estimates, respectively. As the frequency increases, the mean sample coherence decreases (see the next section), and hence the scatter of the phase-lag estimate increases according to Eq.\,\ref{eq21}. When the magnitude of this scatter becomes sufficiently large, the `wings' of the distribution exceed the range of allowed values for the phase-lag estimate, and are thus folded back into the allowed range. Consequently, the distribution becomes increasingly uniform over the interval $(-\pi,\pi]$, and hence its mean converges to zero. This bias reduces as $m$ increases, since, in this case, the scatter of the phase-lag estimates themselves, and hence their bias, decreases (see Eq.\,\ref{eq21}). This is also evident from Fig.\,\ref{fig10}, as the `width' of the distributions is smaller in the case of larger $m$.

The vertical dashed lines in all panels of Figs.\,\ref{figd10}--\ref{figd12} indicate the frequency at which the sample coherence becomes equal to $1.2/(1+0.2m)$ (this value is indicated by the horizontal dotted-dashed lines in the lower panels of the the same figures). We refer to this frequency as the critical frequency, $\nu_{\rm crit}$, and in Sects.\,\ref{sec8} and \ref{sec9} we discuss its importance in detail. Interestingly, we found that, on average, the time-lag bias is similar to its value in the absence of measurement errors at frequencies below $\nu_{\rm crit}$. The same result holds for the time-lag estimates in the case of $m=10$, 30, and 40 as well.

   \begin{figure}[h!]
   \centering
   \includegraphics[width=\hsize]{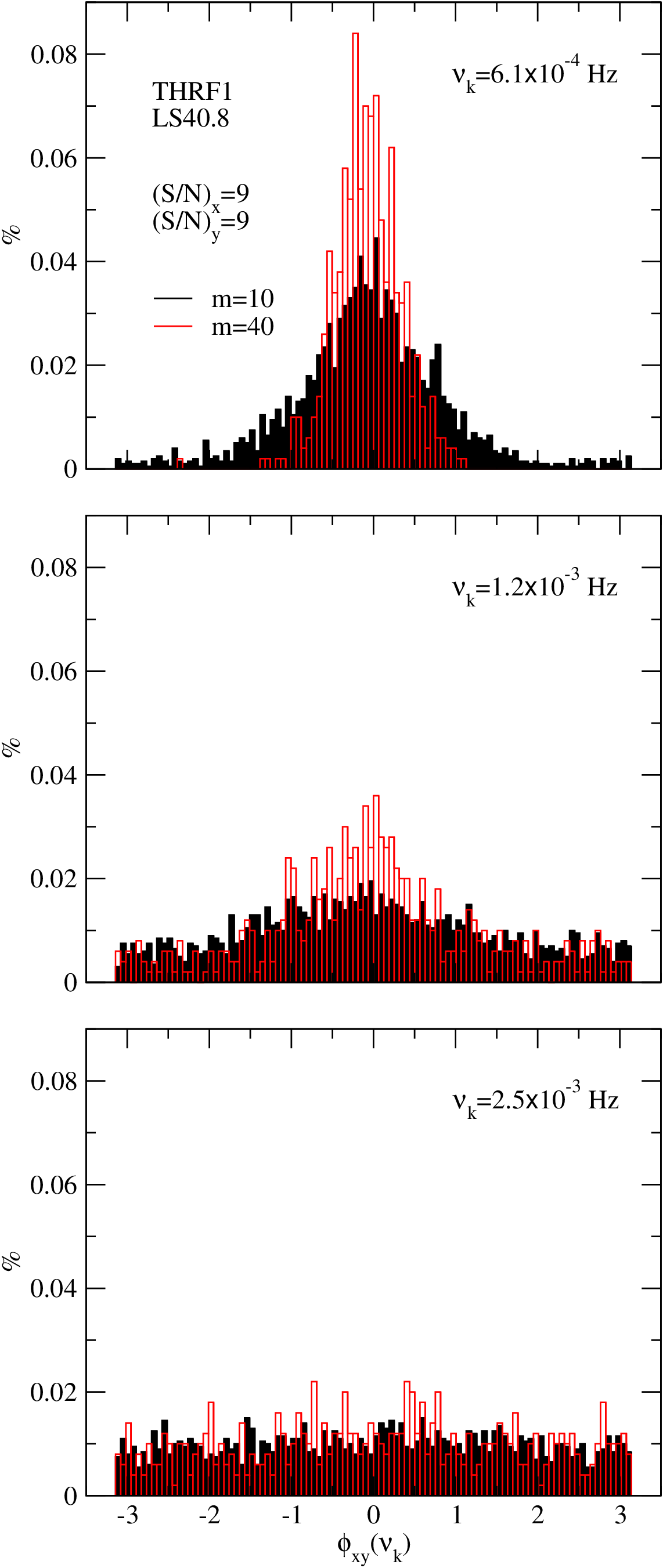}
      \caption{Probability distribution of the averaged phase-lag estimate for the LS40.8 light curves in experiment THRF1 at different frequencies.}
      \label{fig10}
   \end{figure}

\subsection{Effect of measurement errors on the sample coherence} \label{subsec72}

Measurement errors have the intuitive effect of decreasing the intrinsic coherence between two time series. By construction, the intrinsic coherence of all light curves is equal to one at every frequency. As we show in Appendix \ref{appc}, noise decreases the intrinsic coherence at all frequencies. This decrease is minimal at frequencies where the amplitude of the measured signal's intrinsic variations dominates over the amplitude of the noise variations. In the opposite case, the mean sample coherence will be significantly less than unity.

This is exactly what we see in the bottom panels of Figs.\,\ref{figd13}--\ref{figd17}; the sample coherence is always less than unity at high frequencies, even in the highest S/N cases. The decrease depends on the S/N; at a given frequency, the mean sample coherence decreases with decreasing S/N. By fitting various functions to the sample mean coherence we found that, for each numerical experiment and both light curve types, the following simple exponential function describes them well (see Appendix \ref{appc} for a theoretical justification of this fact):
\noindent
\begin{equation} \label{eq25}
\langle\hat{\gamma}^2_{xy}(\nu_k)\rangle=\left(1-\frac{1}{m}\right)\mathrm{exp}[-(\nu/\nu_0)^\alpha]+\frac{1}{m}. 
\end{equation}
\noindent
This is demonstrated in Fig.\,\ref{fig11}, which shows the mean sample coherence for the LS40.8 light curves ($m=20$) in experiment THRF1 for various S/N combinations (we get the same results in the case of experiments CD1 and PLD2 as well). The red dashed lines in this figure indicate the best-fit of the function defined above.

The blue horizontal dashed lines in the same figure indicate the value $1/m$. As we demonstrate in Appendix \ref{appc}, at frequencies where experimental noise dominates the intrinsic variations, the mean sample coherence is expected to be equal to that value. This is indeed what we observe in our simulations, as in the case of the lower S/N light curves the sample coherence indeed converges to the value of $1/m$ (see the top panel in Fig.\,\ref{fig11}).

   \begin{figure}[ht!]
   \centering
   \includegraphics[width=\hsize]{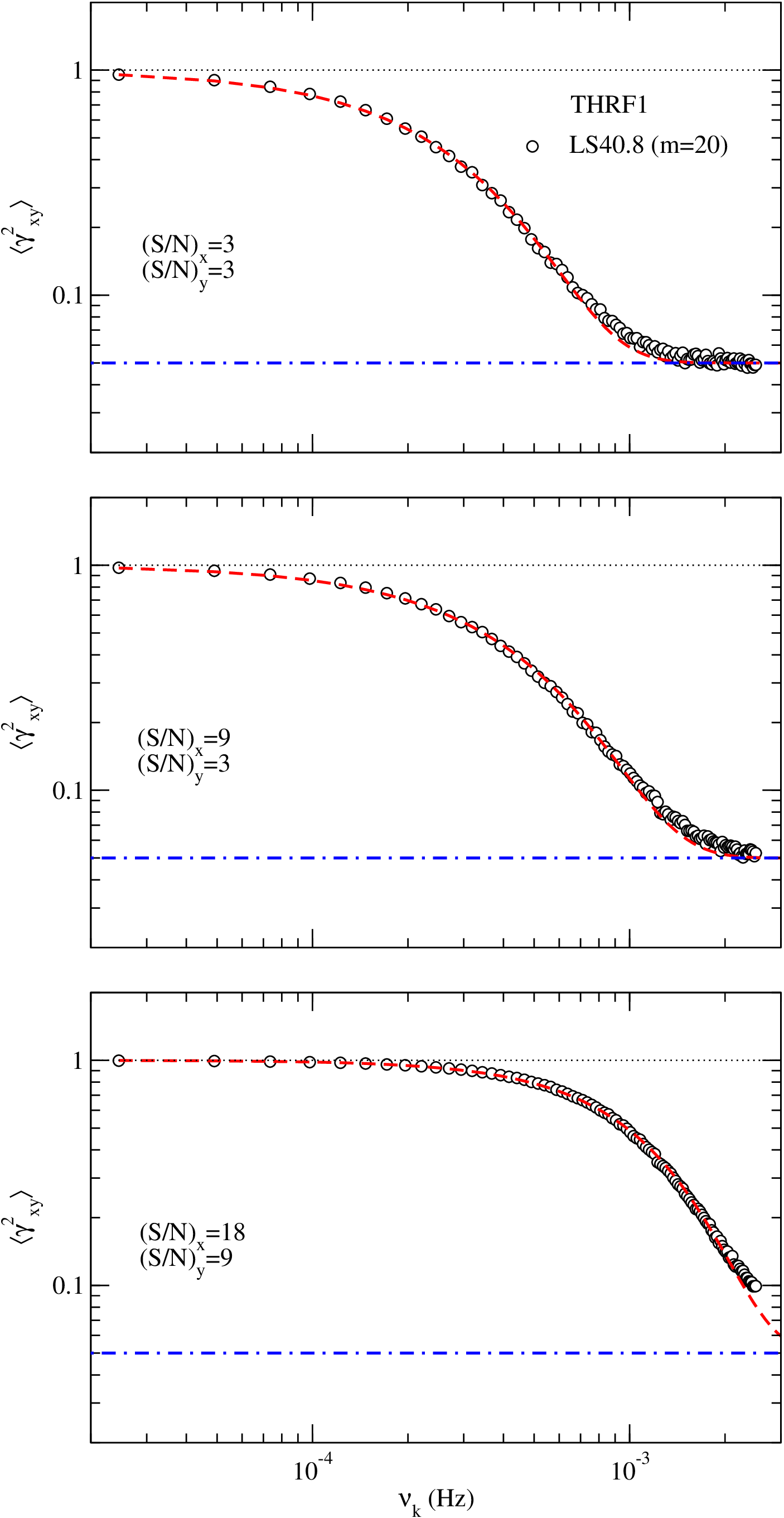}
      \caption{Mean sample coherence ($m=20$) obtained from the LS40.8 light curves in experiment THRF1 for different S/N combinations. The blue horizontal dashed lines indicate the value $1/m$ (see Sect.\,\ref{subsec72} for details).}
      \label{fig11}
   \end{figure}

\section{The error of the time-lag estimates} \label{sec8}

The top panels in Figs.\,\ref{figd13}--\ref{figd17} show the mean sample coherence for the LS40.8 and LS20.4 light curves and various S/N combinations. Columns 1, 2, 3, and 4 show the results for $m=10$, 20, 30, and 40, respectively. The plots show the mean sample coherence for all experiments where we considered the effects of noise (CD1, PLD2, and THRF1). They are difficult to identify, because  their values are practically identical for all S/N combinations. In the middle panels, we plot the so-called error ratio, which we define as $\sigma_{\hat{\tau}}(\nu_k)/\sqrt{\mathrm{Var}[\hat{\tau}(\nu_k)]}$, where $\sigma_{\hat{\tau}}(\nu_k)$ is the mean analytical error estimate (computed using Eq.\,\ref{eq22} by substituting the value of the intrinsic coherence by its estimate), and $\sqrt{\mathrm{Var}[\hat{\tau}(\nu_k)]}$ is the observed $1\sigma$ scatter of the averaged time-lag estimates around their mean. The bottom panels in the same figures show the probability
$p(|\hat{\tau}(\nu_k)-\langle\hat{\tau}(\nu_k)\rangle|\le\sigma_{\hat{\tau}}(\nu_k))$, i.e. the probability that the value of the time-lag estimate falls within the range $[\langle\hat{\tau}(\nu_k)\rangle-\sigma_{\hat{\tau}}(\nu_k),\langle\hat{\tau}(\nu_k)\rangle+\sigma_{\hat{\tau}}(\nu_k)]$ (i.e. within $1\sigma$ of the sample mean).

The black, red, and green lines in the same figures indicate the results obtained from experiments CD1, PLD2, and THRF1, respectively. Continuous and dashed lines correspond to the LS40.8 and LS20.4 light curves. For a given $m$ and S/N combination, the results are almost identical for the different experiments and light curve types. This indicates that the error of the time-lag estimates depends mainly on $m$ and the S/N combination, irrespective of the intrinsic CS and light curve duration.
 
The ratio of the analytical error over the observed time-lag standard deviation is larger than $\approx0.9$ and approaches unity as $m$ increases, at frequencies which are lower than a certain `critical' frequency, say $\nu_{\mathrm{crit}}$. This frequency is indicated by the vertical dashed lines in all panels of Figs.\,\ref{figd13}--\ref{figd17}. In the same frequency range, the probability $p(|\hat{\tau}(\nu_k)-\langle\hat{\tau}(\nu_k)\rangle|\le\sigma_{\hat{\tau}}(\nu_k))$ is close to 0.68 (this value is indicated by the horizontal dotted lines in the bottom panels of the same figures). This is what we would expect if the distribution of the time-lag estimates is approximately Gaussian. 

The critical frequency is independent of the intrinsic CS and on the light curve duration, and depends mainly on the light curve S/N and on $m$; $\nu_{\mathrm{crit}}$ increases with increasing $m$ and S/N. We found that $\nu_{\mathrm{crit}}$ can be estimated by equating the mean sample coherence to the value $1.2/(1+0.2m)$, i.e.
\noindent
\begin{equation} \label{eq26}
\langle\hat{\gamma}^2_{xy}(\nu_{\mathrm{crit}})\rangle=\frac{1.2}{1+0.2m}.
\end{equation}
\noindent
This coherence value is indicated by the blue dotted-dashed horizontal line in the top panels of Figs.\,\ref{figd13}--\ref{figd17}.

At higher frequencies, the analytical error underestimates the true scatter around the mean of the time-lag estimates, and so the error ratio decreases. As we show in Appendix \ref{appc}, the mean sample coherence is always larger than the coherence of the `noisy' processes at all frequencies. Since the analytical time-lag error estimate depends on the sample coherence (see Eqs.\,\ref{eq21} and \ref{eq22} and the comment at the end of Sect.\,\ref{sec6}), it is not surprising that it underestimates the true scatter around the mean, and that the error ratio is smaller than unity. The mean sample coherence becomes significantly larger than the coherence of the `noisy' processes at frequencies $\gtrsim\nu_{\mathrm{crit}}$. Consequently, the analytical error estimate begins to increasingly underestimate the true scatter around the mean and hence the error ratio decreases at the same frequencies.

Interestingly, at frequencies $\lesssim\nu_{\mathrm{crit}}$ the probability $p(|\hat{\tau}(\nu_k)-\langle\hat{\tau}(\nu_k)\rangle|\le\sigma_{\hat{\tau}}(\nu_k))$ also remains constant and close to 0.68 (i.e. the value expected for a Gaussian distribution). At higher frequencies it begins to decrease, indicating that the distribution of the time-lag estimates develops longer `tails' than would be expected from a Gaussian distribution. We investigate the properties of the probability distribution of the time-lag estimates in more detail below.

\section{The probability distribution of the time-lag estimates} \label{sec9}

The results presented above regarding the error of the time-lag estimates imply that, at frequencies below $\nu_{\rm crit}$, their distribution may be, approximately, Gaussian. To further investigate this issue, we estimated the excess kurtosis and skewness of the time-lag estimates distribution, for the three experiments and two light curve types. Figures\,\ref{figd18}--\ref{figd22} show our results (upper and middle panels for the excess kurtosis and skewness, respectively). Black, red, and green lines indicate the results obtained from experiments CD1, PLD2, and THRF1, respectively, while continuous and dashed lines correspond to the LS40.8 and LS20.4 light curves. As before, the lines in each panel overlap, indicating that the distribution of the time-lag estimates depends mainly on $m$ and the S/N, and not on the intrinsic CS or light curve duration.

The vertical lines in all panels indicate $\nu_{\rm crit}$, estimated as explained in the previous section. For a Gaussian distribution, the excess kurtosis and skewness should both be equal to zero. At frequencies lower than $\nu_{\rm crit}$, this is roughly the case for the distribution of the time-lag estimates when $m\gtrsim20$. In fact, the skewness is $\approx0$ at all frequencies, although the scatter of the sample skewness increases at frequencies higher than $\nu_{\rm crit}$. This result indicates that the distribution of the time-lag estimates are symmetric around their mean. The excess kurtosis on the other hand is significantly different from zero at high frequencies, asymptotically reaching the value of $-1.2$. This is the excess kurtosis of the uniform distribution.

The lower panels in the same figures show the probability, $p_{\rm KS}(\nu_k)$, that the time-lag distribution is Gaussian, with a mean and variance equal to the sample mean and variance of the distribution at each frequency. This probability was estimated using the Kolmogorov-Smirnov (KS) test. The dotted lines in the bottom panels show $p_{\rm KS}(\nu_k)=0.01$. This is the typical threshold probability that would normally be considered if one wanted to reject the hypothesis of a Gaussian distribution for the time-lag estimates. The vertical line in the bottom left panels of Figs.\,\ref{figd18}--\ref{figd22} show that this probability is higher than 0.01 at frequencies below the critical frequency, even in the case of $m=10$.

As we discussed above, at frequencies $\lesssim\nu_{\mathrm{crit}}$ the analytical time-lag error estimate differs by $\lesssim10\%$ from the true scatter around the mean. The reason for this effect can be explained by the properties of the mean sample coherence (as discussed in Appendix \ref{appc}). It is not easy to explain why the distribution of the time-lag estimates should approach a Gaussian in the same frequency range (i.e. below $\nu_{\mathrm{crit}}$), although our results indicate that this is the case.

Obviously, the time-lag distribution is not exactly Gaussian. In the case of the $m=10$ and 20 averaged time-lag estimates, the excess kurtosis is certainly larger than zero at all frequencies, although not dramatically so at frequencies $\lesssim\nu_{\mathrm{crit}}$. Likewise, $p_{\mathrm{KS}}(\nu_k)$ is rather low, but not less than 0.01 in the same frequency range. Perhaps the most interesting result for many practical applications is the fact that the $\pm1$ analytical error corresponds to 68\% of the time-lag distribution when $m\gtrsim20$.

\section{Discussion} \label{sec10}

Our investigation was based on both analytical work and extensive numerical simulations. The simulations consisted of generating intrinsically coherent artificial light curve pairs with a prescribed model PSD and CS. The case of unevenly sampled light curves can be treated with maximum likelihood methods \citep[e.g.][]{2010MNRAS.403..196M,2010MNRAS.408.1928M,2013ApJ...777...24Z}, which we will address in a future work.

The artificial light curve characteristics (i.e. duration, sampling rate and time bin size) resembled those offered by AGN observations with present and past X-ray satellites. We assumed the same model PSD for all light curves, with a RMS amplitude of $A=0.01$ and a power-law shape that smoothly `bends' from a slope of $-1$ to $-2$ after a `bend-frequency' $\nu_{\mathrm{b}}=2\times10^{-4}\,\mathrm{Hz}$. This particular shape is characteristic of AGN X-ray PSDs. The values of $A=0.01$ and $\nu_{\mathrm{b}}=2\times10^{-4}\,\mathrm{Hz}$ correspond to the mean values determined by fitting the X-ray PSDs of a sample of $\approx100$ nearby AGN \citep{2012A&A...544A..80G}.

The phases of the model CS correspond to phase-lags that are commonly assumed between AGN X-ray light curves in different energy bands. They are divided into three main categories; constant delays of 10, 150, and $550\,\mathrm{s}$, power-law time-lag spectra with an amplitude of $10^{-3}-10^{-1}\,\mathrm{s}$ and negative index $0.5-1.5$, and time-lag spectra expected in a simple reverberation scenario where the light curves are related by a top-hat response function characterised by a centroid of $t_0=200-2000\,\mathrm{s}$, width $\Delta=200-2000\,\mathrm{s}$ and $f=0.2$. The constant delays correspond to a light-crossing time of $\approx1-100r_{\mathrm{g}}$ and $\approx0.1-10r_{\mathrm{g}}$ for a black hole of mass $M_{\mathrm{BH}}=10^6\,\mathrm{M}_{\odot}$ and $10^7\,\mathrm{M}_{\odot}$, respectively ($r_{\mathrm{g}}=GM_{\mathrm{BH}}/c^3\approx5(M_{\mathrm{BH}}/10^6\,\mathrm{M}_{\odot})\,\mathrm{sec}$ is the gravitational radius of a black hole of mass $M_{\mathrm{BH}}$). Power-law delays with an amplitude of $\approx10^{-3}\,\mathrm{s}$ and a negative index of $\approx1$ are those typically observed between AGN X-ray light curves in different energy bands \citep[e.g.][]{2001ApJ...554L.133P,2011MNRAS.416L..94E}. In addition, modelling of observed time-lag spectra with a top-hat response function requires $t_0\approx\Delta\approx100-300\,\mathrm{s}$ (for AGN with $M_{\mathrm{BH}}\approx10^6\,\mathrm{M}_{\odot}$) and $f\approx0.1-0.3$ \citep[e.g.][]{2011MNRAS.412...59Z,2011MNRAS.416L..94E}. Despite the fact that the light curve characteristics and model time-lag spectra that we considered are directly applicable to AGN studies, most of our results, which we summarise below, should be relevant to all time-lag studies where evenly sampled time series are considered. 

\subsection{Light curve sampling and binning effects} \label{subsec101}

The time-lag estimates are affected by aliasing and binning of the observed time series. If the observed signal is the result of regularly recording the values of a continuous underlying processes at regular intervals $\Delta t$, then time-lags should be estimated at frequencies lower than $\approx\nu_{\mathrm{Nyq}}/5$ (where $\nu_{\mathrm{Nyq}}\equiv1/2\Delta t$ is the Nyquist frequency). This is because aliasing has the effect of decreasing and increasing the imaginary and real parts of the cross-periodogram, respectively. As a result, their ratio (which determines the phase- and time-lag estimates) is decreased, on average, and the time-lag bias increases.

If the observed series are binned (instead of sampled) over time intervals $\Delta t$, then time-lags should be estimated at frequencies $\lesssim\nu_{\mathrm{Nyq}}/2$. This is because binning suppresses the aliasing effect on the cross-periodogram. Our results are consistent with the work of \citet{1998ApJ...493L..71C}, who reported similar findings in the context of XRB studies.

The model PSDs and CS we considered decrease with increasing frequency, and hence aliasing affects the time-lag bias more severely at increasingly higher frequencies. This is, however, expected to be true regardless of the intrinsic CS, as long as the measured signal is a (stationary) random process with finite variance (which is equal to the integrated PSD over all frequencies).

\subsection{Phase-flipping effects} \label{subsec102}

Phase-flipping occurs when the intrinsic phase-lag spectrum exceeds the boundaries of the interval $(-\pi,\pi]$. In such cases, the phase-lag estimate will jump from $\pi$ to $-\pi$ (or vice-versa). This can severely alter the shape of the time-lag spectra; intrinsically constant or power-law-like time-lag spectra can show broad, prominent, `oscillatory' features. In theory, the transitions from $-\pi$ to $\pi$, or vice versa, should be sharp, and hence easy to detect. However, in practice this is not the case, as the time-lag estimates will be significantly biased in the vicinity of frequencies where phase-flipping occurs. The origin of this bias can be traced to the fact that the `wings' of the phase-lag estimate's distribution which exceed the interval $(-\pi,\pi]$ are folded back into the allowed range. This causes the mean of the distribution to shift towards zero. The magnitude of this bias cannot be predicted a priori, since it depends on both the intrinsic `width' of the phase-lag estimate's distribution, as well as the unknown intrinsic phase-lag spectrum.

\subsection{Smoothed vs averaged estimates} \label{subsec103}

It is customary to estimate the cross-periodogram from a single, long time series and then bin it over $m$ neighbouring frequencies (a process called smoothing), or from $m$ individual segments of identical duration and then bin the resulting cross-periodograms at each frequency. This process is necessary to reduce the error of the time-lag estimates, and is in fact necessary to predict their error. This requires an estimation of the coherence function itself (if we estimate the coherence from the `raw' cross-periodogram it will be equal to unity at all frequencies, irrespective of the intrinsic coherence; see P81).

Smoothing of the cross-periodogram can potentially introduce a serious bias to the time-lag estimates. For example, the model CS we considered have a power-law dependence on frequency. This can result in biased estimates of the real and imaginary part when we perform a linear type of smoothing. As a result, in some of the numerical experiments we performed, the time-lag estimates converge to zero at low ($\lesssim2\times10^{-4}\,\mathrm{Hz}$) frequencies. Since the time-lag bias owing to smoothing originates from the cross-periodogram bias, it can only be predicted by prescribing a model CS (and not just a model time-lag spectrum). We therefore suggest the cross-periodogram to be averaged over individual segments, rather than smoothed, before computing time-lag estimates.

\subsection{Measurement error effects} \label{subsec104}

In the presence of measurement errors, the coherence decreases at all frequencies. This effect becomes more severe at frequencies where the amplitude of noise variations dominates over the amplitude of the measured signal's intrinsic variations, and the coherence tends to zero. In the case when the intrinsic coherence (i.e. the coherence in the absence of measurement errors) is unity, we found that the mean sample coherence is well fitted by Eq.\,(\ref{eq25}). The parameters $\nu_0$ and $\alpha$ depend only on the S/N of the light curves. The lower the S/N, the lower $\nu_0$ and the steeper $\alpha$ will be. At frequencies $\gtrsim\nu_{\mathrm{crit}}$ the sample coherence is biased, in the sense that its mean converges to the constant $1/m$, while the coherence of the noisy processes tends to zero.

Measurement errors cause the distribution of the time-lag estimates to become increasingly uniform at high frequencies. Their mean converges to zero, and their scatter is poorly approximated by standard analytical prescriptions. This is because, as the coherence decreases with increasing frequency, the `width' of the phase-lag distribution increases. When this width is sufficiently large, the wings of the distribution exceed the interval $(-\pi,\pi]$, and are thus folded back into the allowed range. This causes the distribution to become increasingly uniform, and its mean to shift towards zero.

We found that measurement errors have a minimal effect on the time-lag bias at frequencies where the sample coherence is $\gtrsim1.2/(1+0.2m)$. Furthermore, the analytical time-lag error estimate will differ from the true scatter by $\lesssim10\%$ in the same frequency range, as long as $m\gtrsim10$. In addition, if $m\gtrsim20$, the $\pm1$ analytical error corresponds to 68\% of the time-lag distribution, and the probability distribution of the time-lag estimates approximates a Gaussian (see Sect.\,\ref{sec9}). These results hold regardless of the intrinsic PSDs and CS.

\subsection{The effects of finite light curve duration} \label{subsec105}

The results mentioned in Sects.\,\ref{subsec101}--\ref{subsec104} above should be applicable in all cases, i.e. they should be independent of the intrinsic PSDs and CS. However, one potentially important issue in correctly determining the intrinsic time-lag spectrum is the cross-periodogram bias. The mean cross-periodogram is not equal to the intrinsic CS, but to its convolution with the so-called F\'{e}jer kernel (see Eq.\,\ref{eq13}). This function approaches the Dirac $\delta$-function in the limit $T\rightarrow\infty$, but when $T$ is finite the bias is non-zero. Its magnitude is difficult to predict, as it depends on the shape of the unknown intrinsic CS.

Owing to the above fact, the time-lag estimates will also be biased. We found that, for the model CS we considered, the time-lag bias is positive, in the sense that the mean time-lag estimate at a given frequency has a smaller magnitude than its intrinsic value. We also found that the relative time-lag bias is $\lesssim15\%$ when $T\gtrsim20\,\mathrm{ks}$, for all model time-lag spectra we considered.

In general, we found that the time-lag bias decreases with increasing light curve duration as $1/\sqrt{T}$. However, it is difficult to determine the normalisation of this relation. For a given intrinsic CS, it depends on the frequency range on which time-lags are estimated as well as the cross-periodogram bias, i.e. the shape and amplitude of the intrinsic CS. In other words, the naive expectation of the relative time-lag bias being negligible if the light curve duration is much larger than the magnitude of the time-lag estimates is generally incorrect. For example, in experiment CD1, where the intrinsic time-lag spectrum is equal to $10\,\mathrm{s}$ at all frequencies, the relative time-lag bias is $\approx20\%$ at the lowest sampled frequency, even if we use $\approx100\,\mathrm{ks}$ light curves (see e.g. the continuous black curve curve in the top left panel of Fig.\,\ref{figd1}).

To illustrate the complexity of the time-lag bias' dependence on the intrinsic CS, we revisited experiment CD1. The solid line in Fig.\,\ref{fig12} shows the relative time-lag bias determined from $\approx100\,\mathrm{ks}$ light curves (this line is identical to the black line in the top left panel of Fig.\,\ref{figd1}). The dashed red line shows the expected relative time-lag bias for the same light curves and intrinsic time-lag spectrum, and an intrinsic PSD whose amplitude is larger by a factor of five. We used Eq.\,\ref{eq13} to estimate the mean cross-periodogram, $\mathrm{E}[I_{xy}(\nu_p)]$, and then determined the expected relative time-lag bias by computing $\{10\,\mathrm{s}-(2\pi\nu_p)^{-1}\mathrm{arg}\{\mathrm{E}[I_{xy}(\nu_p)]\}\}/10\,\mathrm{s}$ at each frequency. The relative bias does not change when compared to the original PSD amplitude. We then repeated the same calculation, but assumed a `bend-frequency' of $2\times10^{-5}\,\mathrm{Hz}$ (i.e. 10 times lower than the original value). The expected relative bias (shown by the blue dashed-dotted line in the same figure) is now increased at each frequency, reaching a value of $\approx30\%$ at the lowest frequency.

These results can be understood by inspecting Eq.\,\ref{eq13}, which determines the cross-periodogram bias. In our simulations, the CS is given by $h_{\mathscr{XY}}(\nu)=(\mu_{\mathscr{Y}}/\mu_{\mathscr{X}})h_\mathscr{X}(\nu)\mathrm{e}^{\mathrm{i}\phi_{\mathscr{XY}}(\nu)}$, where $h_\mathscr{X}(\nu)$ and $\phi_{\mathscr{XY}}(\nu)$ are the intrinsic PSD and phase-lag spectrum, respectively. Since the intrinsic PSD is given by Eq.\,\ref{eq14}, a change in the PSD amplitude has the effect of altering the amplitude of the imaginary and real parts of the intrinsic CS by the same multiplicative factor at each frequency. This however does not change their ratio, which determines the mean of the phase- and time-lag estimates, and hence leaves the relative time-lag bias unchanged. On the other hand, a change in the bend-frequency alters the amplitude of the imaginary and real parts of the intrinsic CS differently at each frequency, resulting in a change of the relative time-lag bias.

We therefore suggest that it is important to always check the expected relative time-lag bias to validate the time-lag analysis in practice. This can be done by assuming a possible model CS using Eq.\,\ref{eq13}. If the intrinsic coherence of the light curves is close to unity, such a model CS may be determined by modelling the PSDs and considering various model time-lag spectra that may be potential candidates for the intrinsic time-lag spectrum.

 \begin{figure}
   \centering
   \includegraphics[width=\hsize]{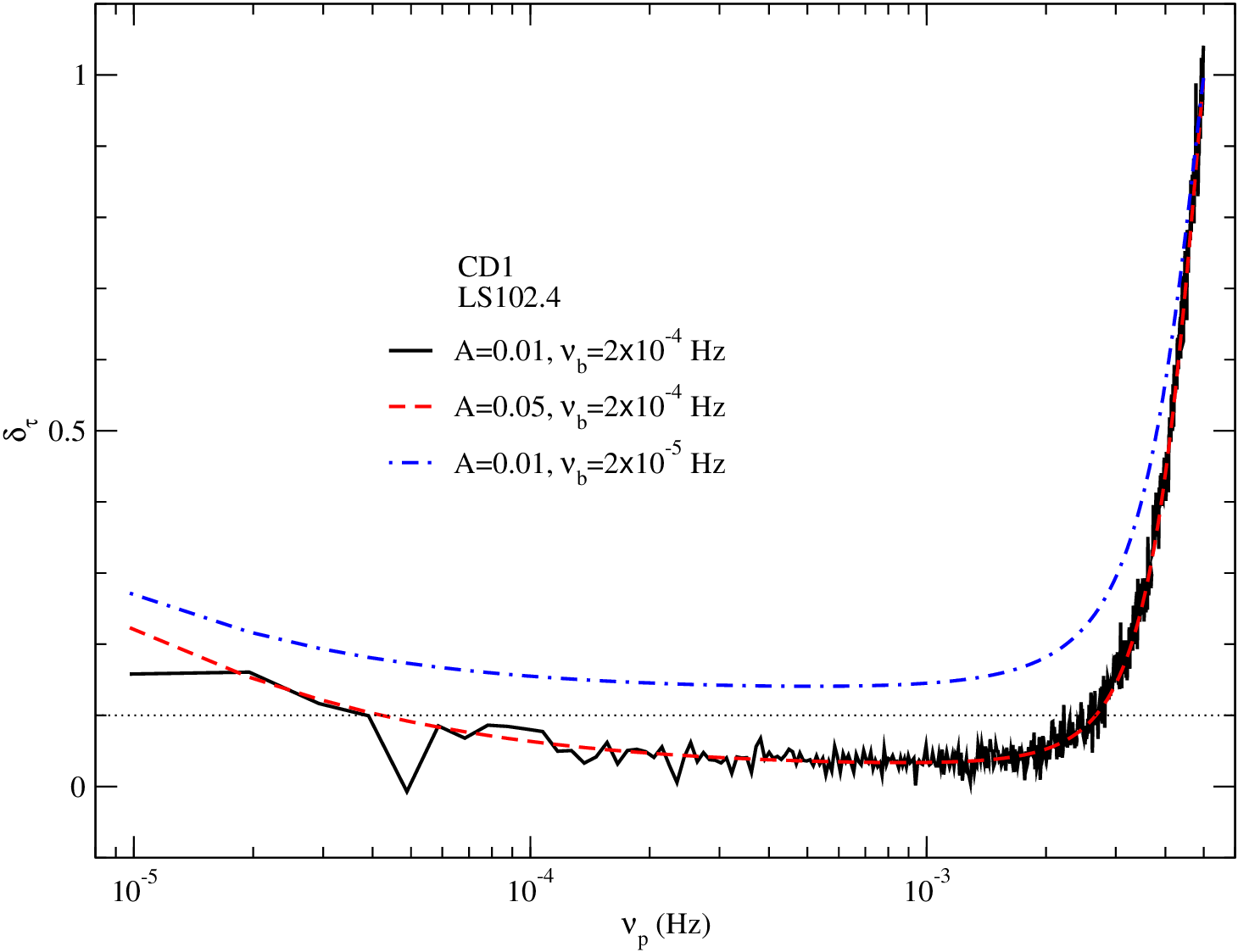}
      \caption{Relative time-lag bias for the LS102.4 light curves and different values of the model PSD parameters, $\{A,\nu_{\mathrm{b}}\}$, in experiment CD1.}
      \label{fig12}
   \end{figure}

\section{Summary} \label{sec11}

We investigated the statistical properties of Fourier-based time-lag estimates. These estimates are based on the cross-periodogram, which is an estimator of the intrinsic CS between two random time series. Unlike the periodogram (the traditional PSD estimator) which is always positive-definite, the cross-periodogram is generally a complex number whose phase is an estimator of the phase-lag spectrum (the time-lag is equal to the phase-lag divided by the angular frequency). Our aims were to quantify the effects of light curve characteristics and measurement errors on the time-lag bias (i.e. the difference between their mean and intrinsic values), to study their distribution, and to investigate whether the standard analytical error prescriptions are accurate approximations of their intrinsic scatter around the mean.

Based on our results, we suggest the following steps when estimating time-lags:
\noindent
\begin{itemize}
\item Estimate the averaged cross-periodogram (Eq.\,\ref{eq18}), using at least $m=20$ pairs of data segments.
\item Use the averaged cross-periodogram to estimate the coherence and time-lags (Eqs.\,\ref{eq23} and \ref{eq20}), along with their errors (Eqs.\,\ref{eq24} and \ref{eq22}).
\item Determine the frequency range for which the sample coherence is greater than $1.2/(1+0.2m)$. If the intrinsic coherence is close to unity,\footnote{I.e. if the observed decrease in the sample coherence at high frequencies is mostly caused by measurement errors.} then the coherence estimates should be well fitted by Eq.\,\ref{eq25}. By determining the best-fit $\nu_0$ and $\alpha$ values of this function, the aforementioned frequency range can then be determined by solving $(1-m^{-1})\mathrm{exp}[-(\nu/\nu_0)^\alpha]+m^{-1}\ge1.2/(1+0.2m)$.
\end{itemize}
\noindent

At the frequency range where the sample coherence is larger than $1.2/(1+0.2m)$, the bias of the averaged time-lag estimates will be similar to its value in the absence of measurement errors, the analytical prescription for the error of the time-lag estimates will be similar to the true scatter around the mean, and their distribution will be roughly approximated by a Gaussian, in the sense discussed in Sect.\,\ref{sec9}.

\begin{acknowledgements}
We thank the referee for his/her suggestions, which significantly improved the quality and clarity of the manuscript. This work was supported by the AGNQUEST project, which is implemented under the Aristeia II action of the Education and Lifelong Learning operational programme of the GSRT, Greece. It was also supported in part by the grant PIRSES-GA-2012-31578 EuroCal.
\end{acknowledgements}


\bibliographystyle{aa}
\bibliography{refs}


\begin{appendix}

\section{The cross-spectrum of two binned time series} \label{appa}

Let us consider the continuous process $\{\mathscr{X}(t),\mathscr{Y}(t)\}$ and the discrete process $\{X(t_s),Y(t_s)\}$, defined according to Eq.\,\ref{eq7} in Sect.\,\ref{sec3}. It is straight-forward to show that the processes have the same mean values. However, this is not true for their CCFs and CS as well.

For notational convenience and without loss of generality, we will henceforth assume that the intrinsic process has a mean value of zero. By definition, the CCF of the discrete process is only defined at lags which are integer multiples of the sampling period, i.e. $\tau_l=l\Delta t_{\mathrm{sam}}$, where $l=0,\pm1,\pm2,\ldots$, and is given by
\noindent
\begin{align}
\nonumber
& R_{XY}(\tau_l)\equiv\mathrm{E}[Y(t_s)X(t_s+\tau_l)] \\
\nonumber
&=\frac{1}{(\Delta t_{\mathrm{bin}})^2}\int_{t_s-(\Delta t_{\mathrm{bin}}/2)}^{t_s+(\Delta t_{\mathrm{bin}}/2)}\int_{t_s-(\Delta t_{\mathrm{bin}}/2)}^{t_s+(\Delta t_{\mathrm{bin}}/2)}\mathrm{E}[\mathscr{Y}(t')\mathscr{X}(t''+\tau_l)]\mathrm{d}t'\mathrm{d}t'' \\
\label{eqa1}
&=\frac{1}{(\Delta t_{\mathrm{bin}})^2}\int_{t_s-(\Delta t_{\mathrm{bin}}/2)}^{t_s+(\Delta t_{\mathrm{bin}}/2)}\int_{t_s-(\Delta t_{\mathrm{bin}}/2)}^{t_s+(\Delta t_{\mathrm{bin}}/2)}R_{\mathscr{XY}}(\tau_l+t''-t')\mathrm{d}t'\mathrm{d}t'',
\end{align}
\noindent
where $R_{\mathscr{XY}}(\tau)\equiv\mathrm{E}[\mathscr{Y}(t)\mathscr{X}(t+\tau)]$ is, by definition, the CCF of the continuous process. According to Eq.\,\ref{eqa1} $R_{XY}(\tau_l)$ and $R_{\mathscr{XY}}(\tau_l)$ are not equal at any value of $\tau_l$. Given that $R_{\mathscr{XY}}(\tau)=\int_{-\infty}^{\infty}h_{\mathscr{XY}}(\nu)\mathrm{e}^{\mathrm{i}2\pi\nu\tau}\mathrm{d}\nu$, Eq.\,\ref{eqa1} becomes
\noindent
\begin{align}
\nonumber
R_{XY}(\tau_l) &=\frac{1}{(\Delta t_{\mathrm{bin}})^2} \int_{t_s-(\Delta t_{\mathrm{bin}}/2)}^{t_s+(\Delta t_{\mathrm{bin}}/2)}\int_{t_s-(\Delta t_{\mathrm{bin}}/2)}^{t_s+(\Delta t_{\mathrm{bin}}/2)} \\
\nonumber
& \quad \left[\int_{-\infty}^{\infty}h_{\mathscr{XY}}(\nu)\mathrm{e}^{\mathrm{i}2\pi\nu(\tau_l+t''t')}\mathrm{d}\nu\right]
\mathrm{d}t'\mathrm{d}t'' \\
\nonumber
&=\int_{-\infty}^{\infty}h_{\mathscr{XY}}(\nu)\left[\frac{1}{(\Delta t_{\mathrm{bin}})^2}\left\lvert\int_{t_s-(\Delta t_{\mathrm{bin}}/2)}^{t_s+(\Delta t_{\mathrm{bin}}/2)}\mathrm{e}^{\mathrm{i}2\pi\nu t}\mathrm{d}t\right\rvert^2\right] \mathrm{e}^{\mathrm{i}2\pi\nu\tau_l}\mathrm{d}\nu \\
\label{eqa2}
&=\int_{-\infty}^{\infty}h_{\mathscr{XY}}(\nu)\mathrm{sinc}^2(\pi\nu\Delta t_{\mathrm{bin}})\mathrm{e}^{\mathrm{i}2\pi\nu\tau_l}\mathrm{d}\nu.
\end{align}
\noindent
Splitting the infinite frequency integration interval into an infinite number of segments with width $1/2\Delta t_{\mathrm{sam}}$, Eq.\,\ref{eqa2} becomes
\noindent
\begin{align}
\nonumber
R_{XY}(\tau_l) &=\sum_{k=-\infty}^{\infty}\int_{(2k-1)/2\Delta t_{\mathrm{sam}}}^{(2k+1)/2\Delta t_{\mathrm{sam}}}h_{\mathscr{XY}}(\nu)\mathrm{sinc}^2(\pi\nu\Delta t_{\mathrm{bin}})\mathrm{e}^{\mathrm{i}2\pi\nu\tau_l}\mathrm{d}\nu \\
\nonumber
&=\int_{-1/2\Delta t_{\mathrm{sam}}}^{1/2\Delta t_{\mathrm{sam}}}\Bigg\{\sum_{k=-\infty}^{\infty}h_{\mathscr{XY}}\left(\nu+\frac{k}{\Delta t_{\mathrm{sam}}}\right)\times \\
\label{eqa3}
&\quad\times\mathrm{sinc}^2\left[\pi\left(\nu+\frac{k}{\Delta t_{\mathrm{sam}}}\right)\Delta t_{\mathrm{bin}}\right]\Bigg\}\mathrm{e}^{\mathrm{i}2\pi\nu\tau_l}\mathrm{d}\nu,
\end{align}
\noindent
since $\rm{e}^{i2\pi k l}=1$ for all $k$, $l$. The relation between the CCF and CS of the discrete process, $h_{XY}(\nu)$ (which is only defined on the interval $|\nu|\le1/2\Delta t_{\mathrm{sam}}$), is
\noindent
\begin{equation} \label{eqa4}
R_{XY}(\tau_l)=\int_{-1/2\Delta t_{\mathrm{sam}}}^{1/2\Delta t_{\mathrm{sam}}}h_{XY}(\nu)\mathrm{e}^{\mathrm{i}2\pi\nu\tau_l}\mathrm{d}\nu.
\end{equation}
\noindent
Comparing Eqs.\,\ref{eqa3} and \ref{eqa4}, we arrive at the desired result:
\noindent
\begin{equation} \label{eqa5}
h_{XY}(\nu)=\sum_{k=-\infty}^{\infty}h_{\mathscr{XY}}\left(\nu+\frac{k}{\Delta t_{\mathrm{sam}}}\right)\mathrm{sinc}^2\left[\pi\left(\nu+\frac{k}{\Delta t_{\mathrm{sam}}}\right)\Delta t_{\mathrm{bin}}\right].
\end{equation}

\section{The mean of the cross-periodogram} \label{appb}

Using Eqs.\,\ref{eq9} and \ref{eq10}, we may write the cross-periodogram of the time series $\{x(t_r),y(t_r)\}$ as follows:
\noindent
\begin{align}
\nonumber
I_{xy}(\nu) &=\frac{\Delta t_{\mathrm{sam}}}{N}\sum_{r=1}^{N}[x(t_r)-\overline{x}]\mathrm{e}^{-\mathrm{i}2\pi\nu r\Delta t_{\mathrm{sam}}}\sum_{k=1}^{N}[y(t_k)-\overline{y}]\mathrm{e}^{\mathrm{i}2\pi\nu k\Delta t_{\mathrm{sam}}} \\
\label{eqb1}
&=\frac{\Delta t_{\mathrm{sam}}}{N}\sum_{r=1}^{N}\sum_{k=1}^{N}[x(t_r)-\overline{x}][y(t_k)-\overline{y}]\mathrm{e}^{-\mathrm{i}2\pi\nu(r-k)\Delta t_{\mathrm{sam}}}.
\end{align}
\noindent  
We can now transform the indices $r$ and $k$ to $s=r-k$ and $l$, where $s$ goes from $\{-(N-1)\}$ to $(N-1)$, and, at fixed $s$, $l$ goes from 1 to $(N-|s|)$. With this transformation, Eq.\,\ref{eqb1} becomes
\noindent
\begin{equation} \label{eqb2}
I_{xy}(\nu)=\Delta t_{\mathrm{sam}}\sum_{s=-(N-1)}^{(N-1)}\frac{1}{N}\sum_{l=1}^{N-|s|} [y(t_l)-\overline{y}][x(t_l+\tau_s)-\overline{x}]\mathrm{e}^{-\mathrm{i}2\pi\nu s\Delta t_{\mathrm{sam}}},
\end{equation}
\noindent
where $(1/N)\sum_{l=1}^{N-|s|} [y(t_l)-\overline{y}][x(t_l+\tau_s)-\overline{x}]$ is, by definition, CCF, $\hat{R}_{xy}(s)$, and $\tau_s=t_j-t_k=s\Delta t_{\mathrm{sam}}$ is the lag. Therefore, Eq.\,\ref{eqb2} can be re-cast into the following form:
\noindent
\begin{equation} \label{eqb3}
I_{xy}(\nu)=\Delta t_{\mathrm{sam}}\sum_{s=-(N-1)}^{(N-1)}\hat{R}_{xy}(s)\mathrm{e}^{-\mathrm{i}2\pi\nu s\Delta t_{\mathrm{sam}}}.
\end{equation}
\noindent
Applying the expectation operator on both sides of Eq.\,\ref{eqb3}, we get
\begin{align}
\nonumber
\mathrm{E}[I_{xy}(\nu)] &=\Delta t_{\mathrm{sam}}\sum_{s=-(N-1)}^{(N-1)}\mathrm{E}[\hat{R}_{xy}(\tau_s)]\mathrm{e}^{-\mathrm{i}2\pi\nu s\Delta t_{\mathrm{sam}}} \\
\label{eqb4}
&= \Delta t_{\mathrm{sam}}\sum_{s=-(N-1)}^{(N-1)}\left(1-\frac{|s|}{N}\right)R_{XY}(s)\mathrm{e}^{-\mathrm{i}2\pi\nu s\Delta t_{\mathrm{sam}}},
\end{align}
\noindent
where we have ignored the effect of estimating the mean values of the observed time series to reach the last equation. $R_{XY}(s)$ is the CCF of the discrete processes $\{X(t_s),Y(t_s)\}$ that was defined in Sect.\,\ref{sec3}. Since the CCF is equal to the inverse Fourier transform of the CS, Eq.\,\ref{eqb4} becomes
\begin{align}
\nonumber
\mathrm{E}[I_{xy}(\nu)] &=\Delta t_{\mathrm{sam}}\sum_{s=-(N-1)}^{(N-1)}\left(1-\frac{|s|}{N}\right)\times \\
\nonumber
&\quad\times\left[\int_{-\nu_{\mathrm{Nyq}}}^{\nu_{\mathrm{Nyq}}}h_{XY}(\nu')\mathrm{e}^{\mathrm{i}2\pi\nu's\Delta t_{\mathrm{sam}}}d\nu'\right]\mathrm{e}^{-\mathrm{i}2\pi\nu s\Delta t_{\mathrm{sam}}} \\
\nonumber
&=\int_{-\nu_{\mathrm{Nyq}}}^{\nu_{\mathrm{Nyq}}}h_{XY}(\nu')\Delta t_{\mathrm{sam}}\sum_{s=-(N-1)}^{(N-1)}\left(1-\frac{|s|}{N}\right)\mathrm{e}^{\mathrm{i}2\pi(\nu'-\nu)s\Delta t_{\mathrm{sam}}}\mathrm{d}\nu' \\
\label{eqb5}
&=\int_{-\nu_{\mathrm{Nyq}}}^{\nu_{\mathrm{Nyq}}}h_{XY}(\nu')F_N(\nu'-\nu)\mathrm{d}\nu',
\end{align}
\noindent
where
\noindent
\begin{align}
\nonumber
F(\nu) &\equiv\Delta t_{\mathrm{sam}}\sum_{s=-(N-1)}^{(N-1)}\left(1-\frac{|s|}{N}\right)\mathrm{e}^{\mathrm{i}2\pi\nu s\Delta t_{\mathrm{sam}}} \\
\label{eqb6}
&=\frac{\Delta t_{\mathrm{sam}}}{N}\left[\frac{\sin(N\pi\nu\Delta t_{\mathrm{sam}})}{\sin(\pi\nu\Delta t_{\mathrm{sam}})}\right]^2
\end{align}
\noindent
is the so-called Fej\'{e}r kernel. One of the properties of this function is that, as $N\rightarrow\infty$, $F_N(\nu)\rightarrow\delta(\nu)$ (the Dirac $\delta$-function). Thus, as $N\rightarrow\infty$ the right-hand side of Eq.\,\ref{eqb5} converges to the intrinsic CS (modified by the effects of discrete binning or sampling according to Eq.\,\ref{eqa5}) value at $\nu$, $h_{XY}(\nu)$. Therefore, the cross-periodogram is an asymptotically unbiased estimator of the CS.

\section{Effects of measurement errors on the coherence} \label{appc}

Let us consider an intrinsically continuous random process $\{\mathscr{X}(t),\mathscr{Y}(t)\}$, whose mean, PSDs and CS are $\{\mu_{\mathscr{X}},\mu_{\mathscr{Y}}\}$, $\{h_{\mathscr{X}}(\nu),h_{\mathscr{Y}}(\nu)\}$ and $h_{\mathscr{XY}}(\nu)$, respectively. Let us also assume that this process is continuously sampled at regular time intervals $t_s=s\Delta t_{\mathrm{sam}}$, where $s=0,\pm1,\pm2,\ldots$, and binned over time bins of size $\Delta t_{\mathrm{bin}}$. In the absence of measurement errors, the measured signal corresponds to a single realisation of a discrete version of the intrinsic process, $\{X(t_s),Y(t_s)\}$, as defined by Eq.\,\ref{eq7}. If there are measurement errors (as is always the case), the observed time series constitute a realisation of the `noisy' process
\noindent
\begin{equation} \label{eqc1}
\{X_{\mathrm{n}}(t_s),Y_{\mathrm{n}}(t_s)\}=\{X(t_s)+\epsilon_X,Y(t_s)+\epsilon_Y\},
\end{equation}
\noindent
where $\epsilon_X$ and $\epsilon_Y$ are a `purely random' processes with zero mean, and variance $\sigma^2_{\epsilon_X}$ and $\sigma^2_{\epsilon_Y}$, respectively. In this case,
\noindent
\begin{align}
\label{eqc2}
h_{X,\mathrm{n}}(\nu) &=h_X(\nu)+h_{\epsilon_X} \quad \text{and} \\
\label{eqc3}
h_{Y,\mathrm{n}}(\nu) &=h_Y(\nu)+h_{\epsilon_Y},
\end{align}
\noindent
where $h_{\epsilon_X}=\sigma^2_{\epsilon_X}\Delta t_{\rm sam}$ and $h_{\epsilon_X}=\sigma^2_{\epsilon_Y}\Delta t_{\rm sam}$ are the PSDs of the purely random processes. Equations\,\ref{eqc2} and \ref{eqc3} describe the well-known fact that the addition of so-called white noise to a process has the effect of adding a constant amount of `power' to its intrinsic PSD. When the measurement errors in one light curve are uncorrelated with the errors in the other, then $h_{XY,\mathrm{n}}(\nu)=h_{XY}(\nu)$. As a result, 
\noindent
\begin{align}
\nonumber
\gamma^2_{XY,\mathrm{n}}(\nu) &\equiv\frac{|h_{XY,\mathrm{n}}(\nu)|^2}{h_{X,\mathrm{n}}(\nu)h_{Y,\mathrm{n}}(\nu)}=\frac{|h_{XY}(\nu)|^2}{[h_X(\nu)+h_{\epsilon_X}][h_Y(\nu)+h_{\epsilon_Y}]} \\
\label{eqc4}
& <\frac{|h_{XY}(\nu)|^2}{h_X(\nu)h_Y(\nu)}\equiv\gamma^2_{XY}(\nu),
\end{align}
\noindent
where $\gamma^2_{XY,\mathrm{n}}(\nu)$ and $\gamma^2_{XY}(\nu)$ is the coherence of the noisy and the `intrinsic' process, respectively (here intrinsic refers to the discrete process $\{X(t_s),Y(t_s)\}$). According to Eq.\,\ref{eqc4}, the presence of (intrinsically uncorrelated) measurement errors decreases the  coherence of the processes under study. In fact, at high frequencies where the intrinsic PSDs are much smaller than $h_{\epsilon_X}$ and $h_{\epsilon_Y}$, the ratio 
$\gamma^2_{XY,n}(\nu)/\gamma^2_{XY}(\nu)$ will be equal to $h_X(\nu)h_Y(\nu)/(h_{\epsilon_X}h_{\epsilon_X})$. In the case of power-law like PSDs, which decrease with increasing frequency, as is typically the case in AGN, this ratio will then tend to zero at high frequencies.

In addition to decreasing the intrinsic coherence, measurement errors will introduce a bias in the sample coherence as well, i.e. the mean of sample coherence will not be equal to $\gamma^2_{XY,\mathrm{n}}(\nu)$. This can be shown by following \citep{1997ApJ...474L..43V}, who showed that
\noindent
\begin{equation} \label{eqc5}
|\hat{h}_{xy}(\nu_k)|^2=|\hat{h}^{(0)}_{xy}(\nu_k)|+|\hat{\varsigma}(\nu_k)|^2,
\end{equation}
\noindent
where $\hat{h}^{(0)}_{xy}(\nu_k)$ is the CS estimate in the absence of measurement errors, and $\hat{\varsigma}(\nu_k)$ is a random complex number with zero mean and variance given by
\noindent
\begin{align}
\nonumber
\mathrm{E}[\left\lvert\hat{\varsigma}(\nu_k)\right\rvert^2] &\sim\frac{1}{m}\left[h_X(\nu_k)h_{\epsilon_Y}+h_Y(\nu_k)h_{\epsilon_X}+h_{\epsilon_X}
h_{\epsilon_Y}\right] \\
\label{eqc6}
&\sim\frac{1}{m}\{[h_X(\nu_k)+h_{\epsilon_X}][h_Y(\nu_k)+h_{\epsilon_Y}]-h_X(\nu_k)h_Y(\nu_k)\}.
\end{align}
\noindent
Since $\hat{\varsigma}(\nu_k)$ is a random complex number, it may be represented by a vector in the complex plane with a random phase distributed uniformly over the interval $(-\pi,\pi]$. As such, $\hat{\varsigma}(\nu_k)$ randomly `perturbs' both the magnitude as well as the direction of the vector $\hat{h}^{(0)}_{xy}(\nu_k)$. On average this perturbation will have no net effect on either the magnitude or direction of $\hat{h}_{xy}(\nu_k)$, hence $\mathrm{E}[|\hat{h}^{(0)}_{xy}(\nu_k)|^2]\sim|h_{XY}(\nu_k)|^2$, although it will obviously increase its variance. This implies that measurement errors will not bias the cross-periodogram (although they will bias the phase-lag estimates at high frequencies, as discussed in Sect.\,\ref{subsec71}).

The mean sample coherence can now be determined. By first applying the expectation operator on both sides of Eq.\,\ref{eq23}, we get
\noindent
\begin{align}
\nonumber
\mathrm{E}[\hat{\gamma}^2_{xy}(\nu_k)] &=\mathrm{E}\left[\frac{|\hat{h}_{xy}(\nu_k)|^2}{\hat{h}_x(\nu_k)\hat{h}_y(\nu_k)}\right] \\
\label{eqc7}
&\sim\frac{\mathrm{E}[|\hat{h}_{xy}(\nu_k)|]^2}{\mathrm{E}[\hat{h}_x(\nu_k)]\mathrm{E}[\hat{h}_y(\nu_k)]}.
\end{align}
Given that $\mathrm{E}[\hat{h}_x(\nu_k)]\sim h_{X,\mathrm{n}}(\nu_k)$ and $\mathrm{E}[\hat{h}_y(\nu_k)]\sim h_{Y,\mathrm{n}}(\nu_k)$, if we substitute Eqs.\,\ref{eqc2}, \ref{eqc3}, \ref{eqc4} and \ref{eqc6} into Eq.\,\ref{eqc7}, we get
\noindent
\begin{align}
\nonumber
\mathrm{E}[\hat{\gamma}^2_{xy}(\nu_k)] &\sim\frac{|h_{XY}(\nu_k)|^2}{[h_X(\nu_k)+h_{\epsilon_X}][h_Y(\nu_k)+h_{\epsilon_Y}]} \\
\nonumber
& \quad +\frac{1}{m}\frac{[h_X(\nu_k)+h_{\epsilon_X}][h_Y(\nu_k)+h_{\epsilon_Y}]-h_X(\nu_k)h_Y(\nu_k)}{[h_X(\nu_k)+h_{\epsilon_X}][h_Y(\nu_k)+h_{\epsilon_Y}]} \\
\nonumber
&=\gamma^2_{XY,\mathrm{n}}(\nu_k) \\
\nonumber
& \quad +\frac{1}{m}-\frac{1}{m}\frac{h_X(\nu_k)h_Y(\nu_k)}{[h_X(\nu_k)+h_{\epsilon_X}][h_Y(\nu_k)+h_{\epsilon_Y}]} \\
\nonumber
&=\gamma^2_{XY,\mathrm{n}}(\nu_k) \\
\nonumber
& \quad +\frac{1}{m}-\frac{1}{m}\frac{\gamma^2_{XY,\mathrm{n}}(\nu_k)}{\gamma^2_{XY}(\nu_k)} \\
\label{eqc8}
&=\left[1-\frac{1}{m}\gamma^{-2}_{XY}(\nu_k)\right]\gamma^2_{XY,\mathrm{n}}(\nu_k)+\frac{1}{m},
\end{align}
\noindent
where $\gamma^2_{XY,\mathrm{n}}(\nu_k)$ is defined by Eq.\,\ref{eqc4}. The above equation holds true for any intrinsic coherence. In our case where $\gamma^2_{XY}(\nu_k)=1$ at all frequencies, Eq.\,\ref{eqc8} reduces to
\noindent
\begin{equation} \label{eqc9}
\mathrm{E}[\hat{\gamma}^2_{xy}(\nu_k)]\sim\left(1-\frac{1}{m}\right)\gamma^2_{XY,\mathrm{n}}(\nu_k)+\frac{1}{m}.
\end{equation}
\noindent
This relation is identical to Eq.\,\ref{eq25} when $\gamma^2_{XY,\mathrm{n}}(\nu)=\mathrm{exp}[-(\nu/\nu_0)^\alpha]$. We found that such a relation fits the mean sample coherence well in the case of experiments CD1, PLD2, and THRF1, and both light curve types we considered in Sects.\,\ref{sec7}--\ref{sec9}. A comparison between this relation and the mean sample coherence is shown in Fig.\,\ref{fig11} for the LS40.8 light curves in experiment THRF1. Equation\,\ref{eqc9} indicates that when $m$ is large, the mean sample coherence will be equal to the coherence of the noisy process.

However, in general, at sufficiently high frequencies the ratio $\gamma^2_{XY,\mathrm{n}}(\nu_k)/\gamma^2_{XY}(\nu_k)$ tends to zero because $\gamma^2_{XY,\mathrm{n}}(\nu_k)$ tends to zero and is always smaller than $\gamma^2_{XY}(\nu_k)$ at all frequencies (Eq.\,\ref{eqc4}). In this case, the first term in the right-hand side of Eq.\,\ref{eqc8} will therefore converge to zero. Consequently,
\noindent
\begin{equation} \label{eqc10}
\mathrm{E}[\hat{\gamma}^2_{xy}(\nu_k)]\sim\frac{1}{m}.
\end{equation}
\noindent
The above equation shows that, in the frequency range where the experimental noise variations dominate over the intrinsic ones (i.e. when $\gamma^2_{XY,\mathrm{n}}(\nu_k)/\gamma^2_{XY}(\nu_k)\ll1$), the mean sample coherence will be equal to $1/m$.

\section{Figures that show our results} \label{appd}

\begin{figure*}
\centering
\includegraphics[width=14.2cm,clip]{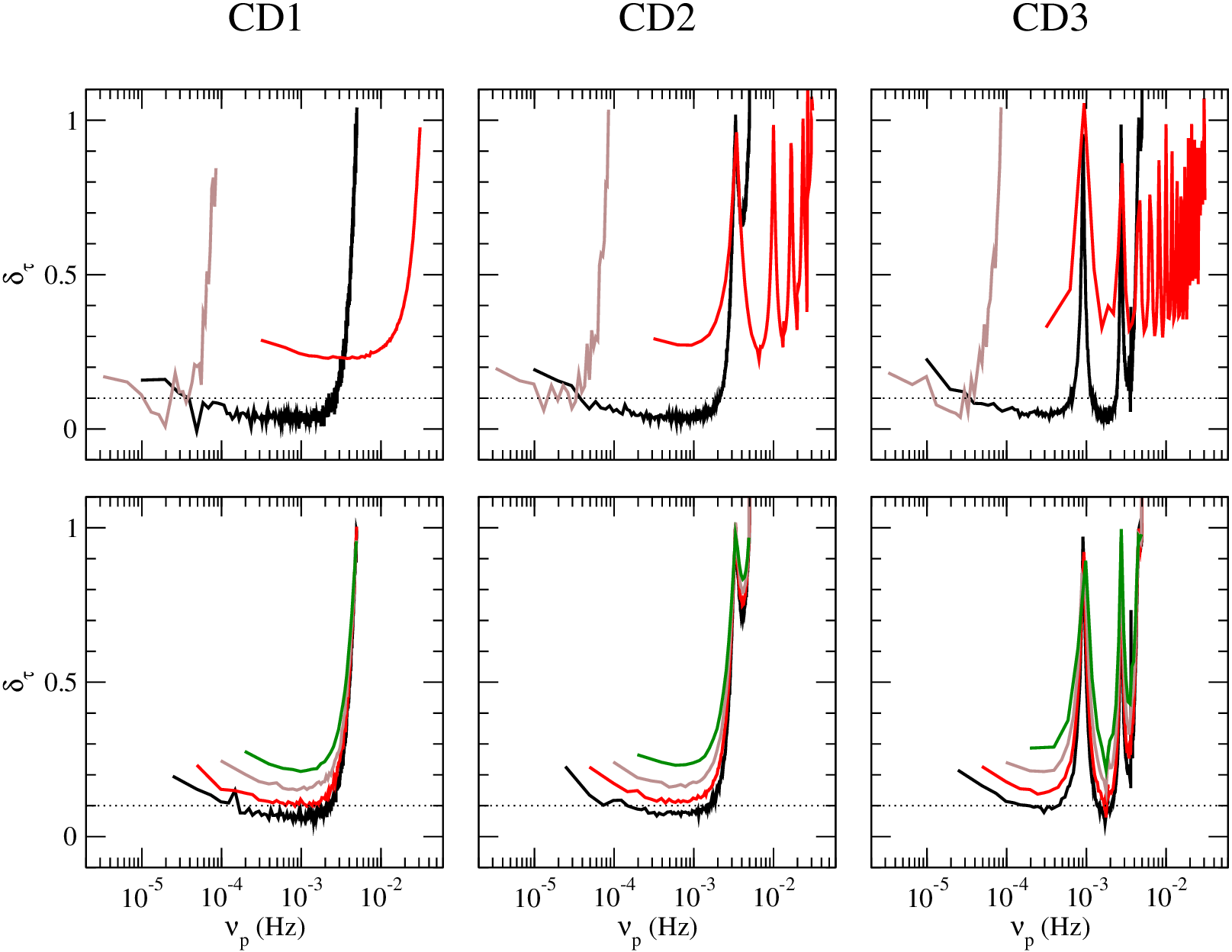}
\caption{Relative time-lag bias for experiments CD1 (first column), CD2 (second column), and CD3 (third column). Top row: Relative time-lag bias for the LS102.4 (black curves), LS3.2 (red curves), and OB (brown curves) light curves. Bottom row: Relative time-lag bias for the LS40.8 (black curves), LS20.4 (red curves), LS10.2 (brown curves), and LS5.1 (green curves) light curves. The horizontal dotted lines in this, and all similar subsequent figures, indicate the 0.1 relative time-lag bias.}
\label{figd1}
\end{figure*}

\begin{figure*}
\centering
\includegraphics[width=14.2cm,clip]{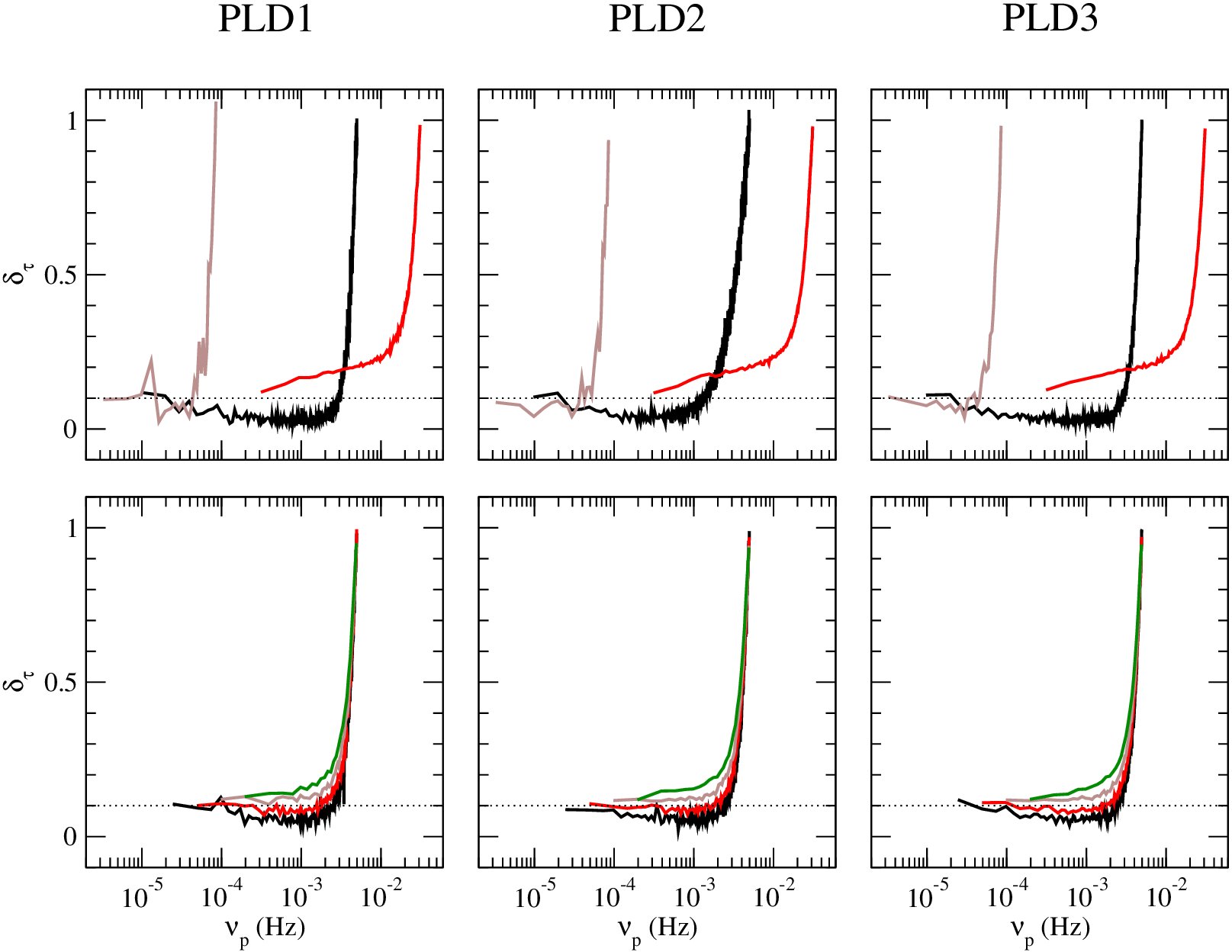}
\caption{As in Fig.\,\ref{figd1}, for experiments PLD1 (first column), PLD2 (second column), and PLD3 (third column).}
\label{figd2}
\end{figure*}

\begin{figure*}
\centering
\includegraphics[width=12.7cm,clip]{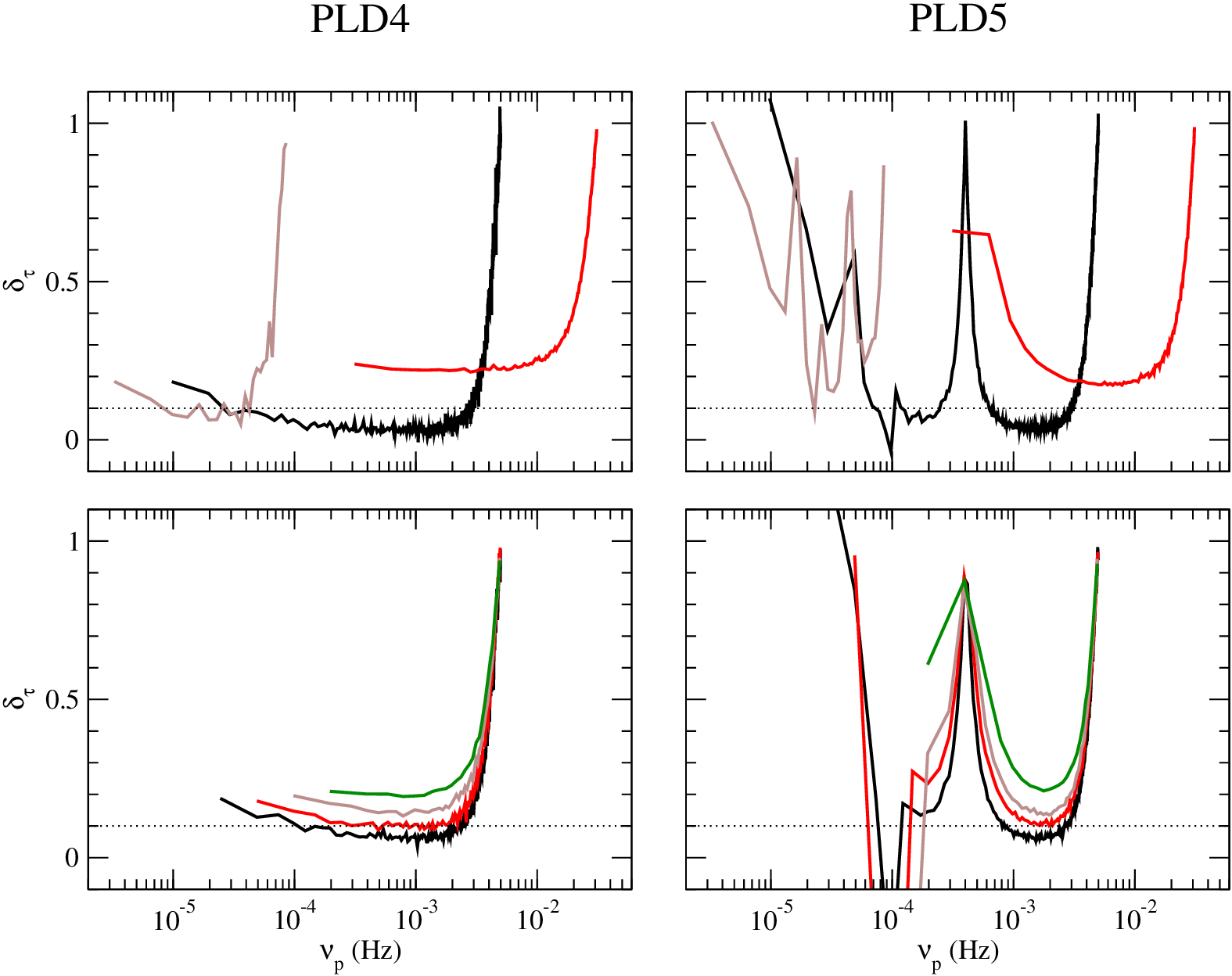}
\caption{As in Fig.\,\ref{figd1}, for experiments PLD4 (first column) and PLD5 (second column).}
\label{figd3}
\end{figure*}

\begin{figure*}
\centering
\includegraphics[width=12.7cm,clip]{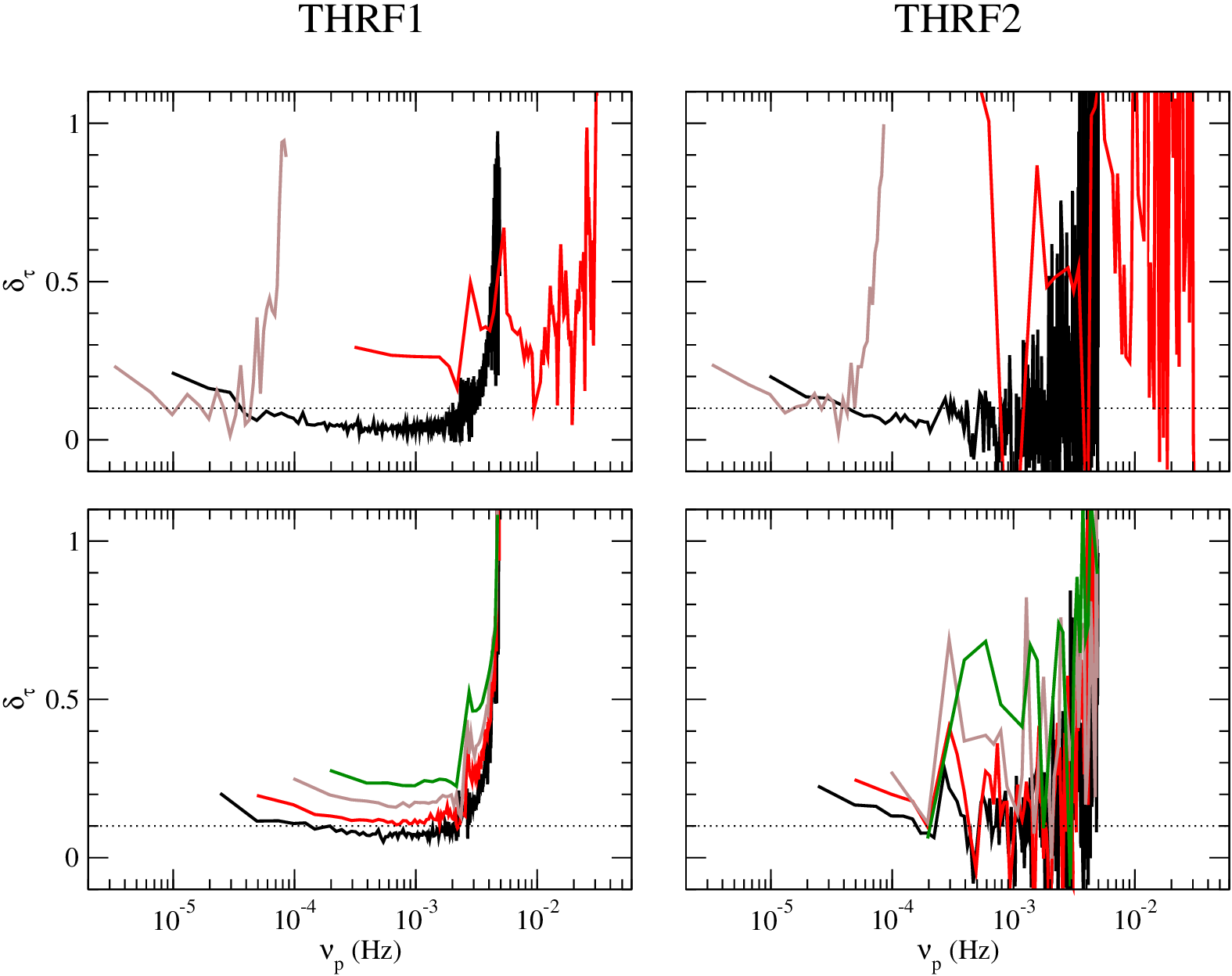}
\caption{As in Fig.\,\ref{figd1}, for experiments THRF1 (first column) and THRF2 (second column).}
\label{figd4}
\end{figure*}

\begin{figure*}
\centering
\includegraphics[width=14cm,clip]{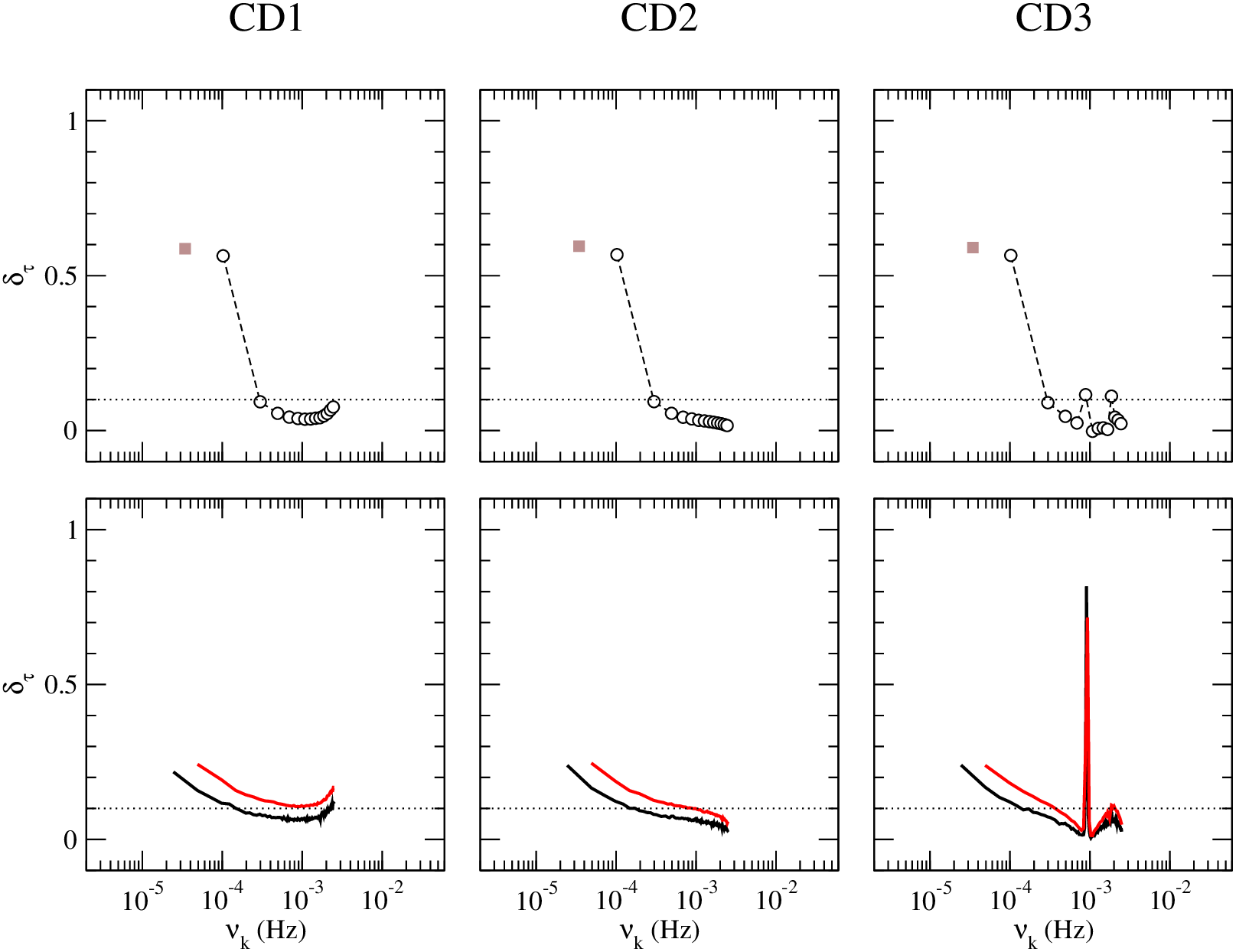}
\caption{Relative time-lag bias for experiments CD1 (first column), CD2 (second column), and CD3 (third column). Top row: Relative time-lag bias for the LS102.4 light curves with a smoothing parameter $m=20$ (open black circles), and OB light curves with a smoothing parameter $m=13$ (filled brown squares). Bottom row: Relative time-lag bias for the LS40.8 (black curves) and LS20.4 (red curves) light curves in the case of $m=20$ segments.}
\label{figd5}
\end{figure*}

\begin{figure*}
\centering
\includegraphics[width=14cm,clip]{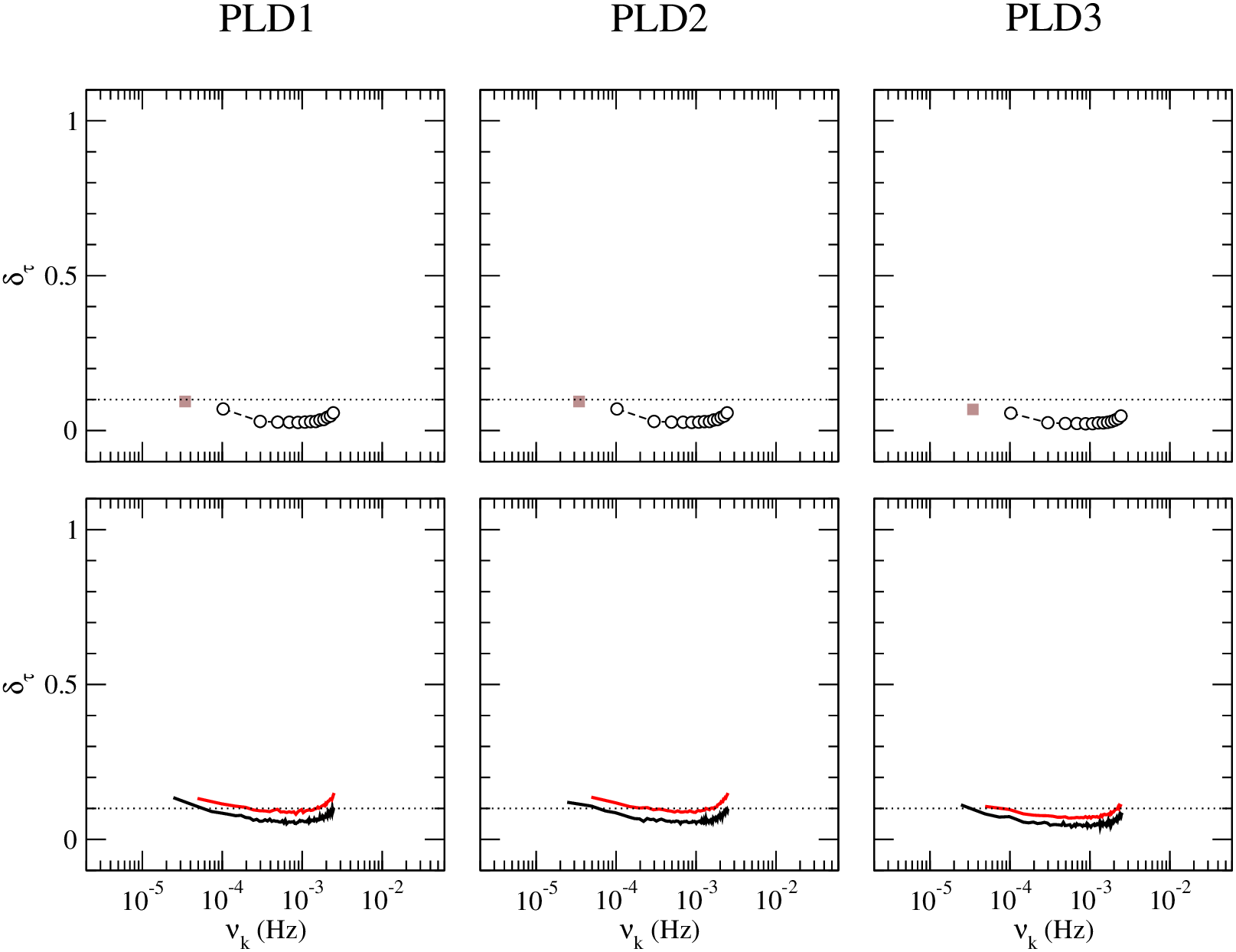}
\caption{As in Fig.\,\ref{figd5}, for experiments PLD1 (first column), PLD2 (second column), and PLD3 (third column).}
\label{figd6}
\end{figure*}

\begin{figure*}
\centering
\includegraphics[width=14cm,clip]{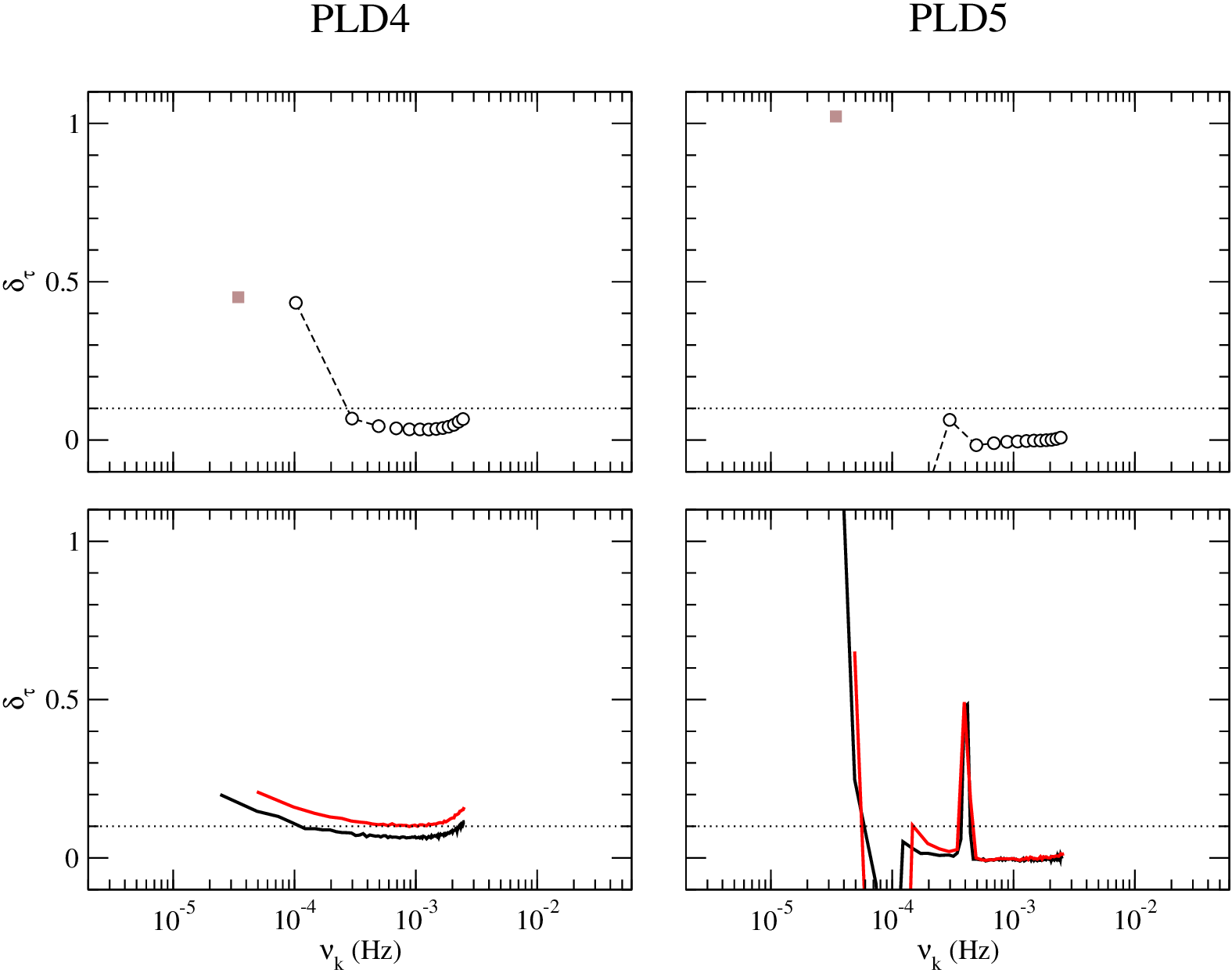}
\caption{As in Fig.\,\ref{figd5}, for experiments PLD4 (first column) and PLD5 (second column).}
\label{figd7}
\end{figure*}

\begin{figure*}
\centering
\includegraphics[width=14cm,clip]{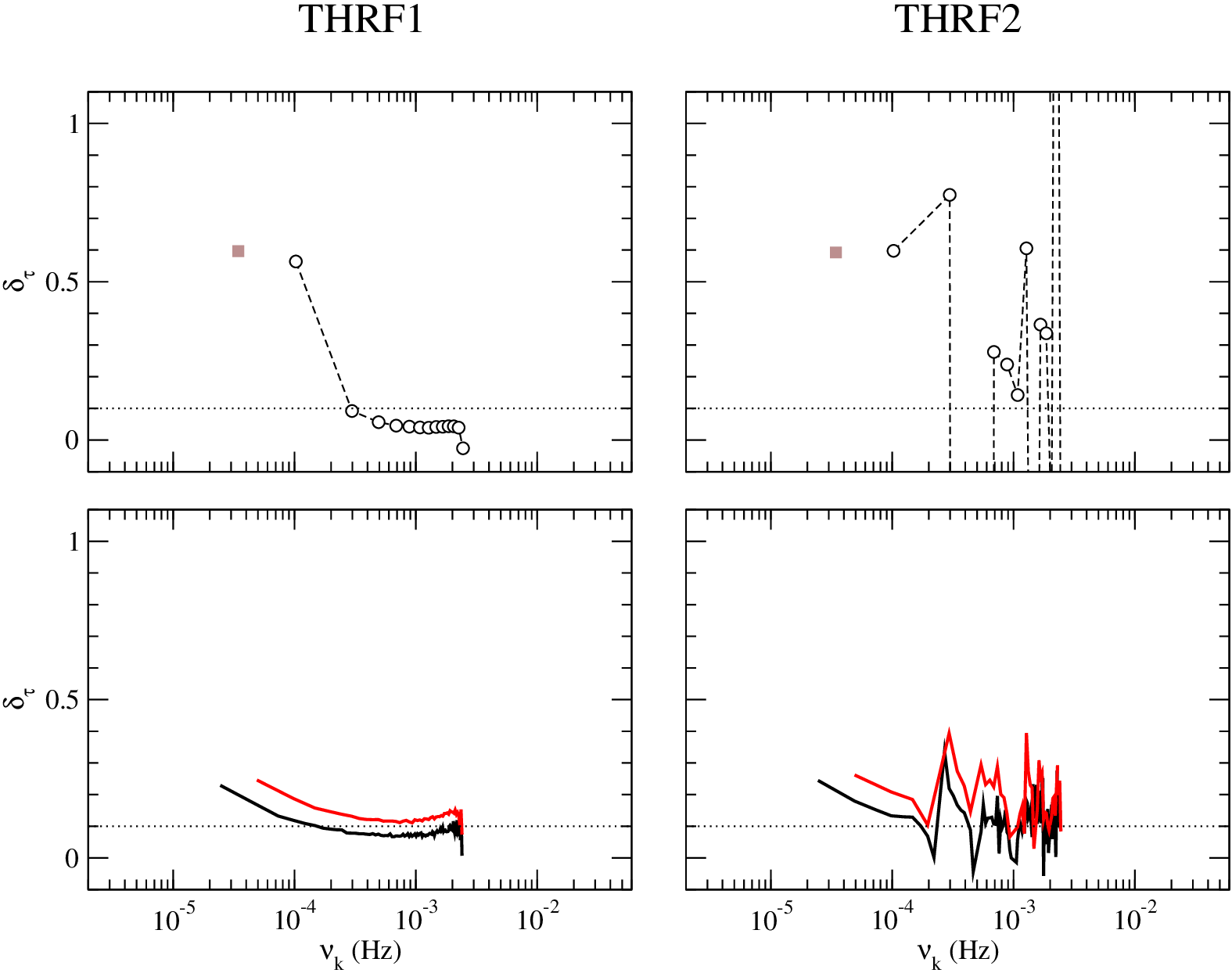}
\caption{As in Fig.\,\ref{figd5}, for experiments THRF1 (first column) and THRF1 (second column).}
\label{figd8}
\end{figure*}

\begin{figure*}
\centering
\includegraphics[width=14cm,clip]{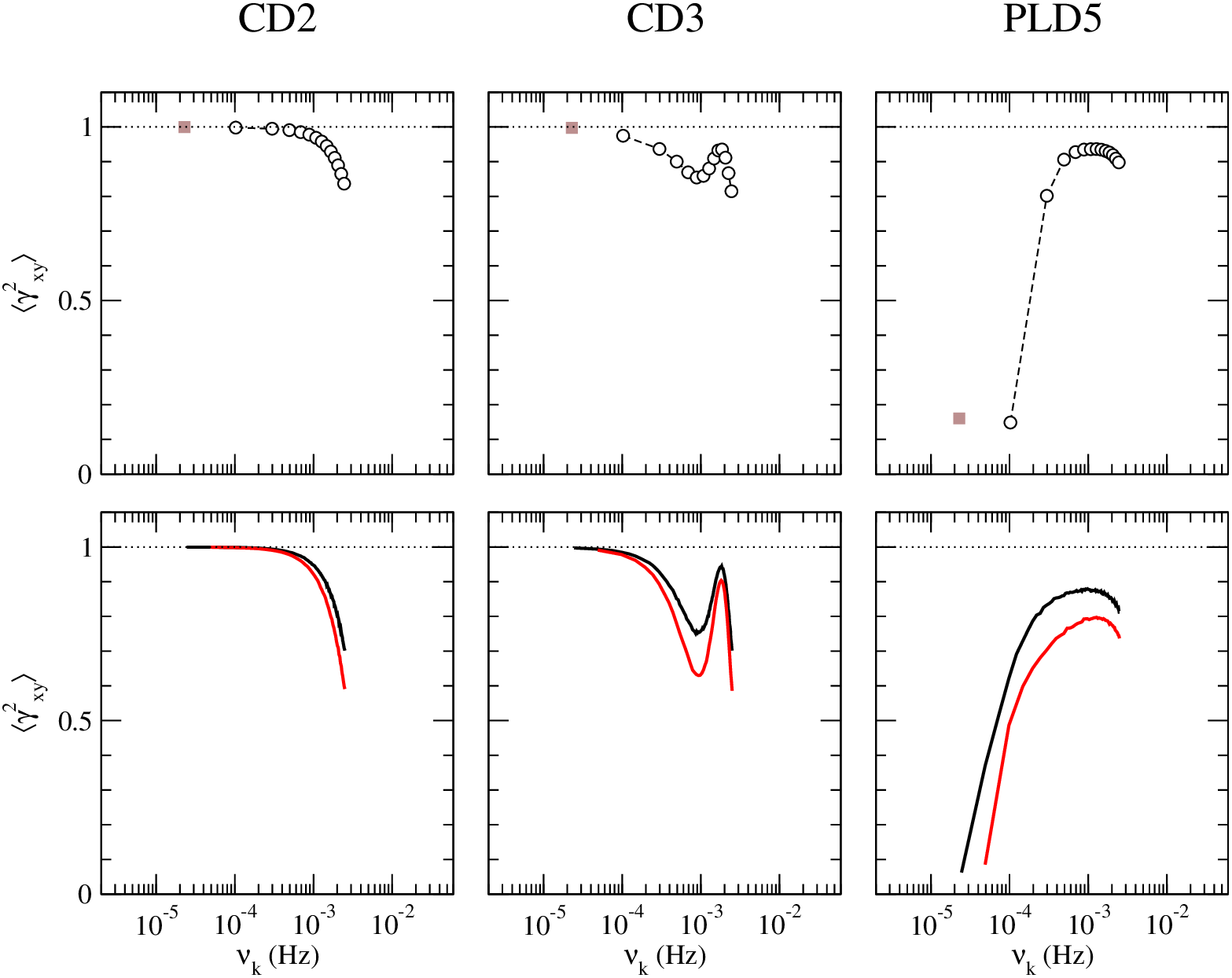}
\caption{Mean sample coherence for experiments CD2 (first column), CD3 (second column), and PLD5 (third column). Top row: Mean sample coherence for the LS102.4 light curves with a smoothing parameter $m=20$ (open black circles), and OB light curves with a smoothing parameter $m=13$ (filled brown squares). Bottom row: Mean sample coherence for the LS40.8 (black curves) and LS20.4 (red curves) light curves estimated from $m=20$ segments.}
\label{figd9}
\end{figure*}

\begin{figure*}
\centering
\includegraphics[width=14cm,clip]{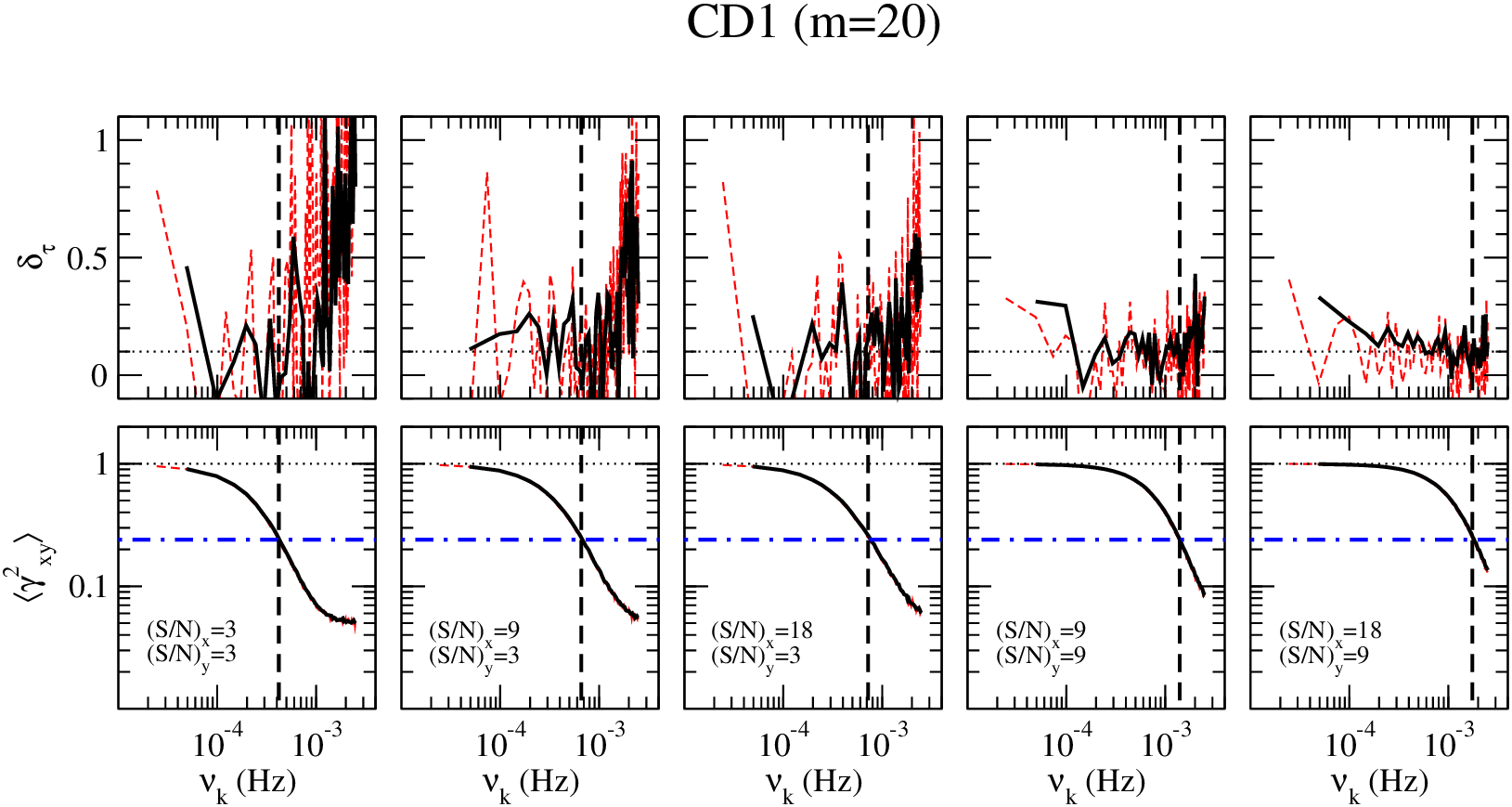}
\caption{Top row: Relative time-lag bias for the LS20.4 (continuous black curves) and LS40.8 (dashed red curves) light curves in experiment CD1. Bottom row: Mean sample coherence for the LS20.4 (continuous black curves) and LS40.8 (dashed red curves) light curves in experiment CD1. Different columns correspond to different $\{\mathrm{(S/N)}_x,\mathrm{(S/N)}_y\}$ combinations, while the estimates were determined from $m=20$ segments in each case. The vertical dashed lines in all panels indicate $\nu_{\rm{crit}}$, while the blue dotted-dashed lines in the lower panels indicate the $1.2/(1+0.2m)$ mean sample coherence value (see Sects.\,\ref{sec8} and \ref{sec9} for details).}
\label{figd10}
\end{figure*}

\begin{figure*}
\centering
\includegraphics[width=14cm,clip]{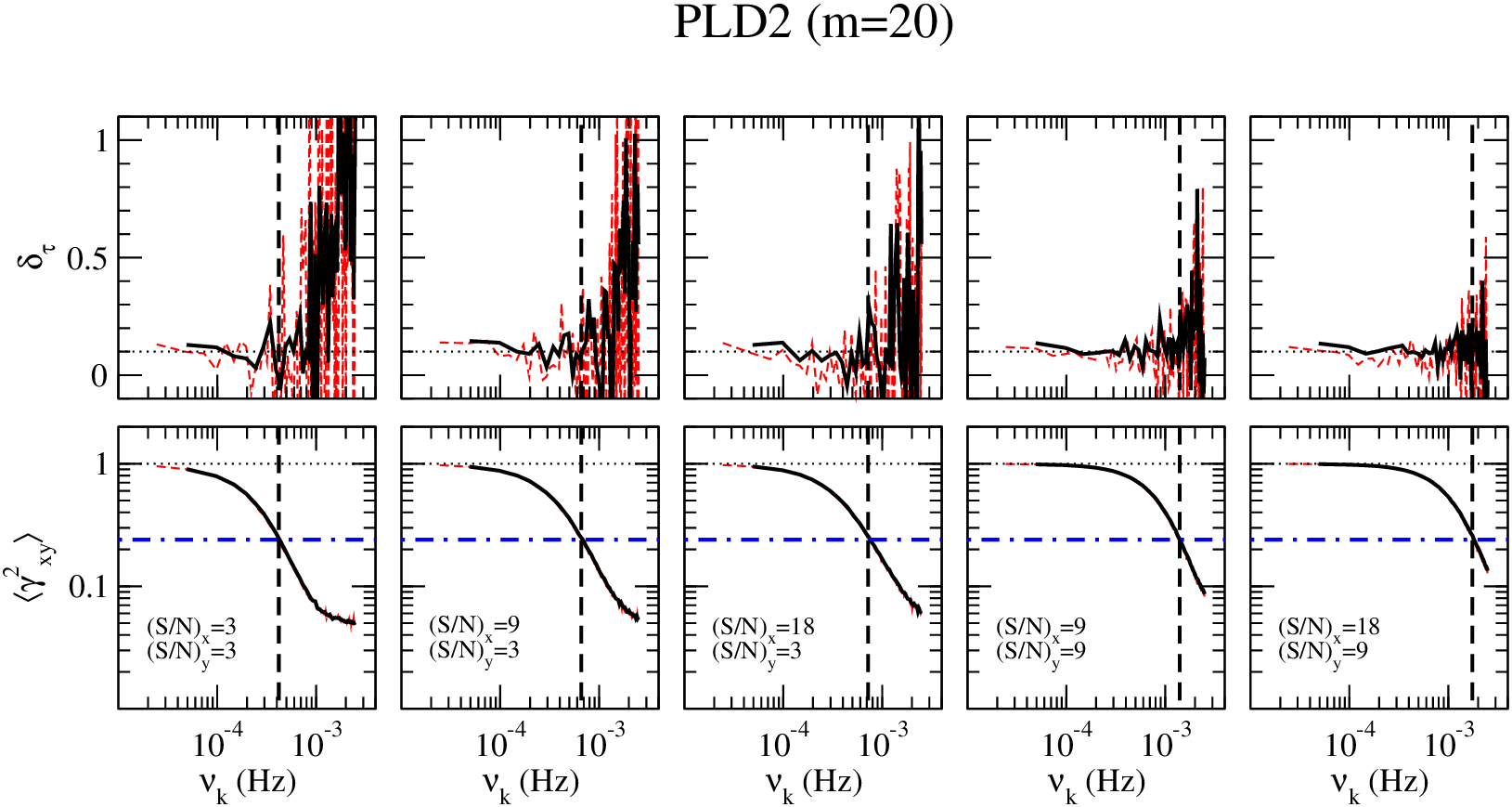}
\caption{As in Fig.\,\ref{figd10}, for experiment PLD2.}
\label{figd11}
\end{figure*}

\begin{figure*}
\centering
\includegraphics[width=14cm,clip]{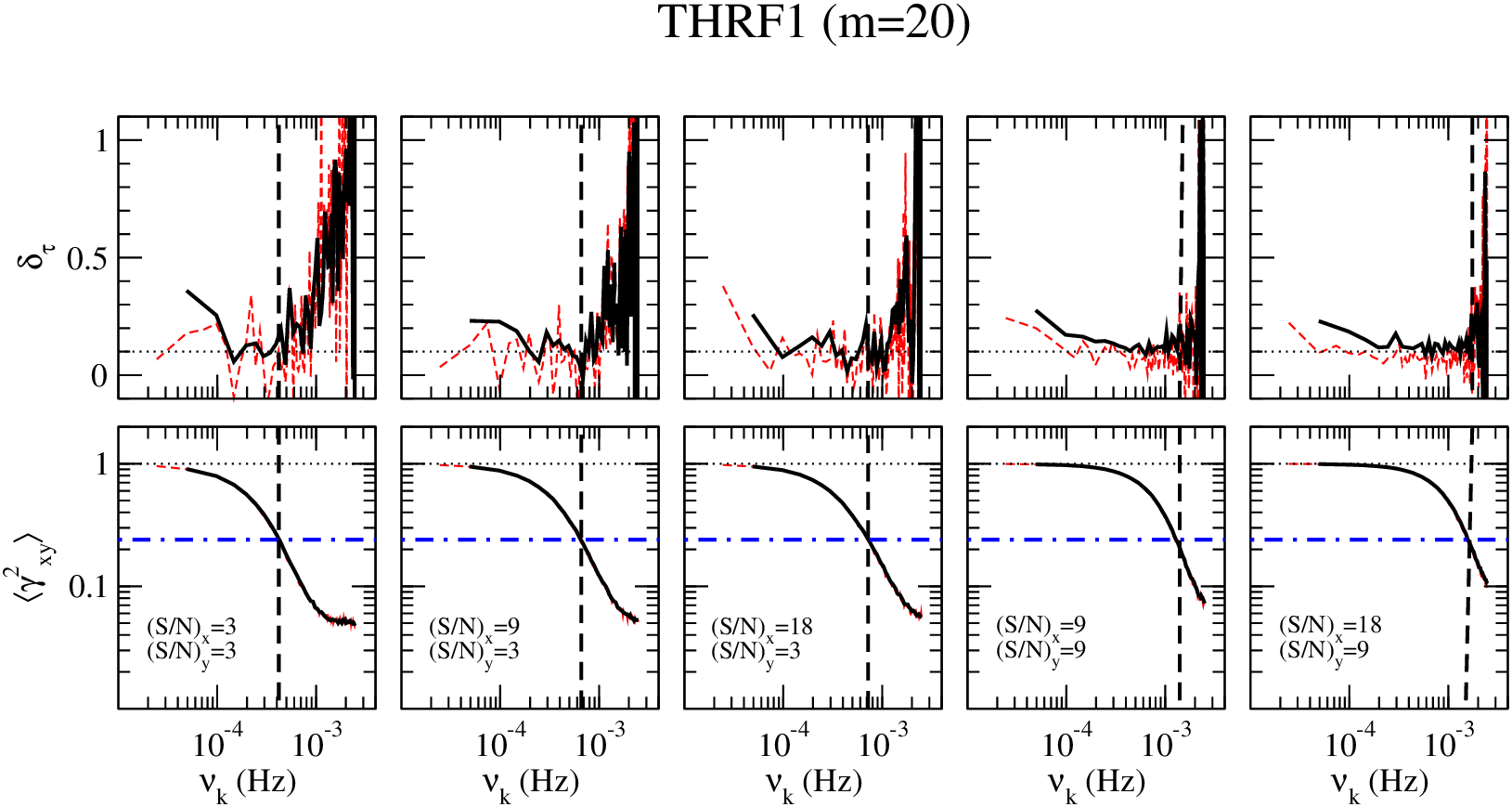}
\caption{As in Fig.\,\ref{figd10}, for experiment THRF1.}
\label{figd12}
\end{figure*}

\begin{figure*}
\centering
\includegraphics[width=14cm,clip]{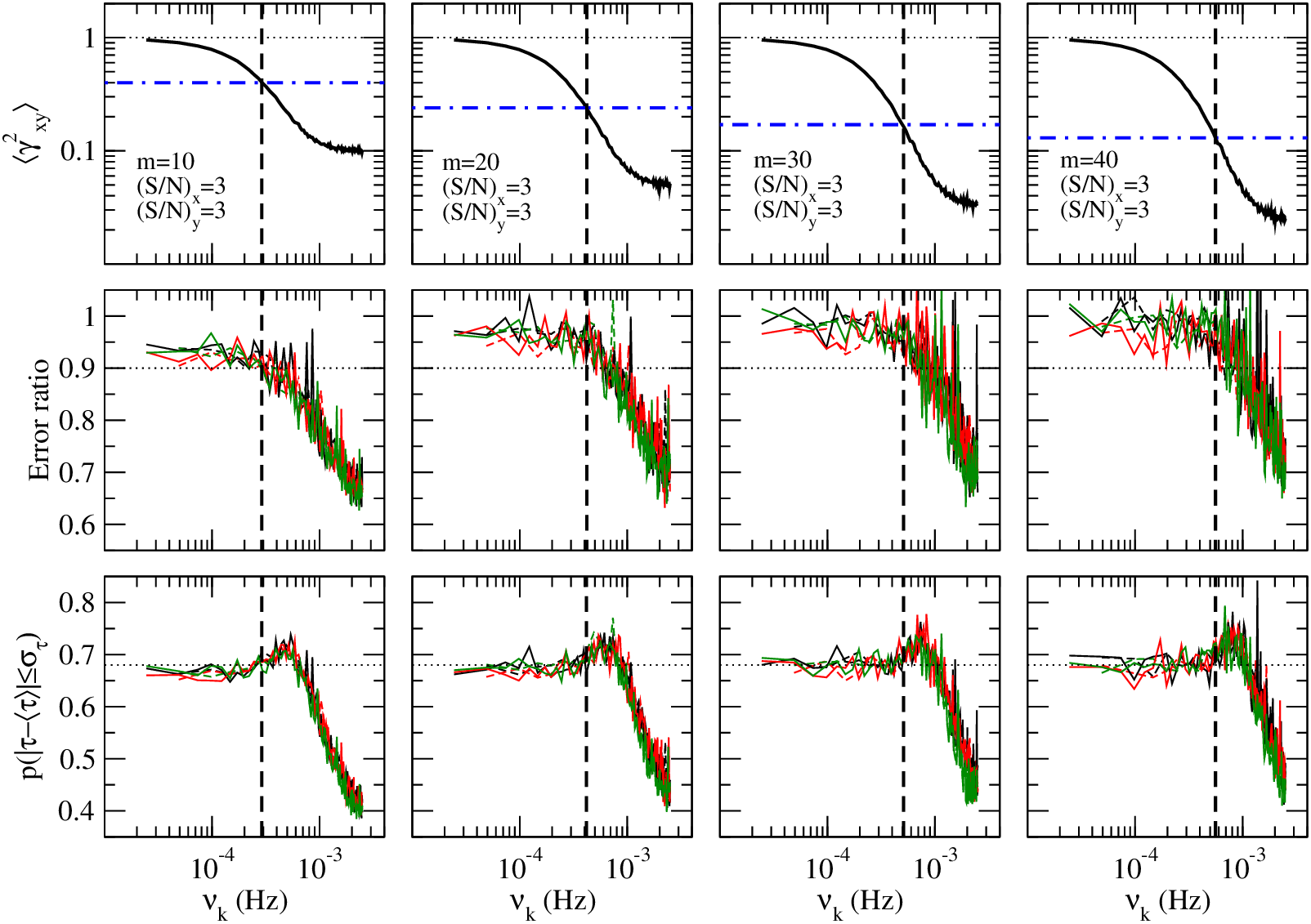}
\caption{The mean sample coherence, time-lag error ratio and probability that the value of the time-lag estimate is within $1\sigma$ of the sample mean (top, middle, and bottom row, respectively) for different values of $m$ and light curve types. In each case, $\mathrm{(S/N)}_x=\mathrm{(S/N)}_y=3$. The vertical dashed lines indicate $\nu_{\mathrm{crit}}$ (see \S\ref{sec8} for details).}
\label{figd13}
\end{figure*}

\begin{figure*}
\centering
\includegraphics[width=14cm,clip]{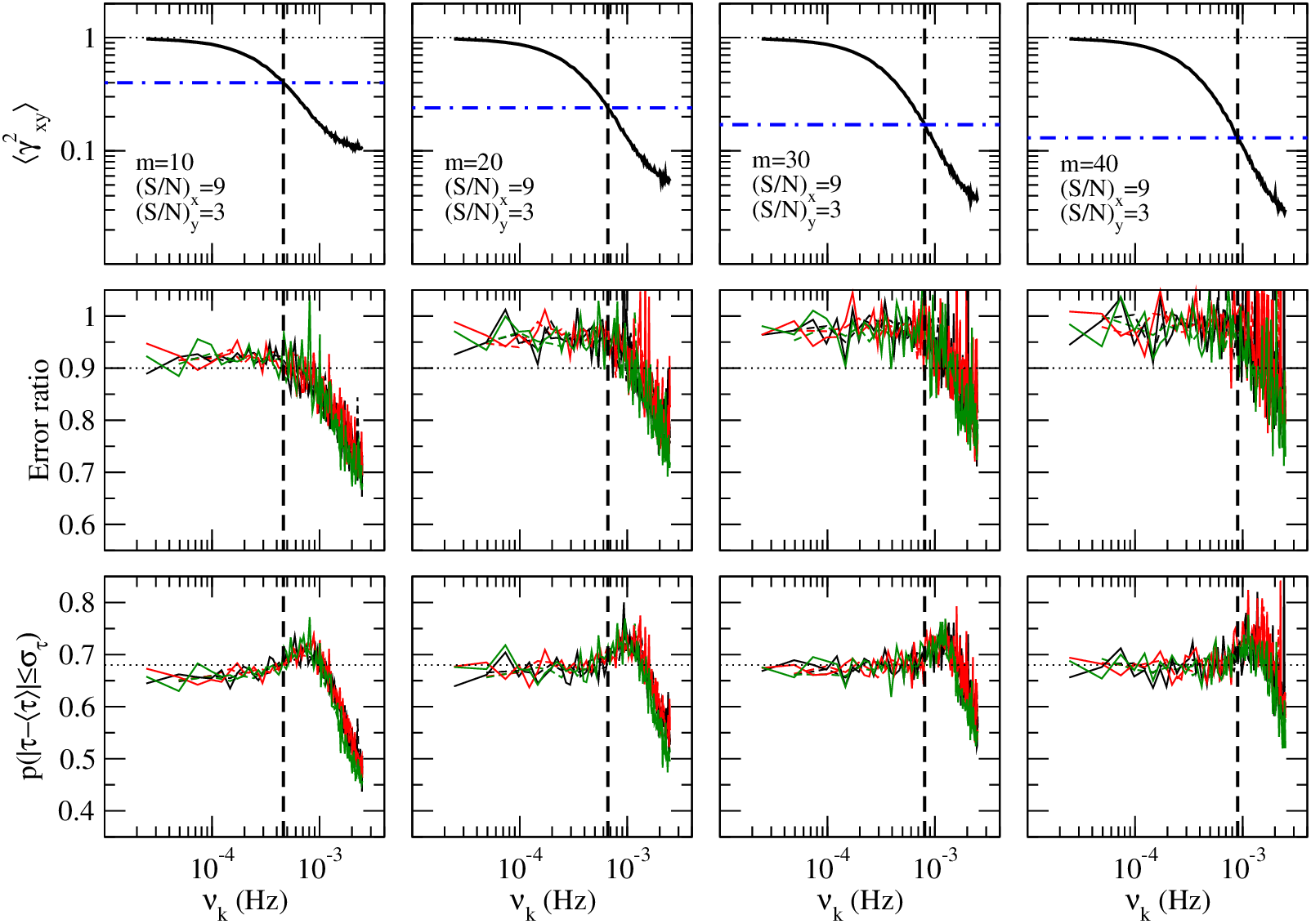}
\caption{As in Fig.\,\ref{figd13}, for $\mathrm{(S/N)}_x=9$ and $\mathrm{(S/N)}_y=3$.}
\label{figd14}
\end{figure*}

\begin{figure*}
\centering
\includegraphics[width=14cm,clip]{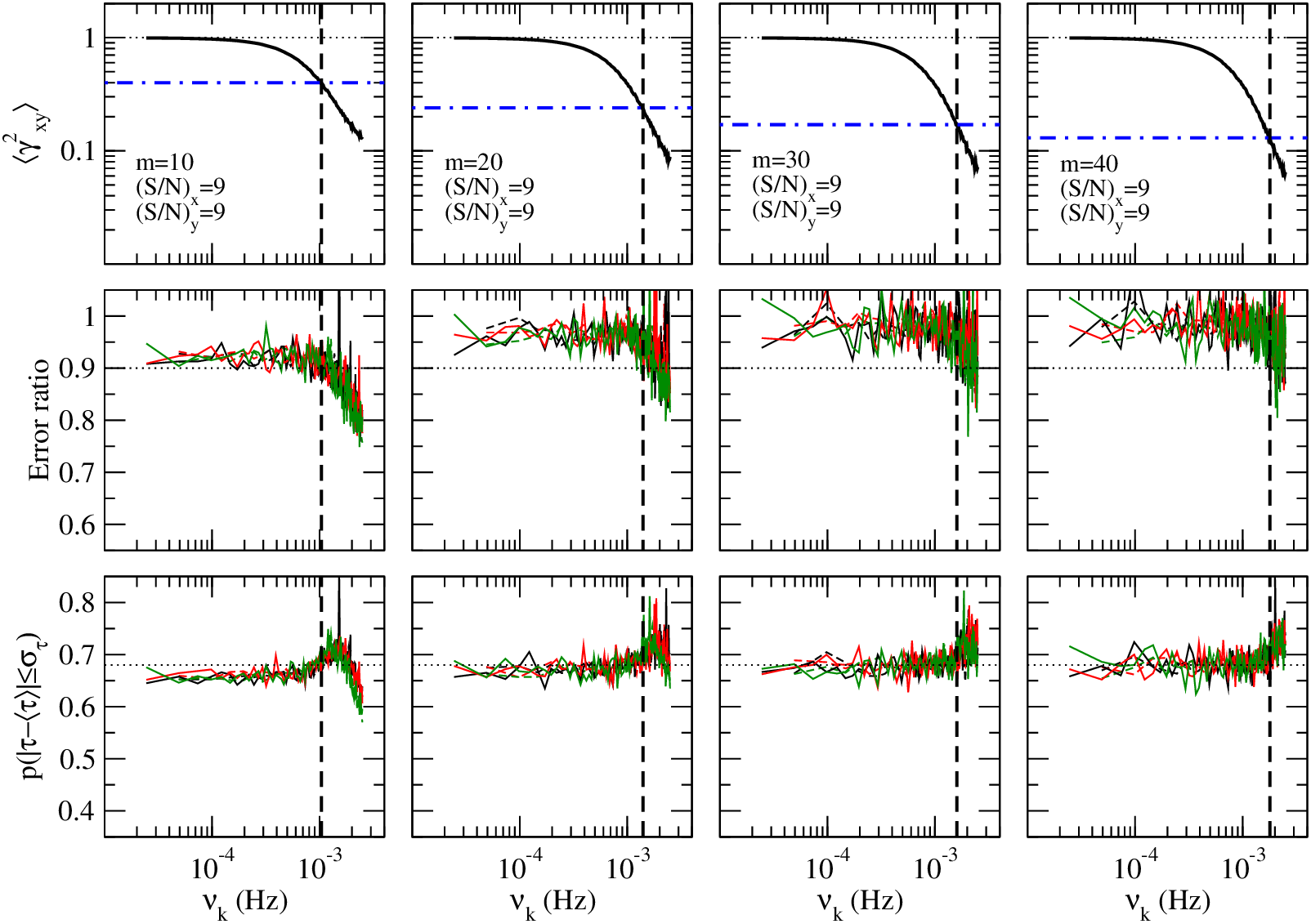}
\caption{As in Fig.\,\ref{figd13}, for $\mathrm{(S/N)}_x=9$ and $\mathrm{(S/N)}_y=9$.}
\label{figd15}
\end{figure*}

\begin{figure*}
\centering
\includegraphics[width=14cm,clip]{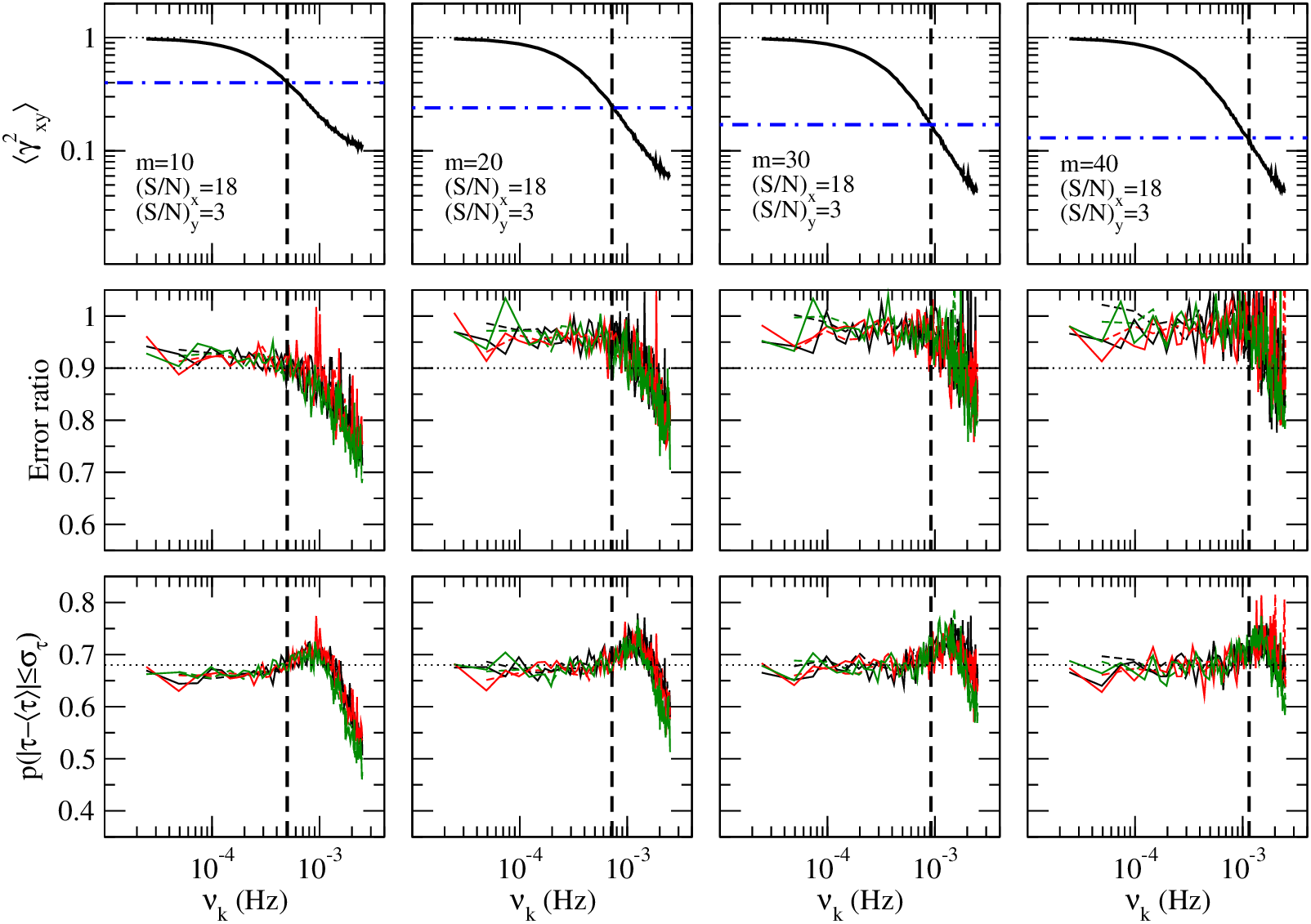}
\caption{As in Fig.\,\ref{figd13}, for $\mathrm{(S/N)}_x=18$ and $\mathrm{(S/N)}_y=3$.}
\label{figd16}
\end{figure*}

\begin{figure*}
\centering
\includegraphics[width=14cm,clip]{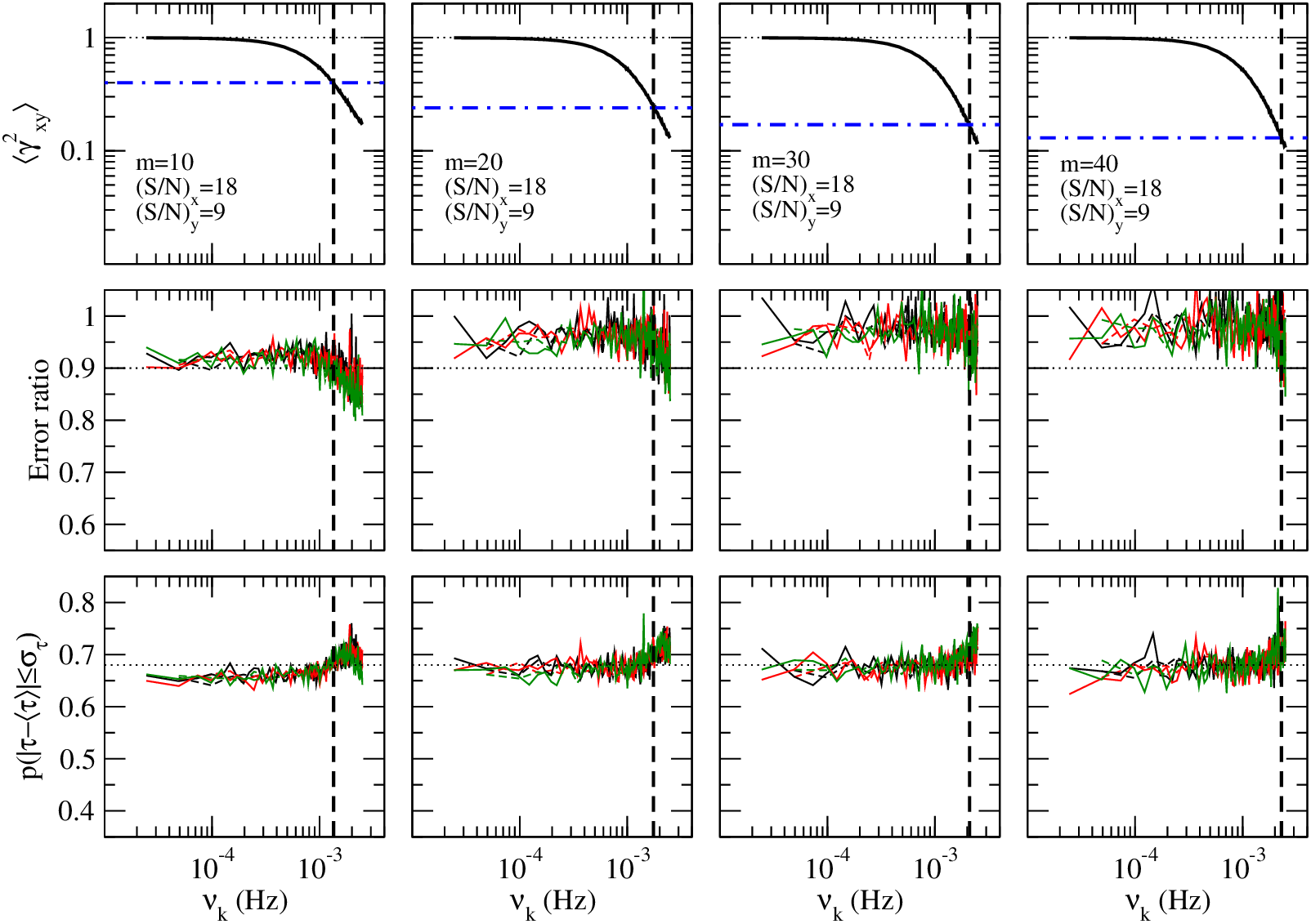}
\caption{As in Fig.\,\ref{figd13}, for $\mathrm{(S/N)}_x=18$ and $\mathrm{(S/N)}_y=9$.}
\label{figd17}
\end{figure*}

\begin{figure*}
\centering
\includegraphics[width=14cm,clip]{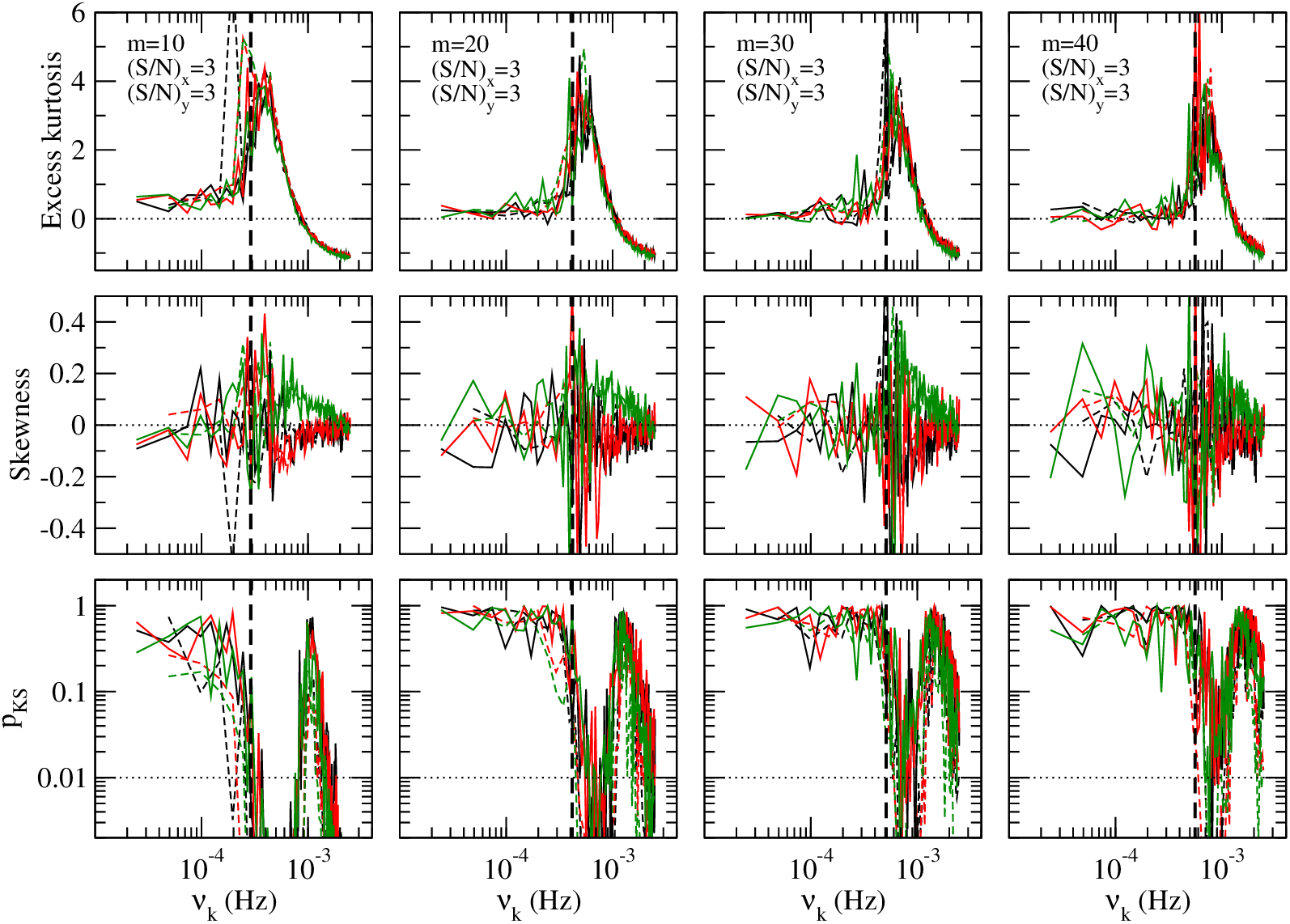}
\caption{The excess kurtosis, skewness and probability that the averaged time-lag estimate is Gaussian-distributed according to the KS test (top, middle, and bottom row, respectively) for different values of $m$ and light curve types. In each case, $\mathrm{(S/N)}_x=\mathrm{(S/N)}_y=3$. The vertical dashed lines indicate $\nu_{\mathrm{crit}}$ (see Sect.\,\ref{sec9} for details).}
\label{figd18}
\end{figure*}

\begin{figure*}
\centering
\includegraphics[width=14cm,clip]{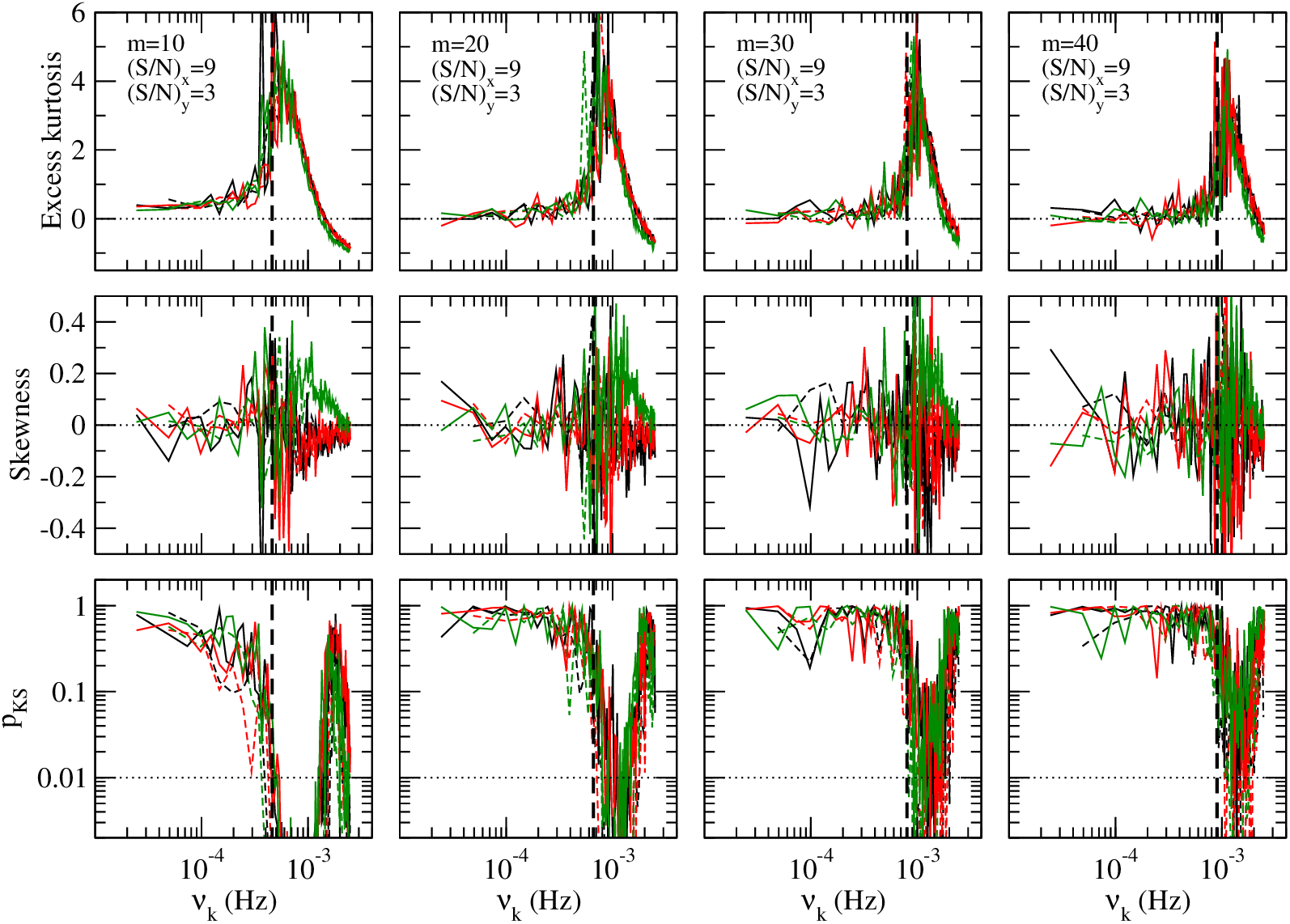}
\caption{As in Fig.\,\ref{figd18}, for $\mathrm{(S/N)}_x=9$ and $\mathrm{(S/N)}_y=3$.}
\label{figd19}
\end{figure*}

\begin{figure*}
\centering
\includegraphics[width=14cm,clip]{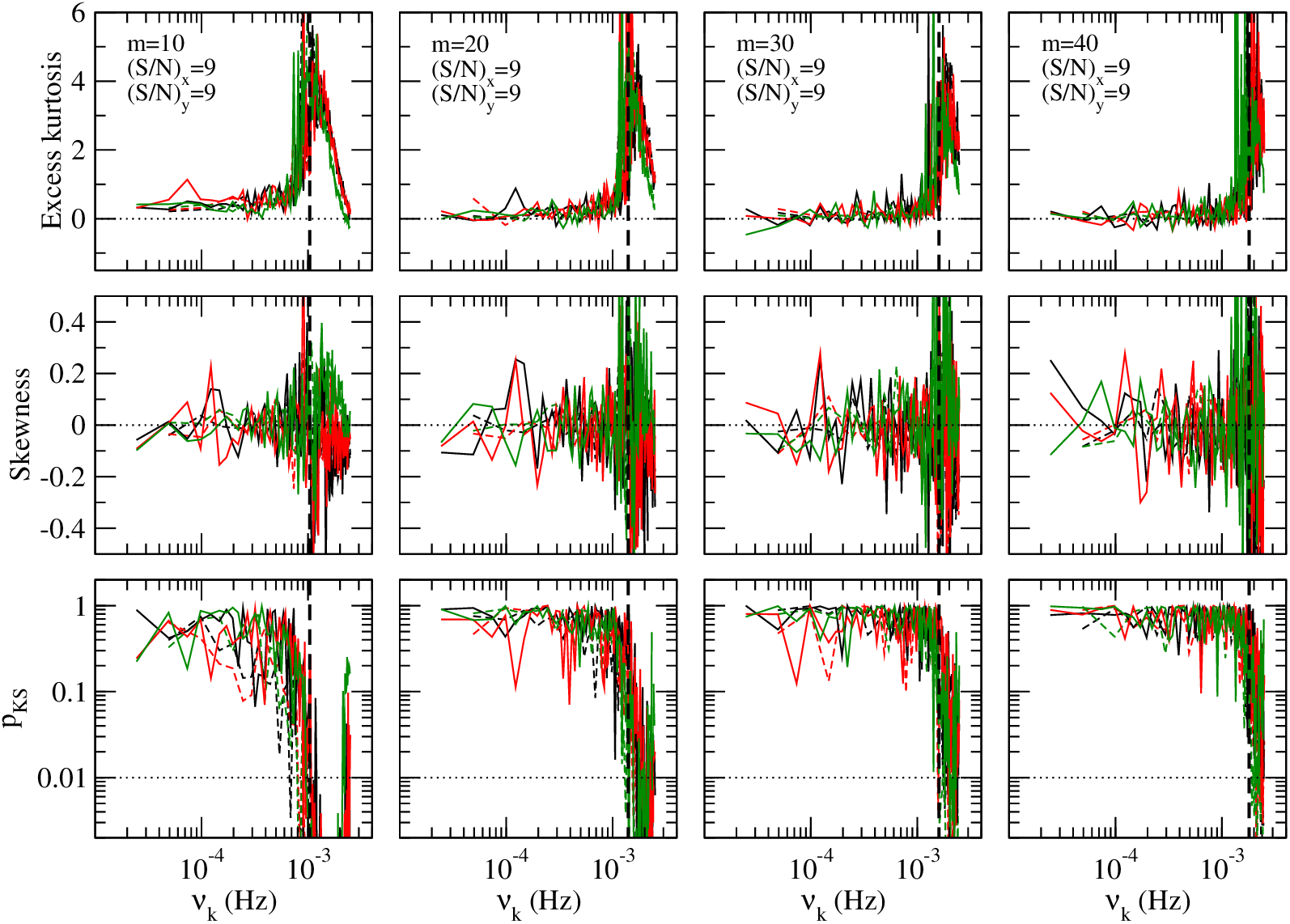}
\caption{As in Fig.\,\ref{figd18}, for $\mathrm{(S/N)}_x=9$ and $\mathrm{(S/N)}_y=9$.}
\label{figd20}
\end{figure*}

\begin{figure*}
\centering
\includegraphics[width=14cm,clip]{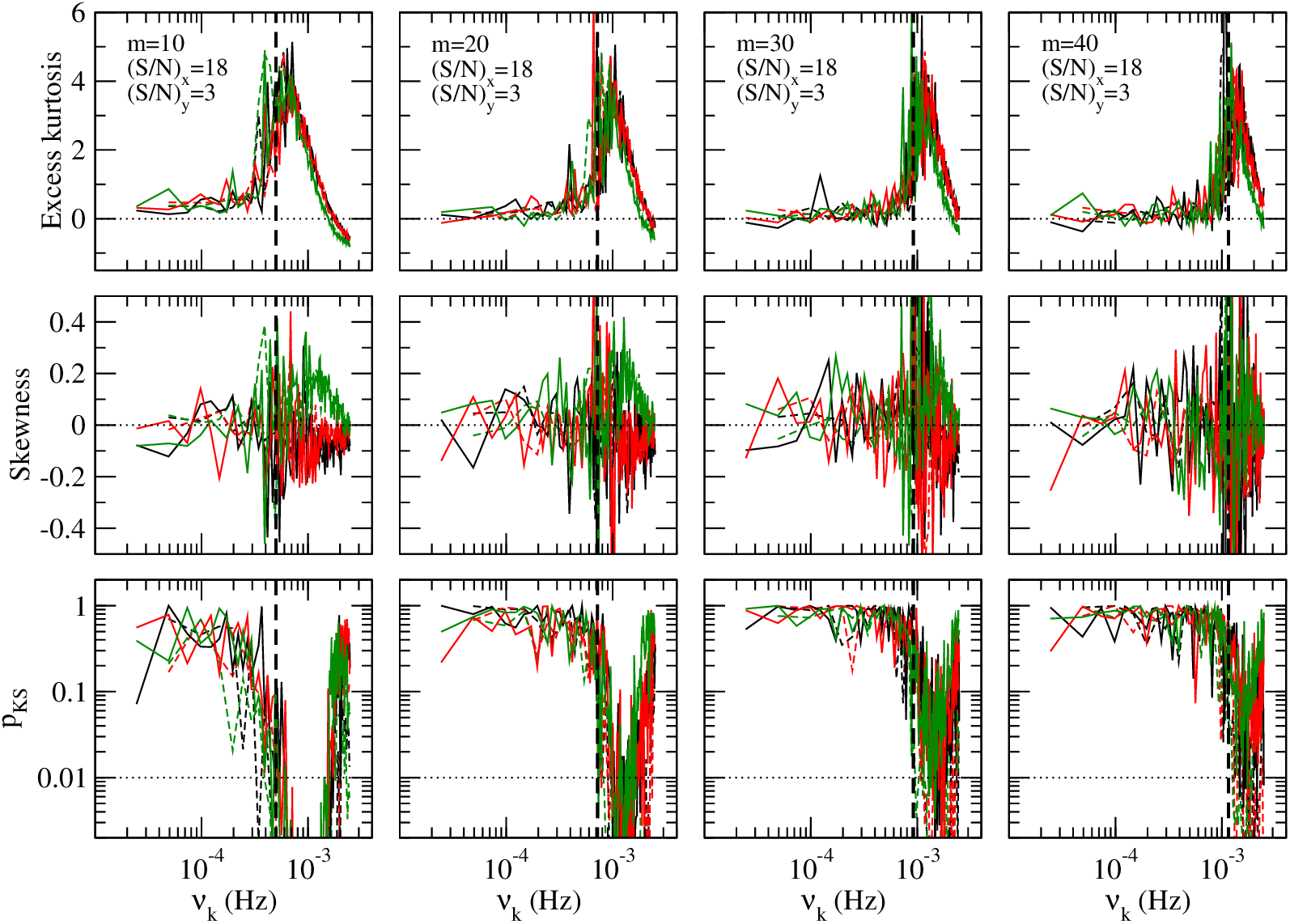}
\caption{As in Fig.\,\ref{figd18}, for $\mathrm{(S/N)}_x=18$ and $\mathrm{(S/N)}_y=3$.}
\label{figd21}
\end{figure*}

\begin{figure*}
\centering
\includegraphics[width=14cm,clip]{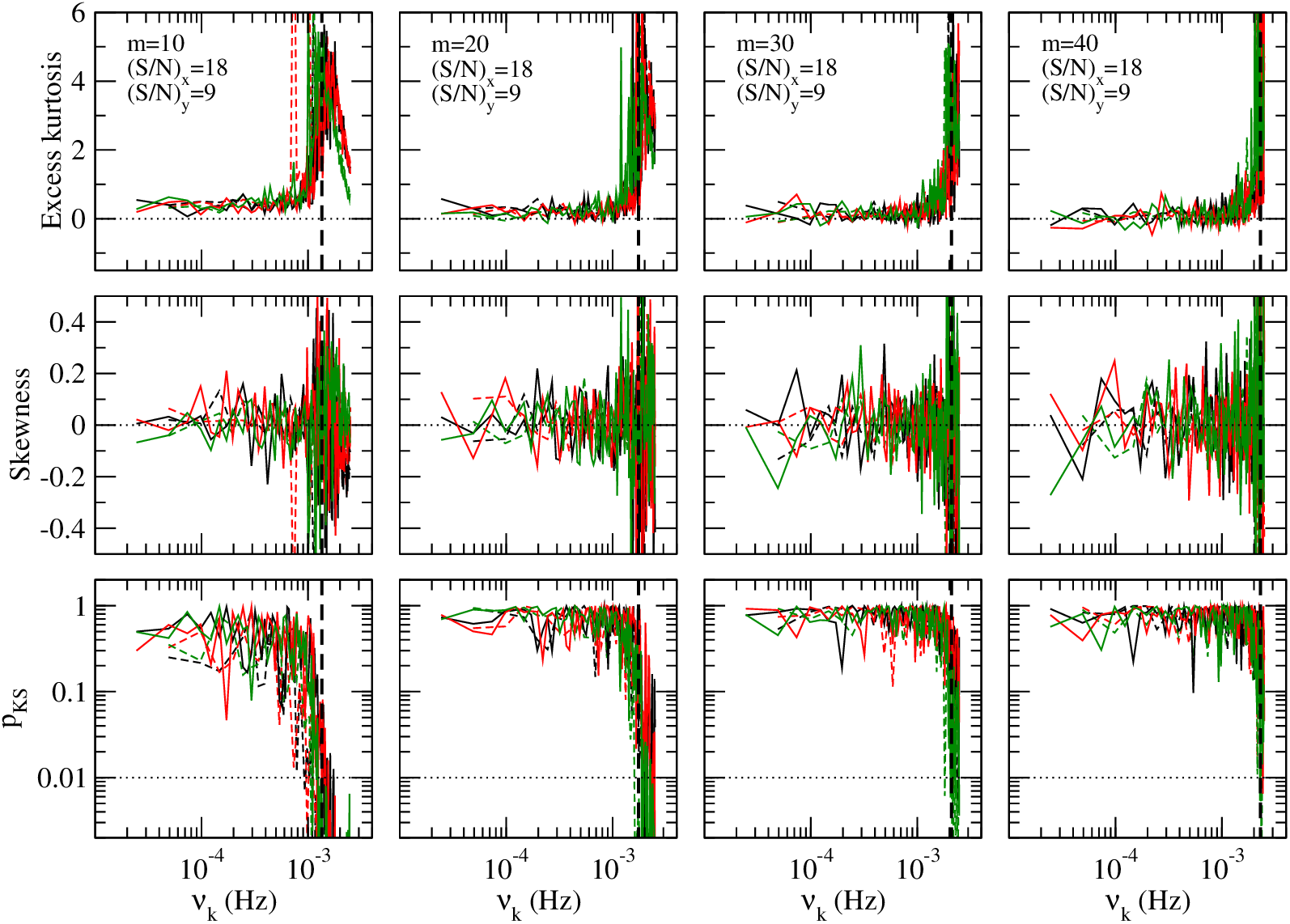}
\caption{As in Fig.\,\ref{figd18}, for $\mathrm{(S/N)}_x=18$ and $\mathrm{(S/N)}_y=9$.}
\label{figd22}
\end{figure*}

\end{appendix}


\end{document}